\documentclass[aps,prx,twocolumn]{revtex4-2}
\usepackage[T1]{fontenc}
\usepackage[latin1]{inputenc}
\usepackage{lmodern}
\expandafter\let\csname equation*\endcsname\relax
\expandafter\let\csname endequation*\endcsname\relax
\usepackage{amsmath}
\usepackage{esint}
\usepackage{amssymb}
\usepackage{graphicx}
\usepackage{xcolor}
\usepackage{natbib}
\usepackage{bm}
\usepackage{bbold}
\usepackage{braket}
\usepackage[mathscr]{euscript}
\usepackage[cal=boondoxo]{mathalfa}
\usepackage{comment}
\usepackage[normalem]{ulem}
\def\nnabla{\boldsymbol{\nabla}} 
\def\ii{{\rm i}}  
\def\GG{{\bf G}}

\def\db{\boldsymbol{\wp}}  
\def\dbu{\hat{\boldsymbol{\wp}}}  
\def\Eb{{\bf E}}  
\def\rb{{\bf r}}

\def\ub{{\bf u}}

\def\hge{\hat{\sigma}_{ge}}  
\def\heg{\hat{\sigma}_{eg}}

\def\kg{k_{\rm 1D}}
\def\ga{\Gamma^{\rm q}_{\rm 1D}}
\def\gap{\Gamma'^{\rm q}}

\def\bra#1{\mathinner{\langle{#1}|}}
\def\ket#1{\mathinner{|{#1}\rangle}}

\def\Rb{\textbf{R}}
\def\ub{\textbf{u}}
\def\vb{\textbf{v}}

\def\Li{\textrm{Li}}

\def\kp{k_z}
\def\kpe{k_{\perp}}

\def\pM{\mathrel{\raise 2pt \hbox{\tiny(}\!\raise 1pt \hbox{+}\settowidth {\dimen03} {+}\hskip-\dimen03 \raise -2.4pt \hbox {$-$} \!\raise 2pt \hbox{\tiny)}}}

\begin{document}
\title{Atomic-waveguide quantum electrodynamics}
\author{Stuart J. Masson}
\email{s.j.masson@columbia.edu}
\affiliation{Department of Physics, Columbia University, New York, NY 10027, USA}
\author{Ana Asenjo-Garcia}
\email{ana.asenjo@columbia.edu}
\affiliation{Department of Physics, Columbia University, New York, NY 10027, USA}

\date{\today}
\begin{abstract}
Atom arrays are a new type of quantum light-matter interface. Here, we propose to employ one-dimensional ordered arrays as atomic waveguides. These arrays support optical guided modes that do not decay into free space. We show that these modes can be harnessed to mediate tunable, long-range interactions between additional ``impurity qubits'' coupled to the chain, without need for photonic structures. The efficient coupling between qubits and atomic waveguides enables the realization of tunable qubit-qubit interactions, which can be short- or long-range, dissipative or coherent, as well as chiral. Moreover, owing to the two-level nature of atoms, these waveguides are intrinsically quantum. In contrast to classical waveguides, where photons do not interact with each other, atomic waveguides display strong non-linearities, which create a tunable dissipative channel for qubit-qubit interactions, and opens the door to the exploration of many-body physics between guided photons. This physics is universal as it only relies on photon interference and can also be observed with other types of quantum emitters, such as those in molecular or solid-state systems.

\end{abstract}
\maketitle

\section{Introduction}

The realization of efficient interactions between photons and atoms is a central challenge in quantum optics. Besides enabling the exploration of exotic many-body physics~\cite{Noh17}, they are also a critical resource to develop practical implementations of quantum information protocols~\cite{Fleischhauer05,Hammerer10}. Deterministic light-matter interactions also form the underpinnings of quantum non-linear optics at the single photon level~\cite{Peyronel12,Chang14}, as well as of metrology and sensing applications~\cite{Ma11,Pezze18}.

To control and enhance the interactions between light and atoms, it is generally believed they must be interfaced with nanophotonic structures. This has propelled the development of the field of cavity quantum electrodynamics (QED) and, more recently, of waveguide QED, where atoms are coupled to one-dimensional (1D) photonic reservoirs, such as fibers~\cite{LeKien05,Vetsch10,Goban12,Gouraud15} and photonic crystal waveguides~\cite{Thompson13,Goban14,Goban15,Hood16}. {Waveguide QED offers efficient light-matter coupling as photons are confined in small volumes and can be almost-deterministically exchanged between distant atoms.}

{Coupling atoms to 1D dielectric structures enables the exploration of fundamentally different paradigms within quantum optics~\cite{Chang18}, without a free-space analog. For instance, engineering of the dispersion relation of the optical modes allows for the generation of bandgaps, frequency regions where photons cannot propagate and become bound. This gives rise to ``slow light'' and localized photonic states~\cite{John90,John91,Sundaresan19}, and enables the realization of almost-arbitrary interactions between atoms~\cite{John96,Douglas15,Chang18}. The intrinsic helicity of the near field leads to directional decay, in the form of chiral light-matter interactions~\cite{Petersen14,Sollner15}. Waveguide QED offers tantalizing possibilities for quantum information storage and processing, such as the realization of quantum memories~\cite{Gorshkov07,Asenjo17PRX} and gates~\cite{Dzsotjan10,Zheng13}, as well as for the preparation of entangled states between distant atoms assisted by collective dissipation~\cite{Stannigel12NJP,GonzalezTudela13,GonzalezTudela15PRL}.} However, {deterministic} interfacing of quantum emitters with nanophotonics in a scalable manner has proven to be technically difficult.

\begin{figure}[b]
\centerline{\includegraphics[width=.9\linewidth]{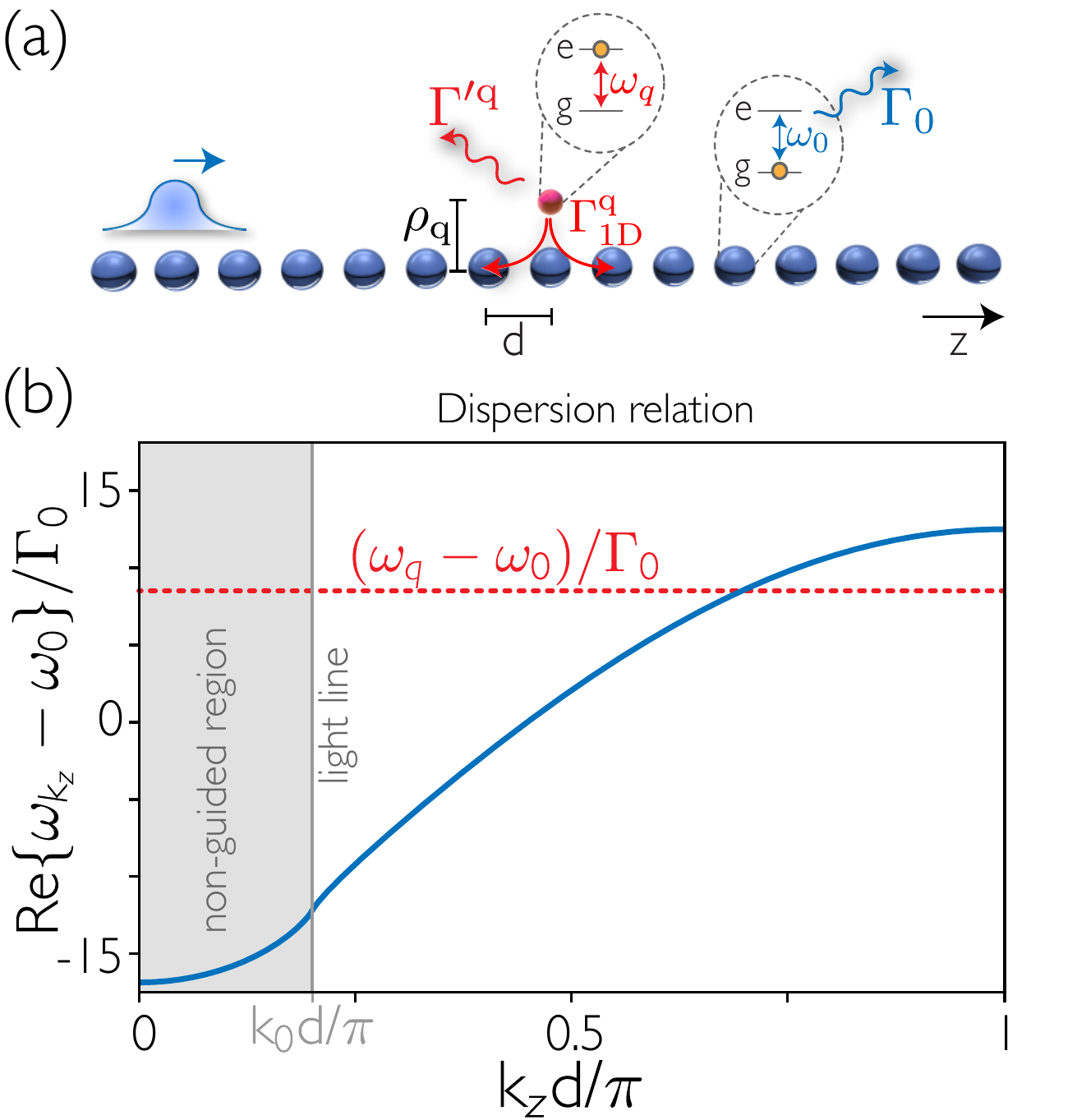}}
\caption{\textbf{A 1D atomic array behaves as a quantum waveguide.} \textbf{(a)} Schematic of the setup under consideration: an ``impurity qubit'' (red) of resonace frequency $\omega_{\rm q}$ is located in the vicinity of a 1D chain of atoms of resonance frequency $\omega_0$ and spontaneous emission rate $\Gamma_0$ (blue), at a distance $\rho_{\rm q}$ from the axis of the array. If the distance $d$ between the array atoms is smaller than $\lambda_0/2$, the chain behaves as a waveguide and supports guided modes that do not decay into free space. The qubit emission rates into the atomic-waveguide mode and into free space are $\ga$ and $\gap$, respectively. \textbf{(b)} Dispersion relation for the single-excitation mode of an infinite, 1D chain of atoms polarized parallel to the chain axis with spacing $d=0.1\lambda_0$. In the region enclosed within the light line (shaded), the chain does not guide light and the mode decays into free space. Beyond the light line ($k_z>k_0=\omega_0/c$), the mode is guided.}\label{Fig1}
\end{figure} 

Here, we suggest an alternative approach: to employ atomic arrays as 1D waveguides to mediate interactions between distant ``impurity qubits''. In ordered arrays, interference in photon emission leads to the emergence of subradiant states, which cannot decay into free space{~\cite{Zoubi10,Bettles15,Bettles16PRA,Asenjo17PRX,Needham19,Kornovan19}}. These states can be understood as guided modes of the atomic chain{~\cite{Zoubi10,Asenjo17PRX,Needham19}}, and can be used to mediate both coherent and dissipative interactions between qubits that are coupled to the atomic waveguide. {The qubit-waveguide interaction can be chiral, and qubit-qubit interactions can be strongly time-delayed even over short physical distances. The coupling between these qubits and the waveguide can be remarkably efficient. Atomic waveguides are an intrinsically quantum reservoir, as a single atom cannot be excited twice. This tunable non-linearity sets a fundamental difference between qubit interactions mediated by an atomic waveguide and those in traditional nanophotonic structures. We demonstrate that qubit-qubit interaction can be primarily unperturbed or strongly damped by the presence of multiple excitations in the chain. Our results provide a comprehensive picture of a new paradigm for light-matter interactions: atomic-waveguide QED.}

\section{Guided photons in atomic chains}

Ordered arrays of atoms support guided photons (in the form of polaritonic spin waves) that do not scatter light into free space~{\cite{Zoubi10,Asenjo17PRX,Needham19,Asenjo19}}. {In this section, we review the physics of guided modes in chains and describe their dispersion relation, setting up the stage for main idea of the manuscript: to harness these states to mediate interactions between coupled impurity qubits. We employ a ``spin model'' to describe dipolar interactions between atoms~\cite{Carmichael00,Clemens03,Bettles15,Bettles16PRL,Sutherland16,Bettles16PRA,Facchinetti16,Asenjo17PRA,Perczel17PRL,Asenjo17PRX,Henriet19,Needham19,MorenoCardoner19,Asenjo19,Kornovan19,Patti20arxiv,Holzinger20PRL,Masson20superradiance}.} We consider an array of  $N$ two-level atoms of resonance frequency $\omega_0$ separated by a distance $d$, as shown in Fig.~\ref{Fig1}(a).  We describe the atoms' dynamics employing a stochastic wavefunction approach~\cite{Carmichael93,QuantumNoiseBook}, where the atomic state evolves under the non-Hermitian Hamiltonian
\begin{equation}\label{hameff}
\mathcal{H}=\hbar\omega_0\sum_{i=1}^N\hat{\sigma}_{ee}^i+\hbar\sum_{i,j=1}^N \left(J^{ij}-\ii \frac{\Gamma^{ij}}{2}\right)\hat{\sigma}_{eg}^i\hat{\sigma}_{ge}^j,
\end{equation}
interrupted by {the action of stochastic} quantum jumps that lower the number of excitations and occur at random times. {Jump operators are found as the eigenstates of the dissipative interaction matrix $\mathbf{\Gamma}$ with elements $\Gamma^{ij}$~\cite{Carmichael00,Clemens03,Clemens04,Masson20superradiance}.} The coherent and dissipative dipolar interaction rates between atoms $i$ and $j$ read~{\cite{Stephen64,Lehmberg70,Carmichael00,Clemens03,Bettles15,Bettles16PRL,Sutherland16,Bettles16PRA,Facchinetti16,Asenjo17PRA,Perczel17PRL,Asenjo17PRX,Henriet19,Needham19,MorenoCardoner19,Asenjo19,Kornovan19,Patti20arxiv,Holzinger20PRL,Masson20superradiance}}
\begin{subequations}\label{rates}
\begin{equation}\label{shiftrate}
J^{ij}=-\frac{\mu_0\omega_0^2}{\hbar}\,\db^*\cdot\text{Re}\,\mathbf{G}_0(\rb_i,\rb_j,\omega_0)\cdot\db, 
\end{equation}
\begin{equation}\label{dissiprate}
\Gamma^{ij} =\frac{2\mu_0\,\omega_0^2}{\hbar}\,\db^*\cdot\text{Im}\,\mathbf{G}_0(\rb_i,\rb_j,\omega_0)\cdot\db,
\end{equation}
\end{subequations}
where $\db$ is the dipole matrix element associated with the atomic transition. The Green's tensor $\mathbf{G}_0(\rb_i,\rb_j,\omega_0)$ is the propagator of the electromagnetic field between atoms $i$ and $j$ in vacuum. It admits the closed expression
\begin{align}
\GG_0(\rb_i,\rb_j,\omega_0) =\frac{1}{4\pi} \left[\mathbb{1}+\frac{1}{k_0^2}\nabla\otimes\nabla\right]\frac{e^{\ii k_0 |\rb_i-\rb_j|}}{|\rb_i-\rb_j|}, \label{Greens_def}
\end{align}
where $k_0=\omega_0/c$. For a single atom, the spontaneous emission rate is $\Gamma_0=(2\mu_0\,\omega_0^2/\hbar)\,\db^*\cdot\text{Im}\,\mathbf{G}_0(\rb_i,\rb_i,\omega_0)\cdot\db=\omega_0^3|\db|^2/3\pi\epsilon_0\hbar c^3$, and the local frequency shift simply renormalizes the resonance frequency $\omega_0$. The Hamiltonian of Eq.~\eqref{hameff} only contains spin degrees of freedom (i.e., the atomic coherence operators $\hge^i=\ket{g_i}\bra{e_i}$ between the ground and excited states, and the population operator $\hat{\sigma}_{ee}^i=\ket{e_i}\bra{e_i}$). In the presence of a driving field of frequency $\omega$, the equations are identical, but with the prescription $\omega_0\rightarrow \omega$.

The non-Hermitian Hamiltonian of Eq.~\eqref{hameff} is that of an open, long-range \textit{XY} model, and is derived within the Born-Markov approximation{~\cite{Gruner96,Dung02}}. This approximation allows for integrating out the electromagnetic degrees of freedom and requires two conditions. First, the spectral response of the reservoir is flat compared to that of the atoms (such that the Green's function is evaluated at the atomic resonance frequency). Second, retardation can be ignored (such that the Hamiltonian is local in time). This approximation is valid in vacuum unless the separation between atoms is extremely large (of the order of a meter for typical optical transitions~\cite{Chang12,Shi15,Guimond16}). 

\begin{figure*}
\centerline{\includegraphics[width=\linewidth]{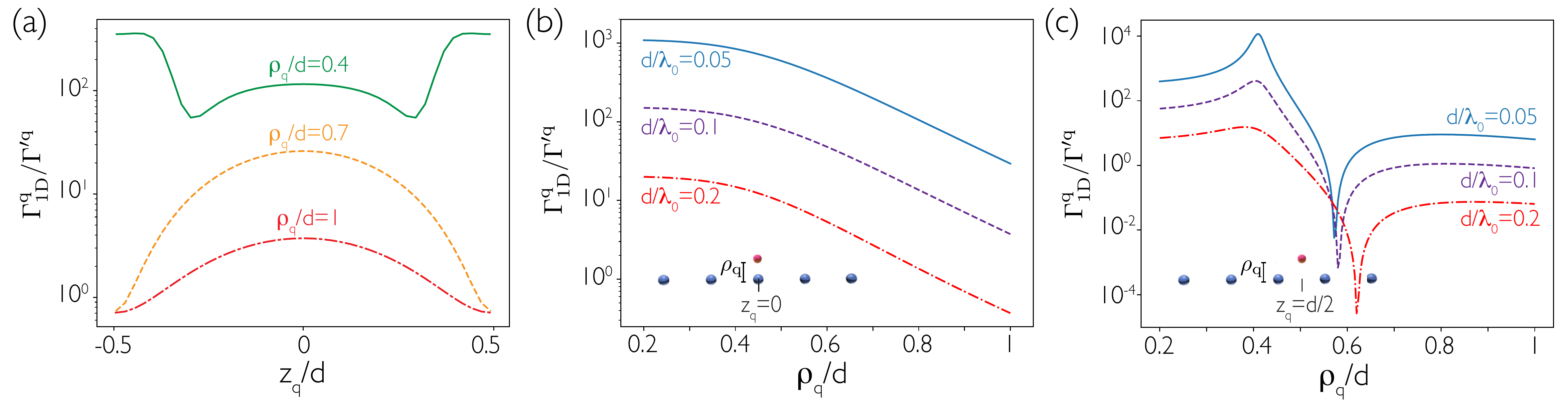}}
\caption{\textbf{Coupling strength between the impurity qubit and the waveguide.} The efficiency is represented by the ratio of the guided ($\ga$) vs non-guided ($\Gamma'^{\rm q}$) qubit decay rates as a function of the qubit position. \textbf{(a)} Bloch modulation along one unit cell of the chain ($z_{\rm q}=0$ is in line with one waveguide atom), for different radial distances $\rho_{\rm q}$, with $d=0.1\lambda_0$. \textbf{(b, c)} Scaling with the radial offset from the chain axis, for a qubit located (b) on top of a waveguide atom at $z_{\rm q}=0$, and (c) in between two waveguide atoms, at $z_{\rm q}=d/2$. For all plots, the detuning between qubit and waveguide atoms is chosen such that the guided wave-vector is $\kg = 0.7\pi/d$ (as shown by the red line in Fig.~\ref{Fig1}).}\label{Fig2}
\end{figure*} 

In the single-excitation regime, guided modes emerge for inter-atomic distances $d<\lambda_0/2$, with $\lambda_0=2\pi c/\omega_0$~\cite{Asenjo17PRX,Needham19}. To demonstrate their waveguiding behavior, we analyze the physics of an infinite chain that extends along the $z$ direction. The eigenstates of the Hamiltonian are spin waves with well defined wave-vector $k_z \in [-\pi/d,\pi/d]$, and are generated by the action of the collective spin operator $\hat{S}^\dagger_{k_z}= (1/\sqrt{N})\sum_j  e^{\ii k_z z_j} \heg^j$ on the ground state $\ket{g}^{\otimes N}${~\cite{Zoubi10,Bettles16PRA,Asenjo17PRX,Henriet19,Needham19}}. These Bloch modes satisfy $\mathcal{H} \,\hat{S}^\dagger_{k_z} \ket{g}^{\otimes N}=\hbar\omega_{k_z}\,\hat{S}_{k_z}^\dagger \ket{g}^{\otimes N}$, where~{\cite{Asenjo17PRX}}
\begin{equation}
\omega_{\kp}=\omega_0-\frac{3\pi\Gamma_0}{k_0} \,\dbu^* \cdot \,\tilde{\GG}_{0} (\kp) \cdot \dbu
\end{equation}
is a complex frequency whose imaginary part describes the decay rate of the spin wave, and whose real part accounts for a frequency shift with respect to the bare atomic resonance. In the above equation, $\tilde{\GG}_{0} (\kp) = \sum_{j} e^{-\ii \kp z_j} \GG_{0} (\rb_j)$ is the discrete Fourier transform of the free-space Green's tensor. Figure~\ref{Fig1}(b) shows the dispersion relation (i.e., the real part of $\omega_{k_z}$ vs $k_z$) of a chain with lattice constant $d=0.1\lambda_0$, for atoms polarized along $z$, the direction of the array~\footnote{The dispersion relation for transverse polarization displays unusual features for $d \lesssim 0.25\lambda_0$, such that for a particular frequency there might be two guided modes of different wave-vector. While we do not believe that the physics changes significantly with respect to the longitudinal polarization case, we do not treat explicitly the transverse polarization in this manuscript.}. For $|k_z|>k_0$, the spin waves have a zero decay rate and are guided modes of the array~\cite{Zoubi10,Asenjo17PRX,Needham19}. For $|k_z|\leq k_0$, the spin waves have a finite lifetime and decay due to photon emission. Guided modes also exist in a finite chain. Due to the presence of boundaries, their decay rate is non-zero and scales as $\sim 1/N^3$~\cite{Asenjo17PRX}. The emergence of guided modes in the single-excitation manifold is not a uniquely quantum property. Waveguiding behavior is also displayed by arrays of classical dipoles, such as subwavelength grating waveguides~\cite{Halir15} and chains of dielectric~\cite{Blaustein07} and metallic~\cite{Alu06} nanoparticles.\\

\section{Coupling of impurity qubits to the atomic waveguide}

The decay rate of an ``impurity qubit'' in the vicinity of the array is altered by the presence of the waveguide [see Fig.~\ref{Fig1}(a)]. The qubit can decay into free space (whose modes are modified by the presence of the waveguide) and into guided modes of the array, exciting spin waves that propagate away from the qubit without scattering. We calculate the decay rates into free space and the guided mode by computing the Green's tensor of the surrounding environment, i.e., the vacuum and the waveguide. Exploiting the cylindrical symmetry of the infinite chain, we find an expression of the Green's tensor in terms of an integral over reciprocal space (see Appendix \ref{derivationappendix}). For a drive frequency $\omega$, the Green's tensor reads 
\begin{align}\label{greenfunctionwaveguide}
\GG(\rb,\rb',\omega)&=\GG_0(\rb,\rb',\omega)\\\nonumber
&+\frac{3\Gamma_0}{32 kd}\int_{-\pi/d}^{\pi/d} d\kp \frac{\ub_{\kp}(\rb)\otimes\vb_{\kp}(\rb')}{\omega-\omega_{\kp}},
\end{align}
where $k=\omega/c$ and the field eigenmodes are given by
\begin{subequations}
\begin{equation}
\ub_{\kp}(\rb)= \sum_g \left[\mathbb{1}+\frac{1}{k^2}\nabla\otimes\nabla\right]\cdot\dbu\,\, e^{\ii(\kp+g)z} H_0^{(1)}(\kpe\rho), 
\end{equation}
\begin{equation}
\vb_{\kp}(\rb')= \sum_g \dbu^*\cdot\left[\mathbb{1}+\frac{1}{k^2}\nabla\otimes\nabla\right]\, e^{-\ii(\kp+g)z'} H_0^{(1)}(\kpe\rho').
\end{equation}
\end{subequations}
In the above expressions, $\dbu = \db/|\db|$, $H_0^{(1)}$ is a Hankel function of the first kind, $\rho$ is the radial distance to the chain axis, and $\kpe=\sqrt{k^2-(\kp+g)^2}$ is the transversal wave-vector. The sum is performed over reciprocal-lattice vectors $g=2\pi n/d$, with $n \in\mathbb{Z}$, and accounts for Umklapp processes {(i.e., scattering terms where the momentum transfer results in a wave-vector that falls outside the first Brillouin zone)}.

The efficiency of the coupling to the waveguide is given by the ratio between the guided ($\ga$) and the free-space ($\gap$) decay rates. The analytical expression for the Green's function provides an elegant way to compute these rates separately. For atoms in free space, the decay rate is given by Eq.~\eqref{dissiprate}. Similarly, we postulate that the decay rate of the qubits is proportional to the imaginary part of the generalized Green's tensor of Eq.~\eqref{greenfunctionwaveguide}. We thus trace out the waveguide atoms, and treat the chain as a bath for the qubits [as the photons were integrated out to derive Eq.~\eqref{hameff}]. This procedure is only exact within the single-excitation subspace (the atoms are spins, not bosons) and under the Born-Markov approximation. This implies that retardation can be ignored (i.e., that the group velocity of the guided mode is not too small) and that the decay rate of the qubit is much smaller than the bandwidth of the waveguide. We discuss how these conditions can be achieved in Section~\ref{implementation}.

The qubit decay rate into free space is given by radiative wave-vectors (i.e. $|k_z|<k$). It reads
\begin{align}
\Gamma'^{\rm q}/\Gamma_0^{\rm q}&=1\\\nonumber
&+\frac{9\pi\Gamma_0}{16 k^2d}\,\text{Im}\int_{-k}^{k} d\kp \frac{\dbu_{\rm q}^*\cdot \ub_{\kp}(\rb_{\rm q}) \otimes \vb_{\kp}(\rb_{\rm q}) \cdot \dbu_{\rm q}}{\omega-\omega_{\kp}},
\end{align}
where $\Gamma_0^{\rm q}$, $\db_{\rm q}$, and $\rb_{\rm q}$ are the qubit's vacuum spontaneous emission rate, dipole matrix element, and position, respectively.

The decay into the guided mode arises from the pole of the Green's function, and is readily found to be (see Appendix \ref{derivationappendix})
\begin{equation}\label{gamma1deqn}
\ga/\Gamma_0^{\rm q}= \frac{9\pi^2\Gamma_0}{8k^2dv_g}|\dbu_{\rm q}^*\cdot \ub_{\kg}(\rb_{\rm q})|^2,
\end{equation}
where $|\kg|>k$ is the guided mode wave-vector (i.e., the wave-vector $k_z$ for which $\omega_{k_z}\equiv\omega$), and $v_g=\partial \omega_{\kp}/\partial \kp|_{\kp=\kg}$ is the group velocity. The decay rate into the guided mode increases for low group velocities, as the mode has more time to interact with the qubit. Close to the bandedge, the group velocity is low, and the decay into the waveguide becomes large (the waveguide behaves more like a photonic crystal than a fiber in this region). Note that $k\simeq k_0$ as $\omega\simeq \omega_0,\omega_q$ except for deviations of the order of $\Gamma_0\ll\omega_0,\omega_q$. 

The qubit interacts efficiently with the atomic-waveguide mode, as shown in Fig.~\ref{Fig2}. The ratio between guided and free-space decay rates, so-called optical depth $\ga/\gap$, can be larger than 1. The optical depth is a relevant quantity for quantum information processing, as it sets the fidelity for multiple protocols, such as photon storage and retrieval~\cite{Gorshkov07}, and quantum gates. The optical depth displays a modulation along $z$ related to the Bloch periodicity, with a contrast that decreases with the distance to the array. We find simple scaling laws for the optical depth when the qubit is placed exactly on top of a waveguide atom [i.e., at $z_q=0$, see Fig~\ref{Fig2}(b)], with $\ga/\Gamma'^{\rm q} \sim 1/d^{3-4}$ for constant $\rho_{\rm q}/d$, and $\ga/\Gamma'^{\rm q} \sim 1/\rho_{\rm q}^6$ for fixed $d$ and $\rho_{\rm q} \gtrsim 0.4d$, below which the coupling rates plateau. Remarkably, a waveguide that is one-atom thick provides an optical depth $\sim30$ times larger than that of a fiber (see Appendix \ref{contourplotappendix}). We corroborate the analytic calculations with numerical simulations, by evolving a finite chain and coupled qubit under the non-Hermitian Hamiltonian of Eq.~\eqref{hameff}. The numerical and analytical results fully agree with each other for qubits in the central part of the chain, where finite size effects are negligible.

Surprisingly, we find a \textit{magic point} where the emission into free space is strongly suppressed. At $\rho_{\rm q}\simeq 0.4 d$ and $z_{\rm q}=d/2$, the optical depth is extremely large while the total linewidth of the qubit remains small, as shown in Fig.~\ref{Fig2}(c). The existence of the magic point is solely due to interference. We study changes in the location of the magic point in Appendix B.\\

\section{Tunable-range interactions between impurity qubits}

\begin{figure}[b]
\centerline{\includegraphics[width=0.9\linewidth]{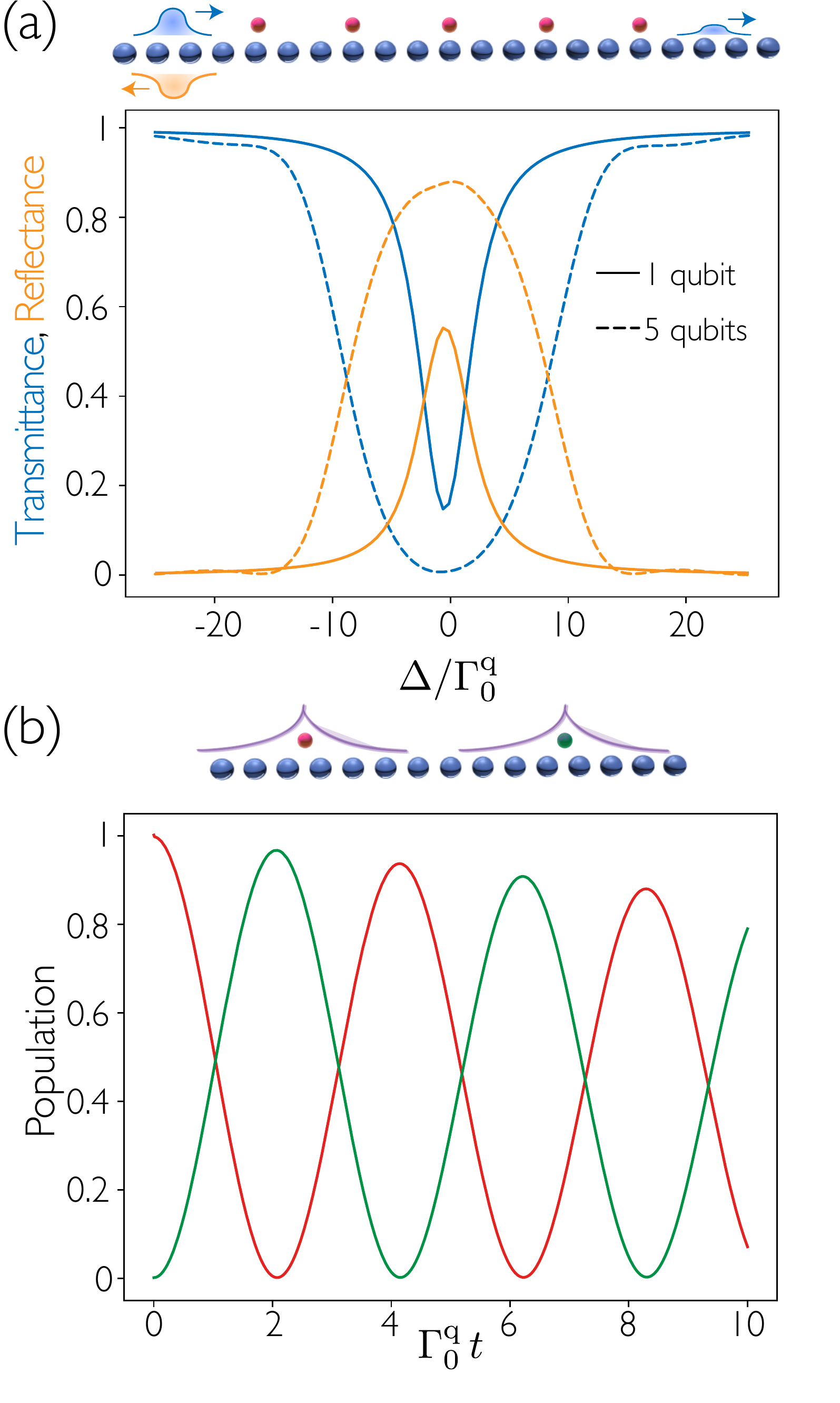}}
\caption{\textbf{Dissipative (a) and coherent (b) long-range interactions between impurity qubits mediated by the atomic waveguide.} \textbf{(a)} Transmission (blue) and reflection spectra (orange) of a  guided mode that interacts with either a single qubit or with five qubits evenly spaced 20 lattice sites apart. In both cases, the qubits' vacuum spontaneous emission rate is $\Gamma_0^{\rm q} = 0.02\Gamma_0$, sit at $\rho_{\text{q}} = d$ and are detuned from the waveguide atoms such that light with wave-vector $\kg = 0.7\pi/d$ is near-resonant with the qubits. The chain consists of $N=4000$ atoms, and the lattice constant is $d=0.1\lambda_0$. \textbf{(b)} Evolution of the excited-state population of qubit 1 (red) and 2 (green) after fully inverting qubit 1 at the initial time. The Rabi oscillations reveal strong coupling mediated by photonic bound states. The resonance frequency of the qubits lies $4.5\Gamma_0$ from the bandedge of the atomic waveguide, with the qubits (of spontaneous emission rate $\Gamma_0^{\rm q} = 0.001\Gamma_0$) placed in between two array atoms at $\rho_{\text{q}} = 0.4d$ and separated by a distance $8d$, for a chain of $N=199$ atoms of lattice constant $d=0.05\lambda_0$.}\label{Fig3}
\end{figure}

\begin{figure*}[t]
    \centerline{\includegraphics[width=.97\textwidth]{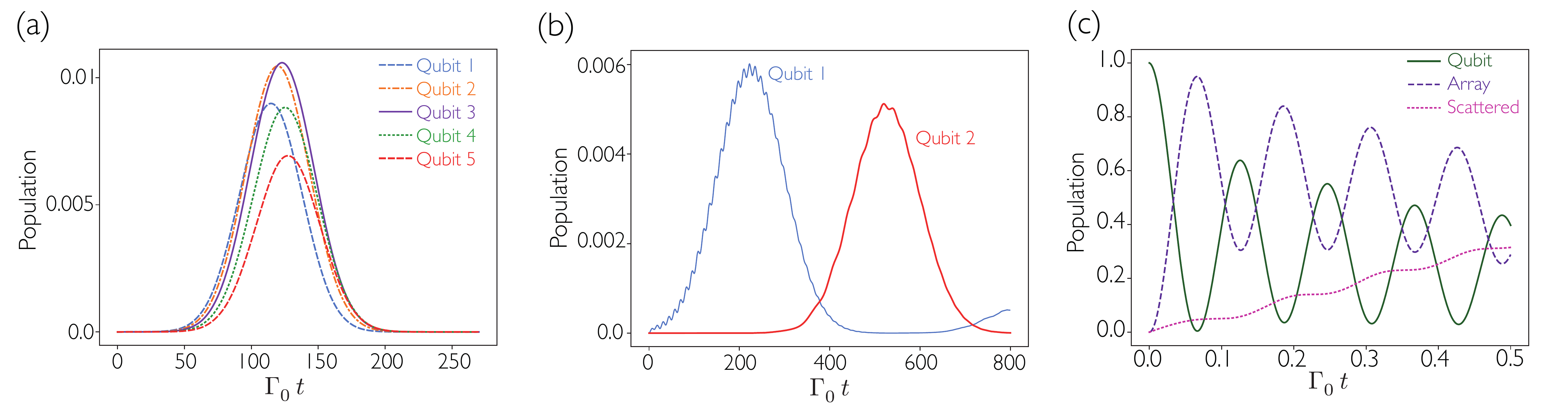}}
    \caption{{\textbf{Non-Markovian effects in atomic-waveguide QED: time-delayed interactions and bound states in the continuum.} \textbf{(a, b)} Time-delayed qubit-qubit interactions. Evolution of the populations of (a) five qubits evenly spaced 20 lattice sites apart as in Fig.~\ref{Fig3}(a) and (b) two qubits spaced 800 sites apart, coupled to an array of 4000 atoms. In both plots, the qubits' vacuum spontaneous emission rate is $\Gamma_0^q = 0.02\Gamma_0$, they sit at $\rho_q = d$ and are detuned from the waveguide atoms such that light with wave-vector (a) $\kg = 0.7\pi/d$, or (b) $\kg = 0.95\pi/d$ is near-resonant with the qubits, and the initial state of the array is a spin wave with frequency shift detuned $14.5\Gamma_0^q$ below the coupled qubits. In all cases, $d = 0.1\lambda_0$. \textbf{(c)}~Bound states in the continuum: evolution of the excited-state populations of qubit and array atoms after fully inverting the qubit at the initial time. The qubit has resonance frequency $\omega_q = \omega_0 + 8.5\Gamma_0$, linewidth $\Gamma_0^q = \Gamma_0$ and position $\rho_q = 0.5d$.}}
    \label{fig9}
\end{figure*}

{Atomic waveguides can be harnessed to mediate qubit-qubit interactions, without the need of photonic structures. The character of these interactions (coherent or dissipative) is modified by tuning the qubit resonance frequency $\omega_\text{q}$. If $\omega_\text{q}$ lies within the band, the interactions are dissipative. If, instead, $\omega_\text{q}$ lies outside the band, the interactions are coherent. This makes atomic waveguides appealing for quantum simulation, where precise control of the form of interactions is of fundamental importance.}

\subsection{Infinite-range interactions: waveguide QED\label{waveguideQED}}

{Dissipative qubit-qubit interactions lead to collective decay and superradiance.} We analyze photon transport through an atomic waveguide with either one or five periodically spaced qubits coupled to it, as shown in Fig.~\ref{Fig3}(a). To calculate transmittance and reflectance spectra, we launch a spin wave of the form
\begin{equation}
\ket{\psi(t=0)} = \sum\limits_{i=1}^N \mathrm{e}^{-\ii\kg \bar{z}_i} \mathrm{e}^{-\bar{z}_i^2/\zeta^2} \hat{\sigma}_{eg}^i \ket{g}^{\otimes N},
\end{equation}
where $\zeta = 300d$ is the spatial spread, $\bar{z}_i$ are the atomic positions relative to the center of the spin wave, and $\kg$ is chosen to determine the relative detuning between qubit and spin wave, i.e., such that  $\Delta\equiv \omega_{\kg}-\omega_q$. We discuss how to prepare such a state in Section \ref{implementation}. The evolution is performed under the Hamiltonian in Eq.~\eqref{hameff}, conditioning the results on no jumps. 

{Given the large optical depth, the impurity qubit behaves as a mirror, reflecting most of the spin wave, as shown in Fig~\ref{Fig3}(a). Impurity qubits with a more complex hyperfine structure can thus be used to realize single-photon transistors and switches, as proposed for a single atom coupled to a metallic 1D reservoir~\cite{Chang07}.} The qubit is positioned 1000 sites from the initial spin wave. We calculate the transmission (reflection) from the population of the array atoms located past (before) the qubit, while the lost norm provides the scattering into free space. The spectra display the traditional Lorentzian lineshape with a width that scales as $\ga+\gap$~\cite{Chang12,Asenjo17PRA}. 

{If there are multiple impurity qubits coupled to the chain, they are bound to interact with the photons (or spin excitations) emitted by all of their neighbors, due to the one-dimensional nature of the waveguide. Impurity qubits whose frequency lies inside the band emit polaritons that propagate without scattering and mediate infinite-range interactions between qubits.} We perform a calculation for five qubits separated by a distance such that $\kg d_\text{q}=14\pi$. In conventional waveguide QED, this corresponds to the mirror configuration, where the qubits behave collectively as a single qubit with a larger dipole moment and superradiantly decay at a rate $5\ga+\gap$~\cite{Chang12,Asenjo17PRA}. Our transmission spectra, shown in Fig.~\ref{Fig3}(a), deviates slightly from a Lorentzian profile due to non-Markovian effects associated with retardation (see Section~\ref{timedelay}). The group velocity of the spin chain is remarkably slow compared to the speed of light in free space, scaling as $v_g\sim (\Gamma_0/k_0) f(k_0d)$ where $f(k_0d)$ decreases with $d$. Atomic waveguides are thus versatile platforms that can be tuned to mediate both Markovian and time-delayed interactions between distant qubits.

\subsection{Finite-range interactions: bandgap physics}

\begin{figure*}[t]
    \centering
    \includegraphics[width=0.95\textwidth]{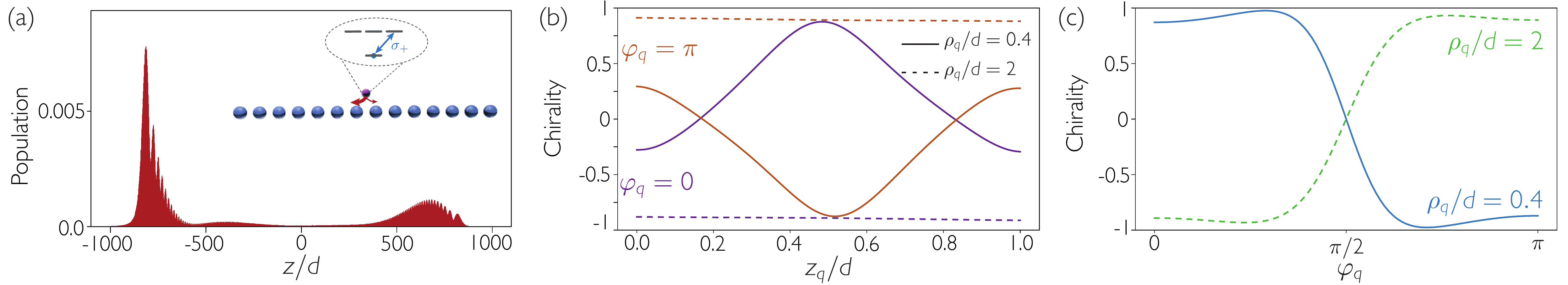}
    \caption{\textbf{Chiral emission from an impurity qubit.} \textbf{(a)}~Atomic population in a chain of 2000 atoms at $\Gamma_0t = 64$ following the inversion of a coupled impurity qubit positioned at $\set{\rho_q,\varphi_q,z_q}=\set{0.4d,0,0.5d}$. \textbf{(b, c)}~Chirality as a function of qubit longitudinal and azimuthal position. Chirality is calculated as $\braket{\hat{P}_\mathrm{left} - \hat{P}_\mathrm{right}}/\braket{\hat{P}_\mathrm{left} + \hat{P}_\mathrm{right}}$, where $\hat{P}_\mathrm{left~(right)}$ is the sum of excited state populations on waveguide atoms to the left (right) of the qubit. Populations are calculated at $\Gamma_0t = 2$ following the inversion of the qubit coupled to a chain of 500 atoms. In (c), $z_q = 0.5d$. In all cases, the chain has inter-atomic spacing $d=0.1\lambda_0$; the qubit has linewidth $\Gamma_0^q = 0.002\Gamma_0$ and detuning such that light with wave-vector $\kg = 0.7\pi/d$ is near-resonant with the qubit. The qubit has three excited levels and is inverted on the transition with dipole matrix element $\db_q=-1(\hat{x} - i\hat{z})/\sqrt{2}$.}
    \label{Fig12}
\end{figure*}

Coherent qubit-qubit interactions mediate spin exchange, {and can be harnessed to realize generic Hamiltonian models for impurity qubits}~\cite{Douglas15}. Interactions are coherent if the qubits' frequency sits beyond the bandedge~\cite{John90,John91,Douglas15,Sanchez17}. In the bandgap, spin waves cannot propagate and form bound states that are spatially localized around the qubit position. Mathematically, it is easy to see that a resonant excitation cannot propagate through the array as there is no pole in the integral of Eq.~\eqref{greenfunctionwaveguide}. Bound states mediate purely coherent, finite-range interactions between qubits, as shown in Fig.~\ref{Fig3}(b). Since the spin-exchange rate is small, we place the qubits at \textit{magic points}, where the free space decay rate is strongly suppressed. This allows us to observe several oscillations before the dynamics is damped. Without the waveguide, given the qubits' separation, they would simply decay.

\subsection{Time-delay in qubit-qubit interactions\label{timedelay}}

Slow group velocity leads to non-Markovian behavior. In particular, retardation in the propagation of the atomic spin wave prevents the waveguide from behaving as a Markovian bath for multiple impurity qubits. Figure~\ref{fig9}(a) shows the effect of retardation in the evolution of the populations of five evenly spaced qubits coupled to the array. As described in Section~\ref{waveguideQED}, we launch a spin wave and evolve the system under the non-Hermitian Hamiltonian of Eq.~\eqref{hameff}. The spin wave is detuned by $14.5\Gamma_0^q$ from the resonance frequency of the qubits, corresponding to a point on the shoulder of Fig.~3(a). Each of the five qubits is excited at slightly different times. The delay in the excitation is small, but significant enough to break the Markovianity of the waveguide \cite{Tufarelli14,Pichler16,Sinha20,Carmele20}.

Tunable time-delayed interactions can be realized exploiting this phenomenon. The delay can be controlled by altering the distance between qubits or the group velocity, which is slowest for large $\kg$ and $d$. Figure~\ref{fig9}(b) shows the evolution of the populations of two impurity qubits spaced 800 sites apart. There is a long delay between each qubit being excited, such that the first qubit is almost completely de-excited before the second atom interacts with the pulse. The small oscillations in the population are due to the slight excitation of bound states in the continuum, as explained in the following section.

\section{Bound states in the continuum\label{BIC}}

The strong coupling of impurity qubits to a finite-bandwidth photonic reservoir gives rise to ``bound states in the continuum''~\cite{Calajo16,Shi16,Calajo19}, a type of dressed atom-photon bound state. This effect can also appear in atomic-waveguide QED, in the region where slow-light effects are relevant (i.e., near the bandedge). If the coupling between the qubit and guided modes is very large, the population scattered into the array is reabsorbed by the qubit on a timescale faster than that required to transport the excitation away. This results in oscillations between the qubit and a bound state of the array, as shown in Fig.~\ref{fig9}(c). This bound state is different to that appearing when the qubit frequency is in the band gap (discussed in the previous section), where the spin excitation is bound because it cannot be guided. Here, the bound state appears due to splitting of the hybridized energy of the qubit and photon. Due to the large interactions, the splitting takes the hybridized energy outside the guided band~\cite{Calajo16}. Bound states in continuum lead to fractional decay, as the qubit scatters into free space, while the array does not.

\section{Chiral quantum optics}

{An impurity qubit with the appropriate hyperfine structure will decay into the waveguide in a directional manner,   even though the chain does not break reciprocity or mirror symmetry. This occurs because the near field of the waveguide is chiral, i.e., has both longitudinal and radial components. Chiral quantum optics has been recently explored for impurities coupled to  dielectric nanofibers and photonic crystals~\cite{Petersen14,Sollner15,Pichler15,Lodahl17,Jones20}, and allows for the realization of ``cascaded'' open systems~\cite{Carmichael93PRL,Gardiner93,Ramos16}, which results in entangled steady states of the impurities, among other applications. To demonstrate chiral emission into the atomic waveguide, we employ a qubit with three excited states, which are coupled to the ground state via optical transitions with $\set{\sigma_-,\pi,\sigma_+}$ polarizations. The quantization axis of the qubit is set to be along the $y$-axis such that the circularly polarized transitions have dipole moments $\pm (\hat{x} - i\hat{z})/\sqrt{2}$. These break the symmetry of coupling into left and right propagating modes, as can be seen from the detailed expressions for the field modes $\textbf{u}_{k_z}$ in Appendix~\ref{derivationappendix}. This is demonstrated in Fig.~\ref{Fig12}(a), where a qubit couples predominantly into left propagating modes.}

{Chiral decay is sensitive to the relative position of the qubit with respect to the chain. We calculate a figure of merit that quantifies the level of chirality as $\braket{\hat{P}_\mathrm{left} - \hat{P}_\mathrm{right}}/\braket{\hat{P}_\mathrm{left} + \hat{P}_\mathrm{right}}$, where $\hat{P}_\mathrm{left~(right)}$ is the sum of excited state populations of waveguide atoms to the left (right) of the qubit. Chirality of $+1 (-1)$ means perfectly chiral decay into left (right) propagating modes. For a qubit close to the waveguide, the chirality is strongly dependent on the relative position of the qubit within a unit cell of the atomic waveguide [Fig.~\ref{Fig12}(b,c)]. In particular, qubits in between two waveguide atoms preferentially emit into left (right) propagating modes when positioned at azimuthal angles $\varphi_q\in(-\pi/2,\pi/2)$ ( $\varphi_q\in(\pi/2,-\pi/2)$). For $\varphi_q=\pm\pi/2$, there is no chirality. For large radial distances $\rho_q$ between the qubit and the waveguide, the sign of chirality flips and the contrast within the unit cell is reduced (as the qubit does not resolve individual atoms). While chirality is stronger at large $\rho_q$, the decay into free space dominates over that to guided modes, thus preventing efficient chiral transport.}
\section{Quantum nonlinearity and photon collisions}

\begin{figure}[t!]
\centerline{\includegraphics[width=.95\linewidth]{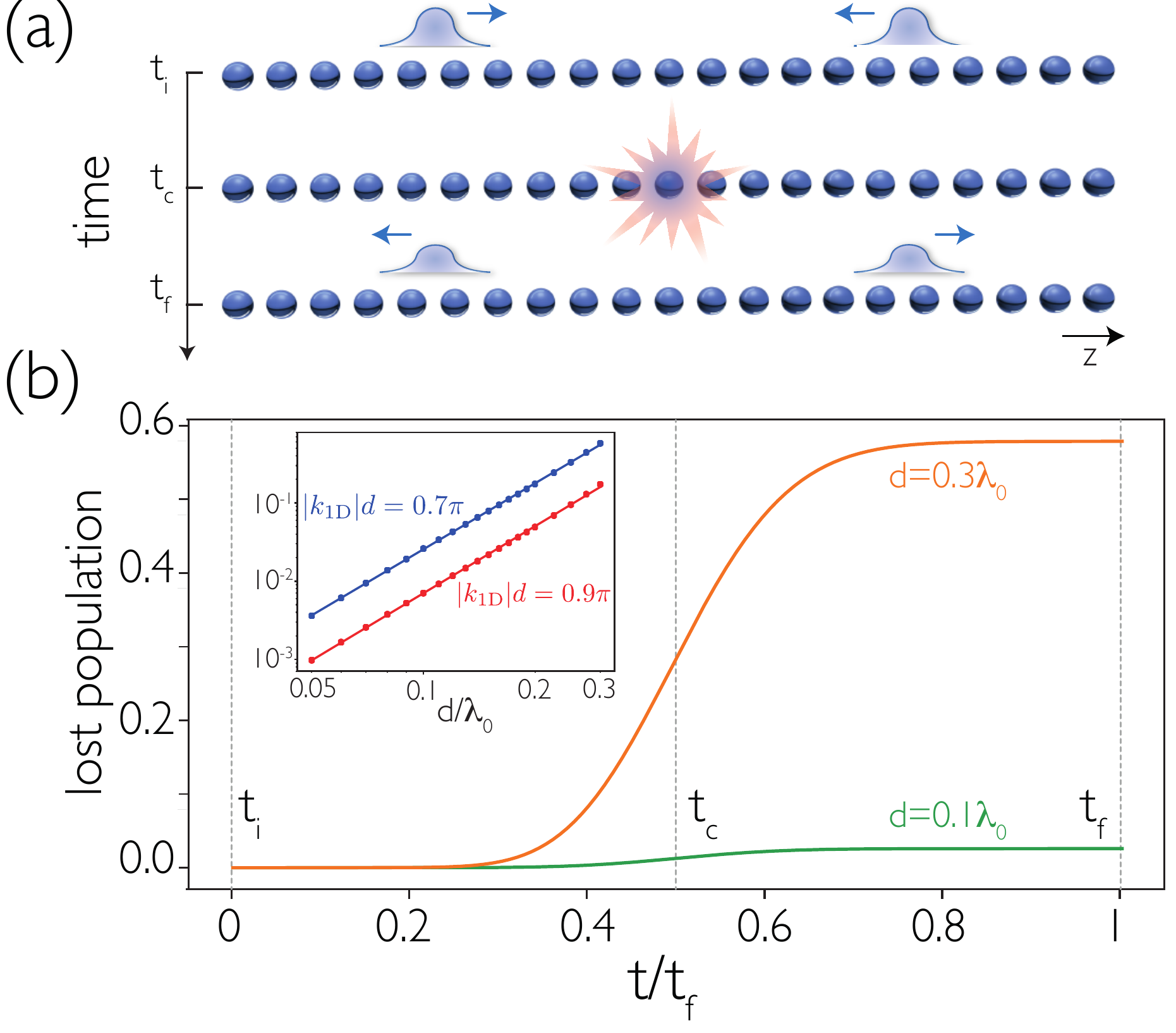}}
\caption{\textbf{Two-photon collision in an atomic waveguide.} \textbf{(a)}~At $t=t_i$, two counter-propagating spin waves are initialized. They collide, leading to population loss (the maximum overlap occurs at $t=t_c$). After the collision has occurred they propagate without loss until they hit the ends of the chain. Before this occurs, they first return to their original positions, at $t=t_f\equiv 2t_c$. \textbf{(b)} Population in the two-excitation manifold as a function of time, for chains of lattice constants $d=0.1\lambda_0$ (green) and $d=0.3\lambda_0$ (orange), for initial wave-vector $\kg =\pm 0.7\pi/d$. Inset: Scaling of lost population in the two-excitation manifold, $\gamma\equiv1-\braket{\hat{\sigma}_{ee}^{(2)}(t_f)}/\braket{\hat{\sigma}_{ee}^{(2)}(0)}$, with inter-atomic distance for $\kg=\pm 0.7\pi/d$ (blue) and $\kg=\pm 0.9\pi/d$ (red). The continuous lines are guides to the eye and scale as $\gamma\sim d^{\sim 2.8}$ for both lines. For both plots $N=200$.}\label{Fig4}
\end{figure}

Interactions between impurity qubits in the presence of multiple photons are impacted by the two-level nature of the atomic-waveguide atoms. As a consequence of this non-linearity, the physics of two-photon transport in an atomic waveguide is qualitatively different from that observed in a classical waveguide. In particular, $(\hat{S}^\dagger_{k_z})^2 \ket{g}^{\otimes N}$ is not an eigenstate of the Hamiltonian. Instead, the true eigenstates behave as fermions (or hardcore bosons), obeying a Pauli exclusion principle in space~\cite{Asenjo17PRX,Albrecht19,Zhang19}. Multiple photons must interact with each other, such that the linear regime of waveguide QED described above is no longer strictly accurate. In this Section, we consider the effect of quantum nonlinearities on waveguide-mediated interactions between qubits. We first analyze interaction-assisted dissipation due to ``collisions'' between counter-propagating photons. We then discuss the impact of these collisions on time-delayed qubit-qubit interactions.

\begin{figure}[b]
    \centering
    \includegraphics[width=.5\textwidth]{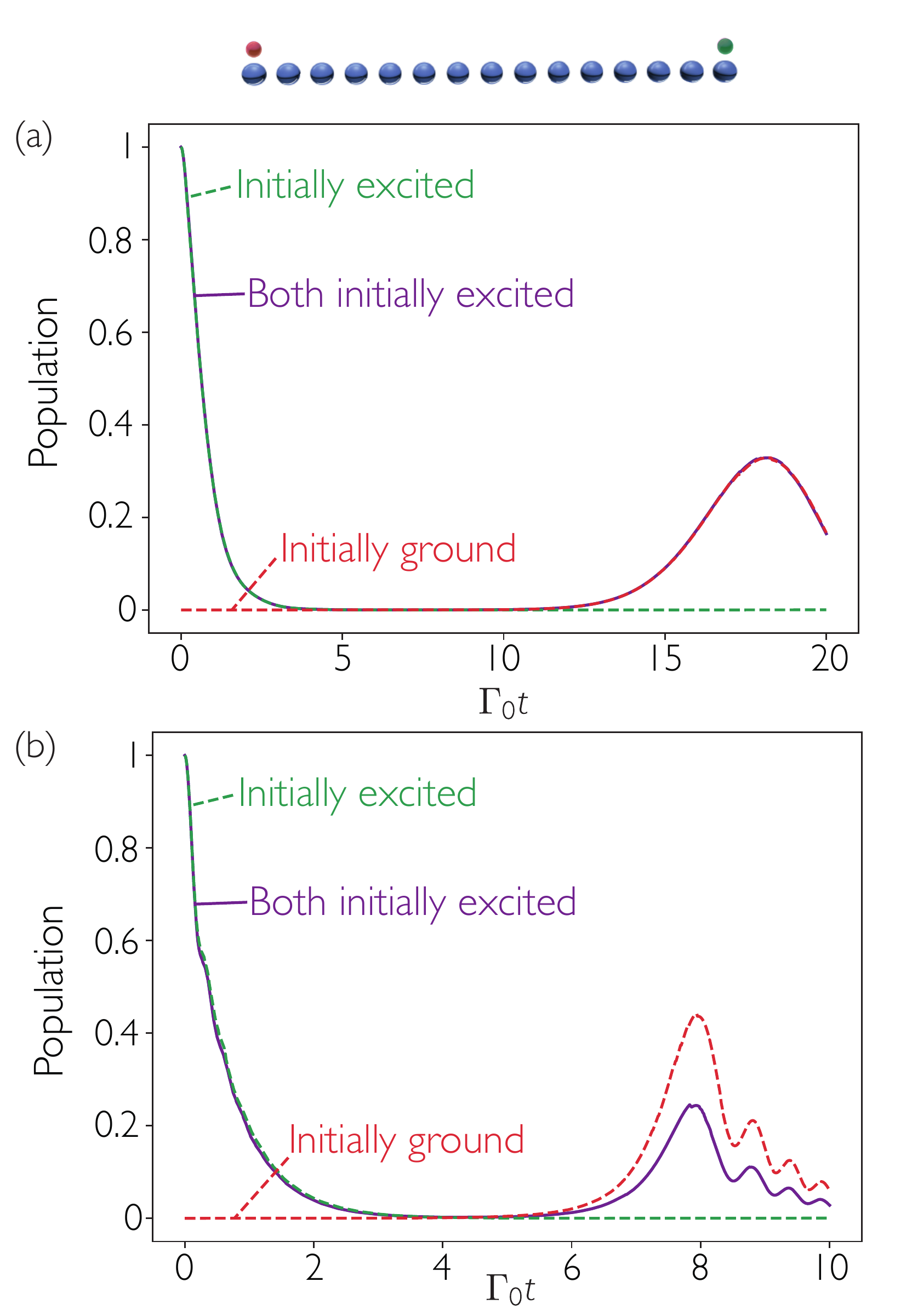}
\caption{{\textbf{Role of photon collisions on multi-excitation qubit-qubit interactions.} \textbf{(a, b)}~Evolution of the excited-state populations of two qubits coupled to a 1D array of $N=120$ atoms with interatomic spacing $d=0.1\lambda_0$. Comparison is given between qubit-qubit interactions when (dashed) one qubit is initially excited and (solid) both qubits are initially excited. When both qubits are initially excited, they have the same temporal evolution due to the symmetry of the system. Qubits are positioned in line with the end atoms and have radial displacement $\rho_q = 0.4d$ and resonance frequency commensurate with the guided mode frequency for \textbf{(a)}~$\kg=0.7\pi/d$ and \textbf{(b)}~$\kg=0.25\pi/d$ respectively. In (a), the qubits have linewidth $\Gamma_0^q = 0.002\Gamma_0$. In (b), the qubits have linewidth $\Gamma_0^q = 0.01\Gamma_0$. All plotted curves are the ensemble average of 100 quantum trajectories.}
    \label{Fig10}}
\end{figure}

The spatial overlap of two photons (spin waves) produces scattering into free space. A photon-photon collision occurs if two spin waves propagate in opposite directions, as shown in Fig.~\ref{Fig4}(a). Dissipation during the collision is due to the spatial distortion of the guided modes. To observe such interaction, we initialize a two-excitation state
\begin{equation}
\ket{\psi^{(2)}(t_i)} = \sum\limits_{i,j=1}^N \mathrm{e}^{\ii\kg (\bar{z}_i - \bar{z}_j)} \mathrm{e}^{- (\bar{z}_i^2+\bar{z}_j^2)/\zeta^2} \hat{\sigma}_{eg}^i\hat{\sigma}_{eg}^j \ket{g}^{\otimes N},
\end{equation}
where $\zeta=15d$ and $\bar{z}_{i,j}$ are the atom positions relative to centers $60$ sites apart. We evolve the wavefunction under the Hamiltonian in Eq.~\eqref{hameff} conditioned on zero jumps. In Fig.~\ref{Fig4}(b), we show the decrease of population ($\braket{\hat{\sigma}_{ee}^{(2)}(t)}=\braket{\psi^{(2)}(t)|\sum_i\hat{\sigma}_{ee}^i|\psi^{(2)}(t)}$) as a function of time. The population loss ranges from less than $1\%$ (for $d=0.1\lambda_0$) to almost $60\%$ (for $d=0.3\lambda_0$) for $|\kg|=0.7\pi/d$; the inset in Fig.~\ref{Fig4}(b) shows that the loss grows with the distance as a power law. This power law is dependent on $\kg$ and $d$, and is discussed in further detail in Appendix~\ref{powerlawappendix}.  The collision generates population in radiative modes (those with $|k_z|<k_0$), leading to the field leaking out of the waveguide. The degree of interaction between two excitations is controlled by tuning a few experimentally accesible parameters (frequency and lattice constant). The decay probability is a function of the overlap between the distorted state and radiative modes, and their scattering rate. Smaller lattice constants lead to larger group velocities and smaller light cones. For large lattice constants, jumps lead to the emission of one or (most probably) two photons. 

For small distances, the probability of photon emission is negligible and the effect of the collision reduces to the acquisition of a global phase dependent on $\kg$ and $d$. This phase accumulation may be harnessed to realize conditioned phase gates between polaritons~\cite{Milburn89,Masalas04,Gorshkov11,Brod16PRA}, which can be used for quantum computation~\cite{Chuang95,Hutchinson04}, and will be explored in future work.

Qubit-qubit interactions are modified by the non-linear nature of the waveguide. To illustrate this, we focus on their impact on time-delayed interactions, explored in Section~\ref{timedelay}. We consider two qubits coupled to slow propagating guided modes. If one qubit is initialized in its excited state, then it decays into the waveguide, and, after some delay as the excitation propagates through the array, the emitted pulse excites the other qubit. However, when both qubits are initially inverted, the two emitted pulses must collide as they propagate along the waveguide, leading to dissipation and phase accumulation. As described above, the geometry of the array and the wave-vector of the propagating photons strongly affect the amount of dissipation during the photon collision. This can have a dramatic impact on time-delayed qubit-qubit interactions, as shown in Fig.~\ref{Fig10}. For photon-photon interactions that are primarily coherent - i.e., for qubit frequencies that are resonant with the guided mode far beyond the light line - the population of the impurity qubits is not significantly altered by the collision, as photons remain guided [Fig.~\ref{Fig10}(a)]. However, for qubits whose resonance frequencies correspond to guided modes close to the light line, photon-photon interactions are strongly dissipative, as shown in Fig.~\ref{Fig10}(b). In this situation, photon loss is unavoidable and excited-state population is not preserved, as the photons are likely to scatter out of the waveguide before reaching the second qubit.

\color{black}

\section{Physical implementations\label{implementation}}

Atomic arrays are not just a toy model, but an experimental reality~\cite{Greif16,Kumar18,Rui20,Kim16,Endres16,Barredo16}. The implementation of an atomic waveguide involves two main challenges: to trap atoms at short distances and to efficiently excite guided modes. Coupling qubits to the waveguide comes with an additional set of difficulties. We discuss strategies for overcoming these challenges below. For most of the discussion, the experimental setup we have in mind consists in neutral atoms trapped in optical lattices~\cite{Bloch05,Bakr09,Bakr10,Greif16,Kumar18,Rui20} or optical tweezers~\cite{Lester15,Labuhn16,Kim16,Endres16,Barredo16,Bernien17,Barredo18,OhlDeMello19,Glicenstein20}, which have recently been suggested as quantum metasurfaces~\cite{Bekenstein20}. {While lattices create grid-like arrays, tweezers allow for almost-arbitrary positioning of the impurity qubits, as non-regular arrays can be created via spatial light modulators or holographic metasurfaces.}

First, we require small inter-atomic separations ($d<\lambda_0/2$). The diffraction limit can be overcome using two different atomic transitions: one to trap and one to drive the optical excitation. As an example, Strontium can be trapped at a magic wavelength with $d=\lambda_0/16.3$~\cite{Olmos13,Perczel17PRL}, driven on the $\lambda_0=2.6~\mu$m $^3P_0\rightarrow {^3D_1}$ transition. The bosonic species lacks hyperfine structure, which prevents additional difficulties~\cite{Hebenstreit17,Asenjo19}. Another possibility would be to use Ytterbium's telecom transition~\cite{Covey19}. Quantum and classical disorder may affect the guiding properties of the waveguide. While guided modes have been shown to be robust against spatial disorder~\cite{Needham19,Asenjo19} disorder may lead to localization for low group velocities. The finite spread of the atomic wavefunction adds an independent decay channel for each atom ($\Gamma'_{\text{trap}}\sim \Gamma_0\eta^2$, where $\eta$ is the Lamb-Dicke parameter~\cite{Guimond19}), but can be reduced using tight traps.

Second, we need to excite guided modes efficiently. The frequency of the external field selects the wave-vector of the spin wave that propagates in the array, and the temporal duration of the laser pulse sets its spatial width. Coupling is possible by focusing the external light into the array edge, either with a lens with high numerical aperture or with a spatial light modulator. One can also employ coupled qubits to inject spin waves into the waveguide. The coupling loss can be alleviated by using a near-field probe, such as a fiber tip close to the array. Finally, a phase could be imprinted, via magnetic or optical fields~\cite{Plankensteiner15,He19arxiv}, into easily-accessible superradiant states.

We require the frequency of the qubits to be distinct from that of the waveguide atoms. The qubit frequency can be tuned with AC Stark shifts through optical tweezer beams. To realize Markovian interactions, the waveguide bandwidth has to be broad compared to the qubit linewidth. One option is to rely on compact chains, as the bandwidth increases as $d$ is reduced. Another is to use different atomic isotopes (e.g., $^{87}$Sr and $^{88}$Sr), as they have similar transition frequencies~\cite{Stellmer14} but different linewidths due to hyperfine structure. For distances $d = \lambda_0/16.3$, the ratio between waveguide bandwidth and qubit linewidth is $\sim 400$. Cold molecules are also interesting candidates, as they have dense frequency spectra and have been recently trapped in tweezer arrays~\cite{Anderegg19}.

This physics can also be observed in arrays of solid state qubits, such as color centers~\cite{Sipahigil16}, rare-earth ions~\cite{Kornher16,Jacob16}, and localized excitonic quantum dots or strain-generated defects in 2D materials~\cite{Palacios17,Proscia18}. While deteministic placement of solid state emitters is becoming a reality, these emitters have their own set of issues, mostly related to inhomogeneous broadening and non-radiative decay.\\

\section{Outlook}

We have demonstrated that atomic waveguides are versatile quantum light-matter interfaces. They support single-photon states that do not leak into free space, with a dispersion relation that can be easily engineered by tuning the inter-atomic separation (and dipole orientation). They also mediate \textit{waveguide-less} long-range interactions between impurity qubits, without the need for interfacing atoms with traditional nanophotonic structures. These qubits interact efficiently with excitations in the chain, allowing for the exploration of different regimes of waveguide QED such as collective decay, bandgap physics (where atomic bound states emerge and mediate coherent finite-range interactions between qubits), non-Markovian dynamics, and chiral quantum optics.

{Atomic arrays allow for dispersion engineering, such that photons acquire a finite mass, propagate with slow group velocity, or are even bestowed with topological properties. Band structure design is crucial to realize almost-arbitrary interactions between impurity qubits, required for quantum simulation, and the processing of quantum information. The optical properties of atomic arrays can be controlled dynamically via external dressing fields, which is hard to achieve in conventional dielectric structures, and may allow for novel schemes in trapping and manipulation of single-photon states. Together with local (i.e., single atom) access through optical tweezers, these ideas open the door to the realization of a new paradigm for controlling light-matter interactions. Atomic waveguides also provide opportunities to explore the rich physics of self-organization~\cite{Domokos02,Black03,Chang13}, as optomechanical degrees of freedom might play an important role in determining the geometry of the array.}

{At the few photon level, atomic waveguides allow for the realization of deterministic and controllable photon-photon interactions enabled by atomic dark states. Interactions form the underpinnings of quantum non-linear optics, and may enable the design of quantum photonic circuitry, including photon transistors and gates~\cite{Milburn89,Masalas04,Gorshkov11,Brod16PRA}. It should be stressed that this physics does not involve Rydberg states, which have, up to now, been the conventional resource to generate photon-photon interactions~\cite{Peyronel12,Chang14,Gorshkov11,Cantu20}. This tunable non-linearity opens the door to the exploration of few-body physics between guided photons and to the realization of conditioned phase gates between counter-propagating photons, which can be used for quantum computation.}

{Finally, atomic waveguides can be harnessed to explore less traditional QED paradigms, such as time-delayed interactions to study the effects of retardation and feedback in interacting quantum systems~\cite{Wiseman94}. They also represent a realizable platform where questions about how baths emerge from finite-sized and mesoscopic systems can be answered.}

\vspace{10pt}
\textbf{Acknowledgments.}-- We are specially grateful to D. E. Chang, H. J. Kimble, M. Lipson, and J. A. Muniz for stimulating discussions. We also thank P. B. Dieterle, R. Guti\'errez J\'auregui, J. Sheng, and K. Sinha for careful reading of the manuscript and insightful comments. This work is financially supported by Programmable Quantum Materials, an Energy Frontier Research Center funded by the U.S. Department of Energy (DOE), Office of Science, Basic Energy Sciences (BES), under award DE-SC0019443; by the Center for Precision Assembly of Superstratic and Superatomic Solids, an NSF MRSEC (Award Number DMR-1420634); and by the National Science Foundation QII-TAQS (Award No. 1936359).

\appendix

\section{Derivation of the impurity qubit decay rates\label{derivationappendix}}

\begin{figure*}[t]
\centerline{\includegraphics[width=.8\linewidth]{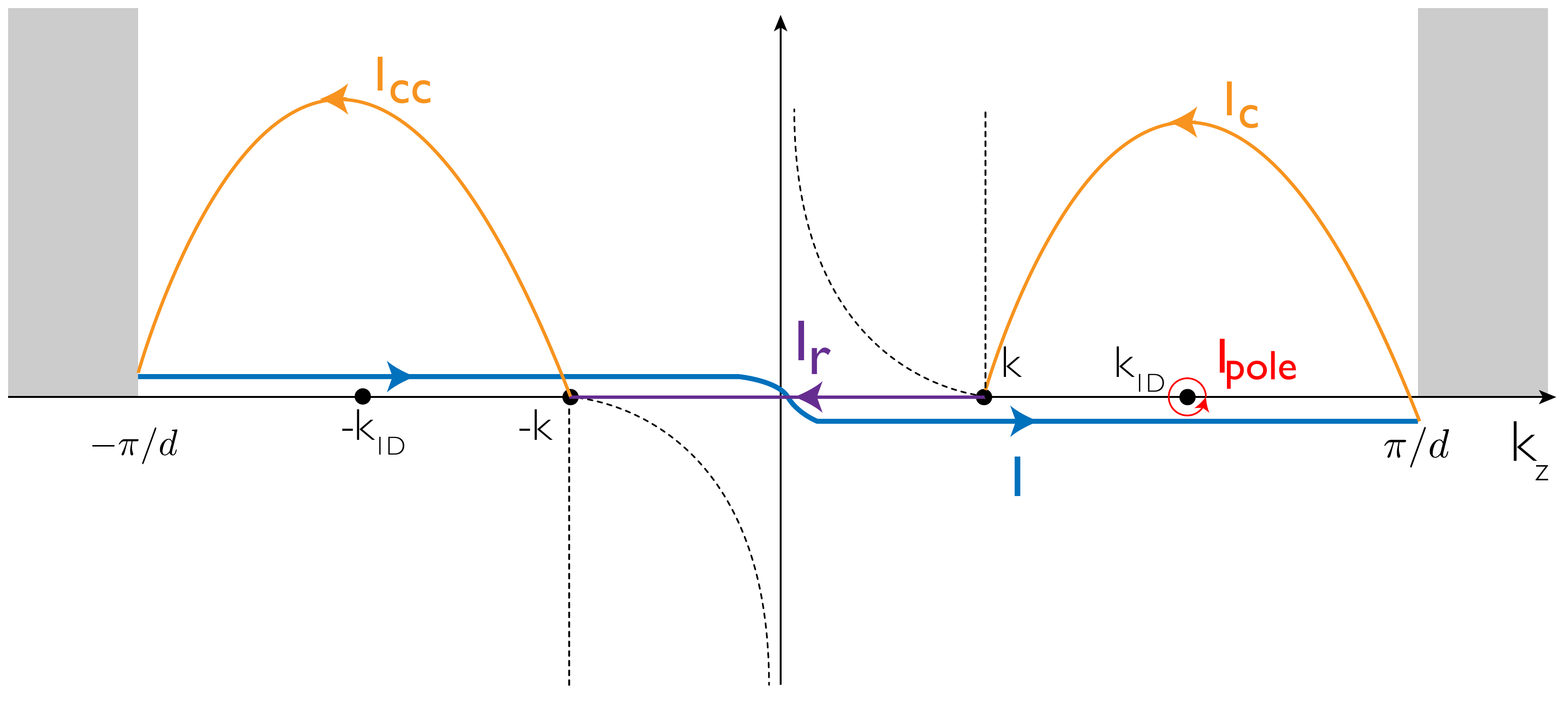}}
\caption{\textbf{Integration contour for Eq.~\eqref{integralequation} depicting the pole.} Branch cuts are shown by dashed lines. The integration is performed in the first Brillouin zone, i.e., $k_z\in[-\pi/d,\pi/d].$}\label{Figcontour}
\end{figure*}

Here we derive the expressions for the qubit decay rates into free space and into the guided mode of the atomic array. The decay rate of an emitter is related to the imaginary part of the Green's tensor evaluated at the emitter's position. We thus begin by finding an expression for the propagator of the electromagnetic field in the presence of the chain. Following the main text, we consider that the atomic array is pumped by a weak coherent field (such that saturation is negligible and the dynamics is confined to the single-excitation manifold). The equation of motion for the expectation value of the atomic coherence operator of atom $j$ is readily found to be~\cite{Asenjo17PRX}
\begin{align}
\dot{\sigma}_{ge}^j&=\ii \Delta \sigma_{ge}^j\\\notag+& \ii\frac{\mu_0\omega^2}{\hbar}\sum_{i=1}^N \db^*\cdot\mathbf{G}_0(\rb_i,\rb_j,\omega)\cdot\db  ~\sigma_{ge}^i+\frac{\ii}{\hbar}\db^*\cdot\Eb_{\rm p}^+ (\rb_j),
\end{align}
where $\mathbf{G}_0(\rb_i,\rb_j,\omega)$ is the vaccuum Green's tensor, $\sigma_{ge}^j \equiv \braket{\hge^j}$, $\Eb_{\rm p}^+=\braket{\hat{\Eb}_{\rm p}^+}$ is the expectation value of the positive-frequency component of the driving field, $\Delta=\omega-\omega_0$ is the detuning between the driving and the atomic resonance frequencies, and $\rb_j=(\Rb_j,z_j)$ is the position of atom $j$. For an infinite chain, we can define a spin-wave operator, $\hat{S}^\dagger_{k_z}=(1/\sqrt{N})\sum_je^{\ii k_z z_j}\heg^j$ that creates an excitation of well-defined longitudinal momentum $k_z$. In the steady state (i.e. $\dot{\sigma}_{ge}^j =0$), the expectation value of the spin-wave annihilation operator reads
\begin{align}
S_{k_z}=-\frac{1}{\hbar\Delta+\mu_0\omega^2 \db^*\cdot\tilde{\mathbf{G}}_0(k_z)\cdot\db} \db^*\cdot\Eb_{\rm p}^+(k_z),
\end{align}
where $\Eb_{\rm p}^+(k_z)=(1/\sqrt{N})\sum_je^{-\ii k_z z_j}\Eb_{\rm p}^+(\Rb_j,z_j)$ is the spatial Fourier transform of the field. Employing the input-output equation~\cite{Asenjo17PRX}
\begin{align}\label{fielddef}
\hat{\Eb}^+(\rb)&=\hat{\Eb}_{\rm p}^+(\rb)+\mu_0 \omega^2\sum_{j=1}^N \GG_0(\rb,\rb_j,\omega)\cdot\db\,\hge^j,
\end{align}
we write the expectation value of the field in any point in space as
\begin{align}
\Eb^+(\rb) = \Eb^+_{\rm p}(\rb) - &\frac{\mu_0\omega^2}{\sqrt{N}} \sum\limits_{k_z} \left[ \Bigg( \sum\limits_j \mathbf{G}_0(\rb,\rb_j)\cdot \db~\mathrm{e}^{ik_zz_j} \right)\notag  \\& \times \frac{\db^*\cdot\Eb^+_{\rm p}(k_z)}{\hbar \Delta + \mu_0\omega^2\,\db^*\cdot\tilde{\mathbf{G}}_0(k_z)\cdot\db} \Bigg].\label{equationtoreferto}
\end{align}
\begin{figure*}[t]
    \centerline{\includegraphics[width=\textwidth]{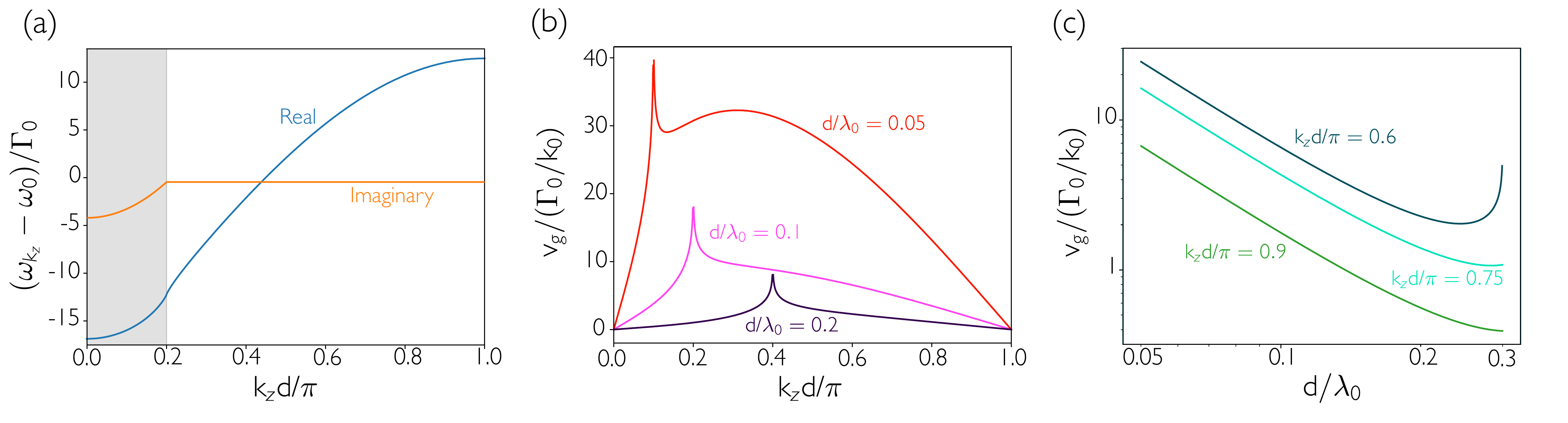}}
    \caption{\textbf{Characterization of the guided mode of a 1D chain of atoms polarized parallel to the chain axis.} \textbf{(a)} Real (blue) and imaginary (orange) part of the dispersion relation of Eq.~\eqref{freqSI}, for a spacing $d=0.1\lambda_0$. In the region enclosed within the light line (shaded), the chain does not guide light and the decay rate of the mode into free space is finite (non-zero imaginary part). \textbf{(b)} Group velocity scaling with $k_z$, for different lattice constants. The light line is approximately in line with the peaks in each curve. \textbf{(c)} Group velocity scaling with lattice constant $d$, for different longitudinal wave-vectors.}
    \label{Fig8}
\end{figure*}
From this equation, we obtain an expression for the Green's tensor of the medium consisting of vacuum modified by the presence of the atomic chain. To do so, we assume that the pump field is generated by a dipole-like source $\mathbf{p}$ at $\rb_p$, which generates the current $\mathbf{j}(\rb,\omega)=-\ii\omega\mathbf{p}\delta(\rb-\rb_\text{p})$, such that
\begin{align}
\Eb^+_{\rm p}(\rb) &= i\mu_0 \omega \int \mathrm{d}\rb'~ \mathbf{G}_0(\rb,\rb',\omega)\cdot\mathbf{j}(\rb',\omega) \\&= \mu_0 \omega^2 \mathbf{G}_0\,(\rb,\rb_\text{p},\omega)\cdot\mathbf{p}.
\end{align}
Substituting the above expression into Eq.~\eqref{equationtoreferto}, and transforming the sum over $k_z$ into an integral over the Brillouin zone, i.e., 
\begin{equation}
\frac{1}{N}\sum\limits_{k_z} \rightarrow \frac{d}{2\pi} \int\limits_{-\pi/d}^{\pi/d} \mathrm{d}k_z,
\end{equation}
we find that $\Eb^+_{\rm p}(\rb)=\mu_0 \omega^2\, \mathbf{G}\,(\rb,\rb_\text{p},\omega)\cdot\mathbf{p}$, with a Green's tensor that now accounts for the presence of the chain and can be written as
\begin{align}
\mathbf{G}(\rb,\rb',\omega) = &\mathbf{G}_0(\rb,\rb',\omega) \\\notag&- \frac{\mu_0\omega^2d}{2\pi\hbar} \int\limits_{-\pi/d}^{\pi/d} \mathrm{d}k_z \frac{\boldsymbol{\alpha}_{k_z}(\rb) \otimes \boldsymbol{\beta}_{k_z} (\rb')}{\Delta + \frac{\mu_0\omega^2}{\hbar}\db^*\cdot\tilde{\mathbf{G}}_0(k_z)\cdot\db},
\end{align}
where we have defined
\begin{subequations}
\begin{equation}
\boldsymbol{\alpha}_{k_z}(\rb) = \sum\limits_j\mathbf{G}_0(\rb,\rb_j,\omega)\cdot\db ~\mathrm{e}^{ik_zz_j},
\end{equation}
\begin{equation}
\boldsymbol{\beta}_{k_z}(\rb) = \sum\limits_j\db^*\cdot\mathbf{G}_0(\rb_j,\rb,\omega) ~\mathrm{e}^{-ik_zz_j}.
\end{equation}
\end{subequations}
We now express the vacuum Green's tensor in cylindrical coordinates, by making use of the integral representation of spherical waves~\cite{Klimov04}
\begin{align}
\frac{\mathrm{e}^{ik|\rb-\rb_j|}}{|\rb-\rb_j|} = \frac{i}{2} \sum\limits_{m=-\infty}^{\infty} \int \mathrm{d} k_z &\mathrm{e}^{im(\phi_j-\phi)} \mathrm{e}^{ik_z(z_j-z)} \\\notag &\times J_m(k_\perp\rho_j) H_m^{(1)}(k_\perp \rho),
\end{align}
where $\rho > \rho_j$, $J_m(\cdot)$ and $H_m^{(1)}(\cdot)$ are Bessel and Hankel functions of the first kind, respectively, and $k_\perp=\sqrt{k^2 - k_z^2}$ is the transversal wave-vector. We choose the waveguide atoms to lie along $z$ with radial and angular coordinates $\rho_j = 0$ and $\phi_j = 0$ for all $j$. This simplifies the sum over azimuthal components in the above expression, as $J_m(k_{\perp}\rho_j)=0$ for $m\neq 0$. Then, the vacuum Green's tensor reduces to
\begin{align}
\mathbf{G}_0(\rb,\rb_j,\omega) = \frac{\ii}{8\pi} &\left[ \mathbb{1} + \frac{1}{k^2} \mathbf{\nnabla}\otimes \mathbf{\nnabla}\right]   \\\notag &~~~\times \int\mathrm{d}k_z \mathrm{e}^{ik_z(z_j-z)} H_0(k_\perp\rho).
\end{align}

Introducing this expression into the equations for $\boldsymbol{\alpha}_{k_z}(\rb)$ and $\boldsymbol{\beta}_{k_z}(\rb)$ and performing the sum over atomic sites, we arrive to the final expression for the total Green's tensor:
\begin{align}
\GG(\rb,\rb',\omega)=\GG_0&(\rb,\rb',\omega) \label{integralequation}\\\notag &+ \frac{3\Gamma_0}{32 kd}\int_{-\pi/d}^{\pi/d} d\kp \frac{\ub_{\kp}(\rb)\otimes\vb_{\kp}(\rb')}{\omega-\omega_{\kp}},
\end{align}
where we have defined the (complex) frequency $\omega_{k_z}=\omega_0-(3\pi\Gamma_0/k)\, \dbu^* \cdot \,\GG_{0} (\kp) \cdot \dbu$, with $\Gamma_0=\omega^3|\db|^2/3\pi\hbar\epsilon_0c^3$ being the spontaneous emission rate of a single waveguide atom in vacuum. In the above equation,
\begin{subequations}
\begin{equation}
\ub_{\kp}(\rb) = \sum_g\left[\mathbb{1}+\frac{1}{k^2}\nnabla\otimes\nnabla\right]\cdot\dbu\,\, e^{\ii(\kp+g)z} H_0^{(1)}(\kpe\rho),
\end{equation}
\begin{equation}
\vb_{\kp}(\rb) = \sum_g\dbu^*\cdot\left[\mathbb{1}+\frac{1}{k^2}\nnabla\otimes\nnabla\right]\,\, e^{-\ii(\kp+g)z} H_0^{(1)}(\kpe\rho),
\end{equation}
\end{subequations}
where the sums are performed over reciprocal-lattice vectors $g=2\pi n/d$ with $n \in\mathbb{Z}$. We note that, for atoms polarized along the direction of the chain, the complex frequency $\omega_{k_z}$ can be written as~\cite{Asenjo17PRX}
\begin{align}\label{freqSI}
\omega_{k_z}&=\omega_0-\frac{3}{2 k^3 d^3} \Big[ \textrm{Li}_{3}\left(e^{\ii (k+k_z)d}\right)+ \textrm{Li}_{3}\left(e^{\ii (k-k_z)d}\right)\notag \\  &-\ii k d   \textrm{Li}_{2}\left(e^{\ii (k+k_z)d}\right)-\ii k d \textrm{Li}_{2}\left(e^{\ii (k-k_z)d}\right) \Big],
\end{align}
where $\Li_s (z) = \sum_{\ell=1}^\infty z^\ell \, \ell^{-s}$ is a polylogarithm function of order $s$.

The decay rate of an impurity qubit placed in the vicinity of the chain is directly related to the imaginary part of the Green's tensor through
\begin{equation}
\Gamma^{\rm q} =\frac{2\mu_0\,\omega^2}{\hbar}\,\db_{\rm q}^*\cdot\text{Im}\,\mathbf{G}(\rb_{\rm q},\rb_{\rm q},\omega)\cdot\db_{\rm q},
\end{equation}
where $\rb_{\rm q}$ and $\db_{\rm q}$ are the qubit position and dipole matrix element, respectively. The integration path for Eq.~\ref{integralequation} is shown in Fig.~\ref{Figcontour}. The integrand displays several branch cuts (associated with the square root and polylogarithm functions), as well as simple poles for $k_z$ such that $\omega_{k_z}=\omega$. We can clearly separate two different contributions to the decay: emission into free space (in the region such that $k_z\in [-k,k]$), and emission into the atomic-waveguide mode (due to the pole at $k_z=\kg$). 

\begin{figure*}[t]
\centerline{\includegraphics[width=\textwidth]{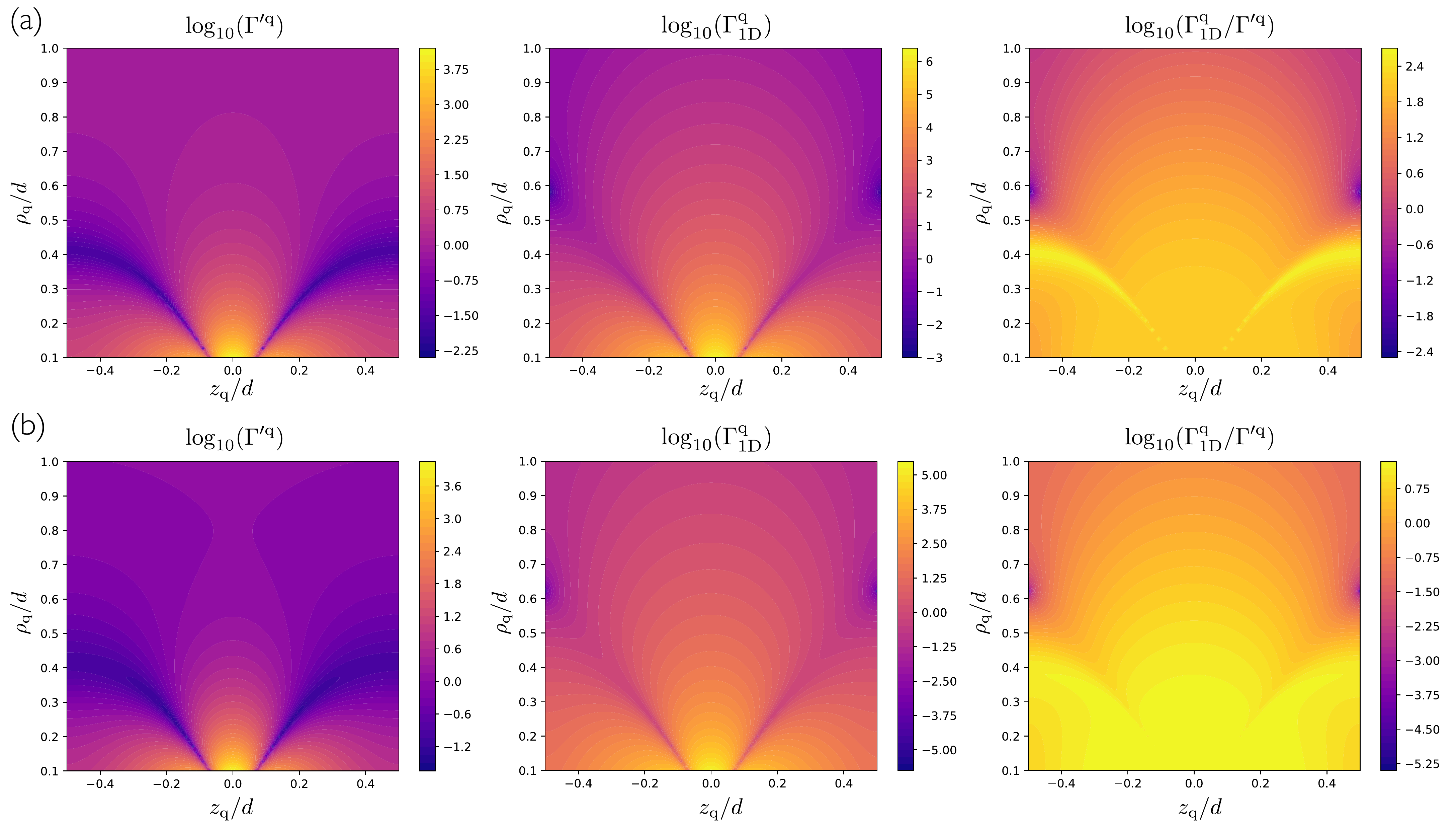}}
\caption{Decay rates into free space, guided mode, and ratio between them, as a function of the qubit radial ($\rho_\text{q}$) and longitudinal ($z_\text{q}$) position for (a) $d=0.1\lambda_0$, and (b) $d=0.2\lambda_0$. For all plots, the detuning between the qubit and the waveguide atoms is such that $\kg=0.7\pi/d$. At $z_\text{q}=0$ the qubit is exactly on top of a waveguide atom.\label{geometryg1d}}
\end{figure*}

\subsection{Free-space decay rate}
The presence of the chain alters the vacuum modes and thus leads to a modified decay rate of the qubit, which is now calculated not only from the vaccuum's Green's tensor, $\mathbf{G}_0$, but also taking into account a contribution to the integral arising from wave-vectors within the light cone, i.e., $k_z\in[-k,k]$. The free-space decay rate is thus readily found to be
\begin{align}
\Gamma'^{\rm q}/\Gamma_0^{\rm q}&=1\\\notag &+\frac{9\pi\Gamma_0}{16 k^2d}\,\text{Im}\int_{-k}^k d\kp \frac{\dbu_{\rm q}^*\cdot \ub_{\kp}(\rb_{\rm q}) \otimes \vb_{\kp}(\rb_{\rm q}) \cdot \dbu_{\rm q}}{\omega-\omega_{\kp}}.
\end{align}
There is also a frequency shift that arises from the real part of the Green's function, which can be calculated numerically by taking the real part of the integrals along $I_r$, $I_c$, and $I_{cc}$, as shown in Fig.~\ref{Figcontour}.

\subsection{Guided-mode decay rate}
For an infinite chain, we can perfectly isolate the decay into the guided mode of the atomic waveguide as it appears as a pole in the integral. Beyond the light line, $\vb_{k_z} = -\ub_{k_z}^\dagger$ and $\omega_{k_z}$ is real (as the guided mode has infinite lifetime, i.e., it is not ``leaky''). This means that the imaginary part of the integral is zero everywhere outside the light cone, except for the poles where $\omega_{k_z}=\omega$. Note that if the driving frequency is detuned from the guided mode band, there is no pole contribution and thus no decay. For longitudinal polarization there are two poles (corresponding to forward and backward propagating guided waves at $\pm \kg$). We close the integral around one of these poles, as shown in Fig.~\ref{Figcontour}, and find
\begin{equation}
\ga/\Gamma_0^{\rm q} = \frac{9\pi\Gamma_0}{16k^2d} \,\mathrm{Im} \ointctrclockwise_{I_\text{pole}} \,dk_z\,\frac{|\dbu_{\rm q}^*\cdot \ub_{k_z}(\rb_{\rm q})|^2}{\omega - \omega_{k_z}}.
\end{equation}
Approximating $\omega_{k_z}\simeq\omega_0 + v_g k_z$, where $v_g$ is the group velocity at frequency $\omega$, and applying Cauchy's residue theorem, we find an expression for the guided decay rate:
\begin{align}
\ga/\Gamma_0^{\rm q}= \frac{9\pi^2\Gamma_0}{8k^2dv_g}|\dbu_{\rm q}^*\cdot \ub_{\kg}(\rb_{\rm q})|^2.
\end{align}

\begin{figure*}[t!]
    \centerline{\includegraphics[width=\textwidth]{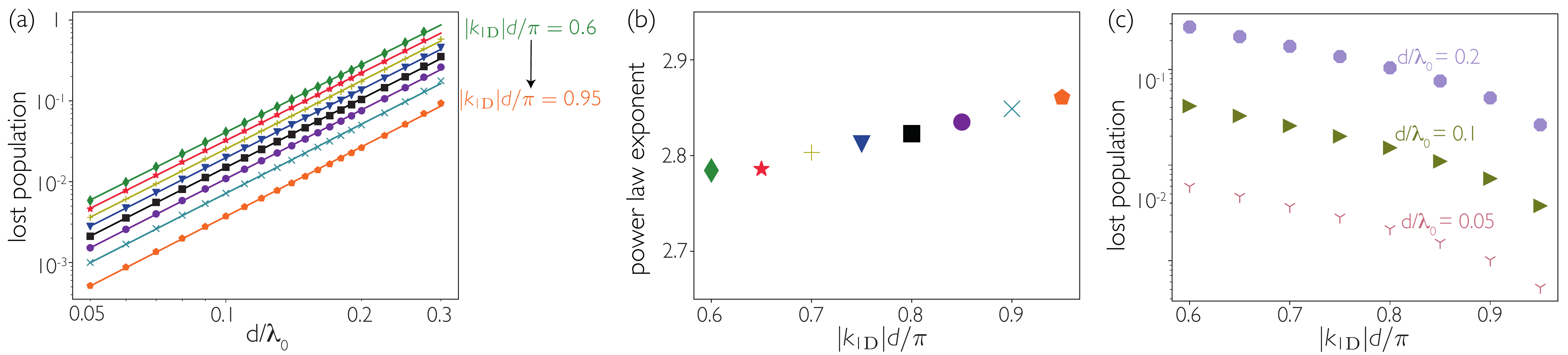}}
    \caption{{\textbf{Loss during collisions in atomic-waveguide QED.} \textbf{(a)} Scaling of lost two-photon population $\gamma=1 - \langle\hat{\sigma}_{ee}^{(2)}(t_f)\rangle / \langle\hat{\sigma}_{ee}^{(2)}(0)\rangle$, with interatomic distance for initial wave-vector $\pm\kg$ values evenly distributed between $0.6\pi/d$ and $0.9\pi/d$. Continuous lines are guides to the eye and are found as fits to a power law $\gamma(d) \simeq A d^B$. \textbf{(b)} Exponents $B$ from power law fits in (a). \textbf{(c)} Scaling of lost two-photon population $\gamma=1-\langle\hat{\sigma}_{ee}^{(2)}(t_f)\rangle\langle\hat{\sigma}_{ee}^{(2)}(0)\rangle$, with initial wave-vector $\pm \kg$ for different $d$. In all cases, initial wavepackets have centres 60 sites apart, and propagate in a chain of $N=120$ atoms until the peaks reach their initial positions.}}
    \label{fig13}
\end{figure*}

\subsection{Functional form of the field modes and scalings of the group velocity}

For waveguide atoms polarized along the direction of the chain, the different polarization components of $\ub_{\kp}(\rb)$ read
\begin{align}
\hat{\rho}\cdot\ub_{\kp}(\rb) &= -\ii \sum_g \,\frac{(\kp+g) k_\perp}{k^2} \,e^{\ii(\kp+g)z} H_1^{(1)}(\kpe\rho),\\
\hat{\phi}\cdot\ub_{\kp}(\rb)  &=0,\\
\hat{z}\cdot\ub_{\kp}(\rb) &= \sum_g \left[1 - \frac{(\kp+g)^2}{k^2} \right] e^{\ii(\kp+g)z} H_0^{(1)}(k_\perp\rho).
\end{align}
The components of $\vb_{\kp}(\rb)$ admit similar expressions.

For an infinite array, we can calculate the group velocity of the guided modes as the derivative of the dispersion relation of Eq.~\eqref{freqSI}, i.e., $v_g=\partial \omega_{\kp}/\partial \kp|_{\kp=\kg}$. We show the group velocity scalings with $d$ and $k_z$ in Fig.~\ref{Fig8}. Note that $k\simeq k_0$ as $\omega\simeq \omega_0,\omega_q$ except for deviations of the order of $\Gamma_0\ll\omega_0,\omega_q$. Beyond the light line [corresponding to the peaks in Fig~\ref{Fig8}(b)] the mode is guided, and the group velocity tends to zero as $k_z$ approaches the edge of the Brillouin zone, though the dependence on $k_z$ is not trivial for small distances. For fixed $k_z d$, $v_g \sim d^{-1.7}$ where the exponent is approximate and varies slightly for different $k_z$.

\section{Spatial dependence of the decay rates\label{contourplotappendix}}

The decay rate into free space is not simply that of a qubit in vacuum ($\Gamma_0^\text{q}$), but is modified by the presence of the atomic waveguide, which alters the vacuum modes. This decay rate displays a non-trivial dependence on the position of the qubit, as shown in Fig.~\ref{fig9}. A similar scaling is followed by the decay rate into the waveguide mode. Generically, both decay rates are enhanced for short radial and longitudinal distances to the waveguide atoms. However, there are \textit{magic points} -- manifested as dark lines in the figures -- where decay isstrongly suppressed due to interference effects. For the free space scattering rate, these lines appear as a narrow band at $\rho_{\rm q}\approx 0.4d$ in between two array atoms (at $z_{\rm q}=\pm 0.5d$) and then move towards the central atom as $\rho_{\rm q}$ decreases. For the guided-mode scattering rate, these positions draw  virtually straight lines that appear at $\rho_{\rm q}\approx0.6d$ in the middle of two array atoms. This translates into a ratio between guided-mode and free-space scattering that is strongly enhanced at $z_{\rm q}=\pm 0.5d$, as shown in Fig.~2 in the main text. This pattern displays only minor changes when altering $\kg$ and $d$.

As we discuss in the main text, a waveguide that is one-atom thick provides an optical depth $\sim30$ times larger than that of a fiber. To estimate these numbers, we have considered a waveguide with lattice constant $d=0.2\lambda_0$ and a qubit frequency such that $\kg=0.7\pi$, placed at $z_{\rm q}=0$, $\rho_{\rm q}=0.1\lambda_0$ [red line in Fig.~2(b) in the main text]. The fiber has radius $k_0r=1.2$ and is made of silicon nitride, with dielectric constant $\epsilon=4$. The qubit is located at $\rho_{\rm q}$ from the surface of the fiber (this leads to $\ga/\gap\simeq 0.3$~\cite{Asenjo17PRX}). Note that the coupling to the atomic waveguide can be increased by placing the qubit frequency closer to the bandedge.

\section{Loss during photon collisions\label{powerlawappendix}}

Dissipation during photon collisions is controlled by both the spacing of the array and the central wave-vectors $\pm\kg$ of the counter-propagating wavepackets. As shown in Fig.~\ref{fig13}(a), loss is minimized for small inter-atomic distance $d$ and $\kg$ close to the edge of the band. Conversely, loss is maximized for large $d$ and $\kg$ close to the light cone. For fixed $\kg$, the loss as a function of inter-atomic spacing can be approximated by a power law, $\gamma(d) \simeq Ad^B$. Generally, the fitted power law is of the form $\gamma(d) \simeq Ad^{2.8}$, though as $|\kg|$ is increased toward the edge of the band, the exponent slowly increases~[Fig.~\ref{fig13}(b)]. In Fig.~\ref{fig13}(c), we plot loss as a function of $\pm\kg$ for fixed $d$. Generally, larger $|\kg|$ leads to lower loss, as the initial spin waves are further away from the light cone. This data does not admit an accurate fit by trivial functions as, for large wave-vectors, the group velocity is extremely slow and both dispersion and single-photon loss due to finite size effects impact the dynamics.

\color{black}


\begin{thebibliography}{120}%
\makeatletter
\providecommand \@ifxundefined [1]{%
 \@ifx{#1\undefined}
}%
\providecommand \@ifnum [1]{%
 \ifnum #1\expandafter \@firstoftwo
 \else \expandafter \@secondoftwo
 \fi
}%
\providecommand \@ifx [1]{%
 \ifx #1\expandafter \@firstoftwo
 \else \expandafter \@secondoftwo
 \fi
}%
\providecommand \natexlab [1]{#1}%
\providecommand \enquote  [1]{``#1''}%
\providecommand \bibnamefont  [1]{#1}%
\providecommand \bibfnamefont [1]{#1}%
\providecommand \citenamefont [1]{#1}%
\providecommand \href@noop [0]{\@secondoftwo}%
\providecommand \href [0]{\begingroup \@sanitize@url \@href}%
\providecommand \@href[1]{\@@startlink{#1}\@@href}%
\providecommand \@@href[1]{\endgroup#1\@@endlink}%
\providecommand \@sanitize@url [0]{\catcode `\\12\catcode `\$12\catcode
  `\&12\catcode `\#12\catcode `\^12\catcode `\_12\catcode `\%12\relax}%
\providecommand \@@startlink[1]{}%
\providecommand \@@endlink[0]{}%
\providecommand \url  [0]{\begingroup\@sanitize@url \@url }%
\providecommand \@url [1]{\endgroup\@href {#1}{\urlprefix }}%
\providecommand \urlprefix  [0]{URL }%
\providecommand \Eprint [0]{\href }%
\providecommand \doibase [0]{https://doi.org/}%
\providecommand \selectlanguage [0]{\@gobble}%
\providecommand \bibinfo  [0]{\@secondoftwo}%
\providecommand \bibfield  [0]{\@secondoftwo}%
\providecommand \translation [1]{[#1]}%
\providecommand \BibitemOpen [0]{}%
\providecommand \bibitemStop [0]{}%
\providecommand \bibitemNoStop [0]{.\EOS\space}%
\providecommand \EOS [0]{\spacefactor3000\relax}%
\providecommand \BibitemShut  [1]{\csname bibitem#1\endcsname}%
\let\auto@bib@innerbib\@empty
%</preamble>
\bibitem [{\citenamefont {Noh}\ and\ \citenamefont {Angelakis}(2017)}]{Noh17}%
  \BibitemOpen
  \bibfield  {author} {\bibinfo {author} {\bibfnamefont {C.}~\bibnamefont
  {Noh}}\ and\ \bibinfo {author} {\bibfnamefont {D.~G.}\ \bibnamefont
  {Angelakis}},\ }\bibfield  {title} {\bibinfo {title} {Quantum simulations and
  many-body physics with light},\ }\href
  {http://stacks.iop.org/0034-4885/80/i=1/a=016401} {\bibfield  {journal}
  {\bibinfo  {journal} {Rep. Prog. Phys.}\ }\textbf {\bibinfo {volume} {80}},\
  \bibinfo {pages} {016401} (\bibinfo {year} {2017})}\BibitemShut {NoStop}%
\bibitem [{\citenamefont {Fleischhauer}\ \emph {et~al.}(2005)\citenamefont
  {Fleischhauer}, \citenamefont {Imamoglu},\ and\ \citenamefont
  {Marangos}}]{Fleischhauer05}%
  \BibitemOpen
  \bibfield  {author} {\bibinfo {author} {\bibfnamefont {M.}~\bibnamefont
  {Fleischhauer}}, \bibinfo {author} {\bibfnamefont {A.}~\bibnamefont
  {Imamoglu}},\ and\ \bibinfo {author} {\bibfnamefont {J.~P.}\ \bibnamefont
  {Marangos}},\ }\bibfield  {title} {\bibinfo {title} {Electromagnetically
  induced transparency: Optics in coherent media},\ }\href
  {https://doi.org/10.1103/RevModPhys.77.633} {\bibfield  {journal} {\bibinfo
  {journal} {Rev. Mod. Phys.}\ }\textbf {\bibinfo {volume} {77}},\ \bibinfo
  {pages} {633} (\bibinfo {year} {2005})}\BibitemShut {NoStop}%
\bibitem [{\citenamefont {Hammerer}\ \emph {et~al.}(2010)\citenamefont
  {Hammerer}, \citenamefont {S\o{}rensen},\ and\ \citenamefont
  {Polzik}}]{Hammerer10}%
  \BibitemOpen
  \bibfield  {author} {\bibinfo {author} {\bibfnamefont {K.}~\bibnamefont
  {Hammerer}}, \bibinfo {author} {\bibfnamefont {A.~S.}\ \bibnamefont
  {S\o{}rensen}},\ and\ \bibinfo {author} {\bibfnamefont {E.~S.}\ \bibnamefont
  {Polzik}},\ }\bibfield  {title} {\bibinfo {title} {Quantum interface between
  light and atomic ensembles},\ }\href
  {https://doi.org/10.1103/RevModPhys.82.1041} {\bibfield  {journal} {\bibinfo
  {journal} {Rev. Mod. Phys.}\ }\textbf {\bibinfo {volume} {82}},\ \bibinfo
  {pages} {1041} (\bibinfo {year} {2010})}\BibitemShut {NoStop}%
\bibitem [{\citenamefont {Peyronel}\ \emph {et~al.}(2012)\citenamefont
  {Peyronel}, \citenamefont {Firstenberg}, \citenamefont {Liang}, \citenamefont
  {Hofferberth}, \citenamefont {Gorshkov}, \citenamefont {Pohl}, \citenamefont
  {Lukin},\ and\ \citenamefont {Vuleti{\'c}}}]{Peyronel12}%
  \BibitemOpen
  \bibfield  {author} {\bibinfo {author} {\bibfnamefont {T.}~\bibnamefont
  {Peyronel}}, \bibinfo {author} {\bibfnamefont {O.}~\bibnamefont
  {Firstenberg}}, \bibinfo {author} {\bibfnamefont {Q.-Y.}\ \bibnamefont
  {Liang}}, \bibinfo {author} {\bibfnamefont {S.}~\bibnamefont {Hofferberth}},
  \bibinfo {author} {\bibfnamefont {A.~V.}\ \bibnamefont {Gorshkov}}, \bibinfo
  {author} {\bibfnamefont {T.}~\bibnamefont {Pohl}}, \bibinfo {author}
  {\bibfnamefont {M.~D.}\ \bibnamefont {Lukin}},\ and\ \bibinfo {author}
  {\bibfnamefont {V.}~\bibnamefont {Vuleti{\'c}}},\ }\bibfield  {title}
  {\bibinfo {title} {Quantum nonlinear optics with single photons enabled by
  strongly interacting atoms},\ }\href {https://doi.org/10.1038/nature11361}
  {\bibfield  {journal} {\bibinfo  {journal} {Nature}\ }\textbf {\bibinfo
  {volume} {488}},\ \bibinfo {pages} {57} (\bibinfo {year} {2012})}\BibitemShut
  {NoStop}%
\bibitem [{\citenamefont {Chang}\ \emph {et~al.}(2014)\citenamefont {Chang},
  \citenamefont {Vuleti{\'c}},\ and\ \citenamefont {Lukin}}]{Chang14}%
  \BibitemOpen
  \bibfield  {author} {\bibinfo {author} {\bibfnamefont {D.~E.}\ \bibnamefont
  {Chang}}, \bibinfo {author} {\bibfnamefont {V.}~\bibnamefont {Vuleti{\'c}}},\
  and\ \bibinfo {author} {\bibfnamefont {M.~D.}\ \bibnamefont {Lukin}},\
  }\bibfield  {title} {\bibinfo {title} {Quantum nonlinear optics ---photon by
  photon},\ }\href {https://doi.org/10.1038/nphoton.2014.192} {\bibfield
  {journal} {\bibinfo  {journal} {Nat. Photon.}\ }\textbf {\bibinfo {volume}
  {8}},\ \bibinfo {pages} {685} (\bibinfo {year} {2014})}\BibitemShut {NoStop}%
\bibitem [{\citenamefont {Ma}\ \emph {et~al.}(2011)\citenamefont {Ma},
  \citenamefont {Wang}, \citenamefont {Sun},\ and\ \citenamefont
  {Nori}}]{Ma11}%
  \BibitemOpen
  \bibfield  {author} {\bibinfo {author} {\bibfnamefont {J.}~\bibnamefont
  {Ma}}, \bibinfo {author} {\bibfnamefont {X.}~\bibnamefont {Wang}}, \bibinfo
  {author} {\bibfnamefont {C.}~\bibnamefont {Sun}},\ and\ \bibinfo {author}
  {\bibfnamefont {F.}~\bibnamefont {Nori}},\ }\bibfield  {title} {\bibinfo
  {title} {Quantum spin squeezing},\ }\href
  {https://doi.org/http://dx.doi.org/10.1016/j.physrep.2011.08.003} {\bibfield
  {journal} {\bibinfo  {journal} {Phys. Rep.}\ }\textbf {\bibinfo {volume}
  {509}},\ \bibinfo {pages} {89 } (\bibinfo {year} {2011})}\BibitemShut
  {NoStop}%
\bibitem [{\citenamefont {Pezz\`e}\ \emph {et~al.}(2018)\citenamefont
  {Pezz\`e}, \citenamefont {Smerzi}, \citenamefont {Oberthaler}, \citenamefont
  {Schmied},\ and\ \citenamefont {Treutlein}}]{Pezze18}%
  \BibitemOpen
  \bibfield  {author} {\bibinfo {author} {\bibfnamefont {L.}~\bibnamefont
  {Pezz\`e}}, \bibinfo {author} {\bibfnamefont {A.}~\bibnamefont {Smerzi}},
  \bibinfo {author} {\bibfnamefont {M.~K.}\ \bibnamefont {Oberthaler}},
  \bibinfo {author} {\bibfnamefont {R.}~\bibnamefont {Schmied}},\ and\ \bibinfo
  {author} {\bibfnamefont {P.}~\bibnamefont {Treutlein}},\ }\bibfield  {title}
  {\bibinfo {title} {Quantum metrology with nonclassical states of atomic
  ensembles},\ }\href {https://doi.org/10.1103/RevModPhys.90.035005} {\bibfield
   {journal} {\bibinfo  {journal} {Rev. Mod. Phys.}\ }\textbf {\bibinfo
  {volume} {90}},\ \bibinfo {pages} {035005} (\bibinfo {year}
  {2018})}\BibitemShut {NoStop}%
\bibitem [{\citenamefont {{Le Kien}}\ \emph {et~al.}(2005)\citenamefont {{Le
  Kien}}, \citenamefont {{Dutta Gupta}}, \citenamefont {Nayak},\ and\
  \citenamefont {Hakuta}}]{LeKien05}%
  \BibitemOpen
  \bibfield  {author} {\bibinfo {author} {\bibfnamefont {F.}~\bibnamefont {{Le
  Kien}}}, \bibinfo {author} {\bibfnamefont {S.}~\bibnamefont {{Dutta Gupta}}},
  \bibinfo {author} {\bibfnamefont {K.~P.}\ \bibnamefont {Nayak}},\ and\
  \bibinfo {author} {\bibfnamefont {K.}~\bibnamefont {Hakuta}},\ }\bibfield
  {title} {\bibinfo {title} {Nanofiber-mediated radiative transfer between two
  distant atoms},\ }\href {https://doi.org/10.1103/PhysRevA.72.063815}
  {\bibfield  {journal} {\bibinfo  {journal} {Phys. Rev. A}\ }\textbf {\bibinfo
  {volume} {72}},\ \bibinfo {pages} {063815} (\bibinfo {year}
  {2005})}\BibitemShut {NoStop}%
\bibitem [{\citenamefont {Vetsch}\ \emph {et~al.}(2010)\citenamefont {Vetsch},
  \citenamefont {Reitz}, \citenamefont {Sagu\'e}, \citenamefont {Schmidt},
  \citenamefont {Dawkins},\ and\ \citenamefont {Rauschenbeutel}}]{Vetsch10}%
  \BibitemOpen
  \bibfield  {author} {\bibinfo {author} {\bibfnamefont {E.}~\bibnamefont
  {Vetsch}}, \bibinfo {author} {\bibfnamefont {D.}~\bibnamefont {Reitz}},
  \bibinfo {author} {\bibfnamefont {G.}~\bibnamefont {Sagu\'e}}, \bibinfo
  {author} {\bibfnamefont {R.}~\bibnamefont {Schmidt}}, \bibinfo {author}
  {\bibfnamefont {S.~T.}\ \bibnamefont {Dawkins}},\ and\ \bibinfo {author}
  {\bibfnamefont {A.}~\bibnamefont {Rauschenbeutel}},\ }\bibfield  {title}
  {\bibinfo {title} {Optical interface created by laser-cooled atoms trapped in
  the evanescent field surrounding an optical nanofiber},\ }\href
  {https://doi.org/10.1103/PhysRevLett.104.203603} {\bibfield  {journal}
  {\bibinfo  {journal} {Phys. Rev. Lett.}\ }\textbf {\bibinfo {volume} {104}},\
  \bibinfo {pages} {203603} (\bibinfo {year} {2010})}\BibitemShut {NoStop}%
\bibitem [{\citenamefont {Goban}\ \emph {et~al.}(2012)\citenamefont {Goban},
  \citenamefont {Choi}, \citenamefont {Alton}, \citenamefont {Ding},
  \citenamefont {Lacro\^ute}, \citenamefont {Pototschnig}, \citenamefont
  {Thiele}, \citenamefont {Stern},\ and\ \citenamefont {Kimble}}]{Goban12}%
  \BibitemOpen
  \bibfield  {author} {\bibinfo {author} {\bibfnamefont {A.}~\bibnamefont
  {Goban}}, \bibinfo {author} {\bibfnamefont {K.~S.}\ \bibnamefont {Choi}},
  \bibinfo {author} {\bibfnamefont {D.~J.}\ \bibnamefont {Alton}}, \bibinfo
  {author} {\bibfnamefont {D.}~\bibnamefont {Ding}}, \bibinfo {author}
  {\bibfnamefont {C.}~\bibnamefont {Lacro\^ute}}, \bibinfo {author}
  {\bibfnamefont {M.}~\bibnamefont {Pototschnig}}, \bibinfo {author}
  {\bibfnamefont {T.}~\bibnamefont {Thiele}}, \bibinfo {author} {\bibfnamefont
  {N.~P.}\ \bibnamefont {Stern}},\ and\ \bibinfo {author} {\bibfnamefont
  {H.~J.}\ \bibnamefont {Kimble}},\ }\bibfield  {title} {\bibinfo {title}
  {Demonstration of a state-insensitive, compensated nanofiber trap},\ }\href
  {https://doi.org/10.1103/PhysRevLett.109.033603} {\bibfield  {journal}
  {\bibinfo  {journal} {Phys. Rev. Lett.}\ }\textbf {\bibinfo {volume} {109}},\
  \bibinfo {pages} {033603} (\bibinfo {year} {2012})}\BibitemShut {NoStop}%
\bibitem [{\citenamefont {Gouraud}\ \emph {et~al.}(2015)\citenamefont
  {Gouraud}, \citenamefont {Maxein}, \citenamefont {Nicolas}, \citenamefont
  {Morin},\ and\ \citenamefont {Laurat}}]{Gouraud15}%
  \BibitemOpen
  \bibfield  {author} {\bibinfo {author} {\bibfnamefont {B.}~\bibnamefont
  {Gouraud}}, \bibinfo {author} {\bibfnamefont {D.}~\bibnamefont {Maxein}},
  \bibinfo {author} {\bibfnamefont {A.}~\bibnamefont {Nicolas}}, \bibinfo
  {author} {\bibfnamefont {O.}~\bibnamefont {Morin}},\ and\ \bibinfo {author}
  {\bibfnamefont {J.}~\bibnamefont {Laurat}},\ }\bibfield  {title} {\bibinfo
  {title} {Demonstration of a memory for tightly guided light in an optical
  nanofiber},\ }\href {https://doi.org/10.1103/PhysRevLett.114.180503}
  {\bibfield  {journal} {\bibinfo  {journal} {Phys. Rev. Lett.}\ }\textbf
  {\bibinfo {volume} {114}},\ \bibinfo {pages} {180503} (\bibinfo {year}
  {2015})}\BibitemShut {NoStop}%
\bibitem [{\citenamefont {Thompson}\ \emph {et~al.}(2013)\citenamefont
  {Thompson}, \citenamefont {Tiecke}, \citenamefont {de~Leon}, \citenamefont
  {Feist}, \citenamefont {Akimov}, \citenamefont {Gullans}, \citenamefont
  {Zibrov}, \citenamefont {Vuleti{\'c}},\ and\ \citenamefont
  {Lukin}}]{Thompson13}%
  \BibitemOpen
  \bibfield  {author} {\bibinfo {author} {\bibfnamefont {J.~D.}\ \bibnamefont
  {Thompson}}, \bibinfo {author} {\bibfnamefont {T.~G.}\ \bibnamefont
  {Tiecke}}, \bibinfo {author} {\bibfnamefont {N.~P.}\ \bibnamefont {de~Leon}},
  \bibinfo {author} {\bibfnamefont {J.}~\bibnamefont {Feist}}, \bibinfo
  {author} {\bibfnamefont {A.~V.}\ \bibnamefont {Akimov}}, \bibinfo {author}
  {\bibfnamefont {M.}~\bibnamefont {Gullans}}, \bibinfo {author} {\bibfnamefont
  {A.~S.}\ \bibnamefont {Zibrov}}, \bibinfo {author} {\bibfnamefont
  {V.}~\bibnamefont {Vuleti{\'c}}},\ and\ \bibinfo {author} {\bibfnamefont
  {M.~D.}\ \bibnamefont {Lukin}},\ }\bibfield  {title} {\bibinfo {title}
  {Coupling a single trapped atom to a nanoscale optical cavity},\ }\href
  {https://doi.org/10.1126/science.1237125} {\bibfield  {journal} {\bibinfo
  {journal} {Science}\ }\textbf {\bibinfo {volume} {340}},\ \bibinfo {pages}
  {1202} (\bibinfo {year} {2013})}\BibitemShut {NoStop}%
\bibitem [{\citenamefont {Goban}\ \emph {et~al.}(2014)\citenamefont {Goban},
  \citenamefont {Hung}, \citenamefont {Yu}, \citenamefont {Hood}, \citenamefont
  {Muniz}, \citenamefont {Lee}, \citenamefont {Martin}, \citenamefont
  {McClung}, \citenamefont {Choi}, \citenamefont {Chang}, \citenamefont
  {Painter},\ and\ \citenamefont {Kimble}}]{Goban14}%
  \BibitemOpen
  \bibfield  {author} {\bibinfo {author} {\bibfnamefont {A.}~\bibnamefont
  {Goban}}, \bibinfo {author} {\bibfnamefont {C.~L.}\ \bibnamefont {Hung}},
  \bibinfo {author} {\bibfnamefont {S.~P.}\ \bibnamefont {Yu}}, \bibinfo
  {author} {\bibfnamefont {J.~D.}\ \bibnamefont {Hood}}, \bibinfo {author}
  {\bibfnamefont {J.~A.}\ \bibnamefont {Muniz}}, \bibinfo {author}
  {\bibfnamefont {J.~H.}\ \bibnamefont {Lee}}, \bibinfo {author} {\bibfnamefont
  {M.~J.}\ \bibnamefont {Martin}}, \bibinfo {author} {\bibfnamefont {A.~C.}\
  \bibnamefont {McClung}}, \bibinfo {author} {\bibfnamefont {K.~S.}\
  \bibnamefont {Choi}}, \bibinfo {author} {\bibfnamefont {D.~E.}\ \bibnamefont
  {Chang}}, \bibinfo {author} {\bibfnamefont {O.}~\bibnamefont {Painter}},\
  and\ \bibinfo {author} {\bibfnamefont {H.~J.}\ \bibnamefont {Kimble}},\
  }\bibfield  {title} {\bibinfo {title} {Atom--light interactions in photonic
  crystals},\ }\href {http://dx.doi.org/10.1038/ncomms4808} {\bibfield
  {journal} {\bibinfo  {journal} {Nat. Commun.}\ }\textbf {\bibinfo {volume}
  {5}} (\bibinfo {year} {2014})}\BibitemShut {NoStop}%
\bibitem [{\citenamefont {Goban}\ \emph {et~al.}(2015)\citenamefont {Goban},
  \citenamefont {Hung}, \citenamefont {Hood}, \citenamefont {Yu}, \citenamefont
  {Muniz}, \citenamefont {Painter},\ and\ \citenamefont {Kimble}}]{Goban15}%
  \BibitemOpen
  \bibfield  {author} {\bibinfo {author} {\bibfnamefont {A.}~\bibnamefont
  {Goban}}, \bibinfo {author} {\bibfnamefont {C.-L.}\ \bibnamefont {Hung}},
  \bibinfo {author} {\bibfnamefont {J.~D.}\ \bibnamefont {Hood}}, \bibinfo
  {author} {\bibfnamefont {S.-P.}\ \bibnamefont {Yu}}, \bibinfo {author}
  {\bibfnamefont {J.~A.}\ \bibnamefont {Muniz}}, \bibinfo {author}
  {\bibfnamefont {O.}~\bibnamefont {Painter}},\ and\ \bibinfo {author}
  {\bibfnamefont {H.~J.}\ \bibnamefont {Kimble}},\ }\bibfield  {title}
  {\bibinfo {title} {Superradiance for atoms trapped along a photonic crystal
  waveguide},\ }\href {https://doi.org/10.1103/PhysRevLett.115.063601}
  {\bibfield  {journal} {\bibinfo  {journal} {Phys. Rev. Lett.}\ }\textbf
  {\bibinfo {volume} {115}},\ \bibinfo {pages} {063601} (\bibinfo {year}
  {2015})}\BibitemShut {NoStop}%
\bibitem [{\citenamefont {Hood}\ \emph {et~al.}(2016)\citenamefont {Hood},
  \citenamefont {Goban}, \citenamefont {Asenjo-Garcia}, \citenamefont {Lu},
  \citenamefont {Yu}, \citenamefont {Chang},\ and\ \citenamefont
  {Kimble}}]{Hood16}%
  \BibitemOpen
  \bibfield  {author} {\bibinfo {author} {\bibfnamefont {J.~D.}\ \bibnamefont
  {Hood}}, \bibinfo {author} {\bibfnamefont {A.}~\bibnamefont {Goban}},
  \bibinfo {author} {\bibfnamefont {A.}~\bibnamefont {Asenjo-Garcia}}, \bibinfo
  {author} {\bibfnamefont {M.}~\bibnamefont {Lu}}, \bibinfo {author}
  {\bibfnamefont {S.-P.}\ \bibnamefont {Yu}}, \bibinfo {author} {\bibfnamefont
  {D.~E.}\ \bibnamefont {Chang}},\ and\ \bibinfo {author} {\bibfnamefont
  {H.~J.}\ \bibnamefont {Kimble}},\ }\bibfield  {title} {\bibinfo {title}
  {Atom{\textendash}atom interactions around the band edge of a photonic
  crystal waveguide},\ }\href {https://doi.org/10.1073/pnas.1603788113}
  {\bibfield  {journal} {\bibinfo  {journal} {Proc. Natl. Acad. Sci. USA}\
  }\textbf {\bibinfo {volume} {113}},\ \bibinfo {pages} {10507} (\bibinfo
  {year} {2016})}\BibitemShut {NoStop}%
\bibitem [{\citenamefont {Chang}\ \emph {et~al.}(2018)\citenamefont {Chang},
  \citenamefont {Douglas}, \citenamefont {Gonz\'alez-Tudela}, \citenamefont
  {Hung},\ and\ \citenamefont {Kimble}}]{Chang18}%
  \BibitemOpen
  \bibfield  {author} {\bibinfo {author} {\bibfnamefont {D.~E.}\ \bibnamefont
  {Chang}}, \bibinfo {author} {\bibfnamefont {J.~S.}\ \bibnamefont {Douglas}},
  \bibinfo {author} {\bibfnamefont {A.}~\bibnamefont {Gonz\'alez-Tudela}},
  \bibinfo {author} {\bibfnamefont {C.-L.}\ \bibnamefont {Hung}},\ and\
  \bibinfo {author} {\bibfnamefont {H.~J.}\ \bibnamefont {Kimble}},\ }\bibfield
   {title} {\bibinfo {title} {Colloquium: Quantum matter built from nanoscopic
  lattices of atoms and photons},\ }\href
  {https://doi.org/10.1103/RevModPhys.90.031002} {\bibfield  {journal}
  {\bibinfo  {journal} {Rev. Mod. Phys.}\ }\textbf {\bibinfo {volume} {90}},\
  \bibinfo {pages} {031002} (\bibinfo {year} {2018})}\BibitemShut {NoStop}%
\bibitem [{\citenamefont {John}\ and\ \citenamefont {Wang}(1990)}]{John90}%
  \BibitemOpen
  \bibfield  {author} {\bibinfo {author} {\bibfnamefont {S.}~\bibnamefont
  {John}}\ and\ \bibinfo {author} {\bibfnamefont {J.}~\bibnamefont {Wang}},\
  }\bibfield  {title} {\bibinfo {title} {Quantum electrodynamics near a
  photonic band gap: Photon bound states and dressed atoms},\ }\href
  {https://doi.org/10.1103/PhysRevLett.64.2418} {\bibfield  {journal} {\bibinfo
   {journal} {Phys. Rev. Lett.}\ }\textbf {\bibinfo {volume} {64}},\ \bibinfo
  {pages} {2418} (\bibinfo {year} {1990})}\BibitemShut {NoStop}%
\bibitem [{\citenamefont {John}\ and\ \citenamefont {Wang}(1991)}]{John91}%
  \BibitemOpen
  \bibfield  {author} {\bibinfo {author} {\bibfnamefont {S.}~\bibnamefont
  {John}}\ and\ \bibinfo {author} {\bibfnamefont {J.}~\bibnamefont {Wang}},\
  }\bibfield  {title} {\bibinfo {title} {Quantum optics of localized light in a
  photonic band gap},\ }\href {https://doi.org/10.1103/PhysRevB.43.12772}
  {\bibfield  {journal} {\bibinfo  {journal} {Phys. Rev. B}\ }\textbf {\bibinfo
  {volume} {43}},\ \bibinfo {pages} {12772} (\bibinfo {year}
  {1991})}\BibitemShut {NoStop}%
\bibitem [{\citenamefont {Sundaresan}\ \emph {et~al.}(2019)\citenamefont
  {Sundaresan}, \citenamefont {Lundgren}, \citenamefont {Zhu}, \citenamefont
  {Gorshkov},\ and\ \citenamefont {Houck}}]{Sundaresan19}%
  \BibitemOpen
  \bibfield  {author} {\bibinfo {author} {\bibfnamefont {N.~M.}\ \bibnamefont
  {Sundaresan}}, \bibinfo {author} {\bibfnamefont {R.}~\bibnamefont
  {Lundgren}}, \bibinfo {author} {\bibfnamefont {G.}~\bibnamefont {Zhu}},
  \bibinfo {author} {\bibfnamefont {A.~V.}\ \bibnamefont {Gorshkov}},\ and\
  \bibinfo {author} {\bibfnamefont {A.~A.}\ \bibnamefont {Houck}},\ }\bibfield
  {title} {\bibinfo {title} {Interacting qubit-photon bound states with
  superconducting circuits},\ }\href
  {https://doi.org/10.1103/PhysRevX.9.011021} {\bibfield  {journal} {\bibinfo
  {journal} {Phys. Rev. X}\ }\textbf {\bibinfo {volume} {9}},\ \bibinfo {pages}
  {011021} (\bibinfo {year} {2019})}\BibitemShut {NoStop}%
\bibitem [{\citenamefont {John}\ and\ \citenamefont {Quang}(1996)}]{John96}%
  \BibitemOpen
  \bibfield  {author} {\bibinfo {author} {\bibfnamefont {S.}~\bibnamefont
  {John}}\ and\ \bibinfo {author} {\bibfnamefont {T.}~\bibnamefont {Quang}},\
  }\bibfield  {title} {\bibinfo {title} {Quantum optical spin-glass state of
  impurity two-level atoms in a photonic band gap},\ }\href
  {https://doi.org/10.1103/PhysRevLett.76.1320} {\bibfield  {journal} {\bibinfo
   {journal} {Phys. Rev. Lett.}\ }\textbf {\bibinfo {volume} {76}},\ \bibinfo
  {pages} {1320} (\bibinfo {year} {1996})}\BibitemShut {NoStop}%
\bibitem [{\citenamefont {Douglas}\ \emph {et~al.}(2015)\citenamefont
  {Douglas}, \citenamefont {Habibian}, \citenamefont {Hung}, \citenamefont
  {Gorshkov}, \citenamefont {Kimble},\ and\ \citenamefont {Chang}}]{Douglas15}%
  \BibitemOpen
  \bibfield  {author} {\bibinfo {author} {\bibfnamefont {J.~S.}\ \bibnamefont
  {Douglas}}, \bibinfo {author} {\bibfnamefont {H.}~\bibnamefont {Habibian}},
  \bibinfo {author} {\bibfnamefont {C.~L.}\ \bibnamefont {Hung}}, \bibinfo
  {author} {\bibfnamefont {A.~V.}\ \bibnamefont {Gorshkov}}, \bibinfo {author}
  {\bibfnamefont {H.~J.}\ \bibnamefont {Kimble}},\ and\ \bibinfo {author}
  {\bibfnamefont {D.~E.}\ \bibnamefont {Chang}},\ }\bibfield  {title} {\bibinfo
  {title} {Quantum many-body models with cold atoms coupled to photonic
  crystals},\ }\href {http://dx.doi.org/10.1038/nphoton.2015.57} {\bibfield
  {journal} {\bibinfo  {journal} {Nat. Photon.}\ }\textbf {\bibinfo {volume}
  {9}},\ \bibinfo {pages} {326} (\bibinfo {year} {2015})}\BibitemShut {NoStop}%
\bibitem [{\citenamefont {Petersen}\ \emph {et~al.}(2014)\citenamefont
  {Petersen}, \citenamefont {Volz},\ and\ \citenamefont
  {Rauschenbeutel}}]{Petersen14}%
  \BibitemOpen
  \bibfield  {author} {\bibinfo {author} {\bibfnamefont {J.}~\bibnamefont
  {Petersen}}, \bibinfo {author} {\bibfnamefont {J.}~\bibnamefont {Volz}},\
  and\ \bibinfo {author} {\bibfnamefont {A.}~\bibnamefont {Rauschenbeutel}},\
  }\bibfield  {title} {\bibinfo {title} {Chiral nanophotonic waveguide
  interface based on spin-orbit interaction of light},\ }\href
  {https://doi.org/10.1126/science.1257671} {\bibfield  {journal} {\bibinfo
  {journal} {Science}\ }\textbf {\bibinfo {volume} {346}},\ \bibinfo {pages}
  {67} (\bibinfo {year} {2014})}\BibitemShut {NoStop}%
\bibitem [{\citenamefont {S{\"o}llner}\ \emph {et~al.}(2015)\citenamefont
  {S{\"o}llner}, \citenamefont {Mahmoodian}, \citenamefont {Hansen},
  \citenamefont {Midolo}, \citenamefont {Javadi}, \citenamefont {Kir{\v
  s}ansk{\.e}}, \citenamefont {Pregnolato}, \citenamefont {El-Ella},
  \citenamefont {Lee}, \citenamefont {Song}, \citenamefont {Stobbe},\ and\
  \citenamefont {Lodahl}}]{Sollner15}%
  \BibitemOpen
  \bibfield  {author} {\bibinfo {author} {\bibfnamefont {I.}~\bibnamefont
  {S{\"o}llner}}, \bibinfo {author} {\bibfnamefont {S.}~\bibnamefont
  {Mahmoodian}}, \bibinfo {author} {\bibfnamefont {S.~L.}\ \bibnamefont
  {Hansen}}, \bibinfo {author} {\bibfnamefont {L.}~\bibnamefont {Midolo}},
  \bibinfo {author} {\bibfnamefont {A.}~\bibnamefont {Javadi}}, \bibinfo
  {author} {\bibfnamefont {G.}~\bibnamefont {Kir{\v s}ansk{\.e}}}, \bibinfo
  {author} {\bibfnamefont {T.}~\bibnamefont {Pregnolato}}, \bibinfo {author}
  {\bibfnamefont {H.}~\bibnamefont {El-Ella}}, \bibinfo {author} {\bibfnamefont
  {E.~H.}\ \bibnamefont {Lee}}, \bibinfo {author} {\bibfnamefont {J.~D.}\
  \bibnamefont {Song}}, \bibinfo {author} {\bibfnamefont {S.}~\bibnamefont
  {Stobbe}},\ and\ \bibinfo {author} {\bibfnamefont {P.}~\bibnamefont
  {Lodahl}},\ }\bibfield  {title} {\bibinfo {title} {Deterministic
  photon--emitter coupling in chiral photonic circuits},\ }\href
  {https://doi.org/10.1038/nnano.2015.159} {\bibfield  {journal} {\bibinfo
  {journal} {Nat. Nanotechnol.}\ }\textbf {\bibinfo {volume} {10}},\ \bibinfo
  {pages} {775} (\bibinfo {year} {2015})}\BibitemShut {NoStop}%
\bibitem [{\citenamefont {Gorshkov}\ \emph {et~al.}(2007)\citenamefont
  {Gorshkov}, \citenamefont {Andr\'e}, \citenamefont {Fleischhauer},
  \citenamefont {S\o{}rensen},\ and\ \citenamefont {Lukin}}]{Gorshkov07}%
  \BibitemOpen
  \bibfield  {author} {\bibinfo {author} {\bibfnamefont {A.~V.}\ \bibnamefont
  {Gorshkov}}, \bibinfo {author} {\bibfnamefont {A.}~\bibnamefont {Andr\'e}},
  \bibinfo {author} {\bibfnamefont {M.}~\bibnamefont {Fleischhauer}}, \bibinfo
  {author} {\bibfnamefont {A.~S.}\ \bibnamefont {S\o{}rensen}},\ and\ \bibinfo
  {author} {\bibfnamefont {M.~D.}\ \bibnamefont {Lukin}},\ }\bibfield  {title}
  {\bibinfo {title} {Universal approach to optimal photon storage in atomic
  media},\ }\href {https://doi.org/10.1103/PhysRevLett.98.123601} {\bibfield
  {journal} {\bibinfo  {journal} {Phys. Rev. Lett.}\ }\textbf {\bibinfo
  {volume} {98}},\ \bibinfo {pages} {123601} (\bibinfo {year}
  {2007})}\BibitemShut {NoStop}%
\bibitem [{\citenamefont {Asenjo-Garcia}\ \emph
  {et~al.}(2017{\natexlab{a}})\citenamefont {Asenjo-Garcia}, \citenamefont
  {Moreno-Cardoner}, \citenamefont {Albrecht}, \citenamefont {Kimble},\ and\
  \citenamefont {Chang}}]{Asenjo17PRX}%
  \BibitemOpen
  \bibfield  {author} {\bibinfo {author} {\bibfnamefont {A.}~\bibnamefont
  {Asenjo-Garcia}}, \bibinfo {author} {\bibfnamefont {M.}~\bibnamefont
  {Moreno-Cardoner}}, \bibinfo {author} {\bibfnamefont {A.}~\bibnamefont
  {Albrecht}}, \bibinfo {author} {\bibfnamefont {H.~J.}\ \bibnamefont
  {Kimble}},\ and\ \bibinfo {author} {\bibfnamefont {D.~E.}\ \bibnamefont
  {Chang}},\ }\bibfield  {title} {\bibinfo {title} {Exponential improvement in
  photon storage fidelities using subradiance and ``selective radiance'' in
  atomic arrays},\ }\href {https://doi.org/10.1103/PhysRevX.7.031024}
  {\bibfield  {journal} {\bibinfo  {journal} {Phys. Rev. X}\ }\textbf {\bibinfo
  {volume} {7}},\ \bibinfo {pages} {031024} (\bibinfo {year}
  {2017}{\natexlab{a}})}\BibitemShut {NoStop}%
\bibitem [{\citenamefont {Dzsotjan}\ \emph {et~al.}(2010)\citenamefont
  {Dzsotjan}, \citenamefont {S\o{}rensen},\ and\ \citenamefont
  {Fleischhauer}}]{Dzsotjan10}%
  \BibitemOpen
  \bibfield  {author} {\bibinfo {author} {\bibfnamefont {D.}~\bibnamefont
  {Dzsotjan}}, \bibinfo {author} {\bibfnamefont {A.~S.}\ \bibnamefont
  {S\o{}rensen}},\ and\ \bibinfo {author} {\bibfnamefont {M.}~\bibnamefont
  {Fleischhauer}},\ }\bibfield  {title} {\bibinfo {title} {Quantum emitters
  coupled to surface plasmons of a nanowire: A Green's function approach},\
  }\href {https://doi.org/10.1103/PhysRevB.82.075427} {\bibfield  {journal}
  {\bibinfo  {journal} {Phys. Rev. B}\ }\textbf {\bibinfo {volume} {82}},\
  \bibinfo {pages} {075427} (\bibinfo {year} {2010})}\BibitemShut {NoStop}%
\bibitem [{\citenamefont {Zheng}\ and\ \citenamefont
  {Baranger}(2013)}]{Zheng13}%
  \BibitemOpen
  \bibfield  {author} {\bibinfo {author} {\bibfnamefont {H.}~\bibnamefont
  {Zheng}}\ and\ \bibinfo {author} {\bibfnamefont {H.~U.}\ \bibnamefont
  {Baranger}},\ }\bibfield  {title} {\bibinfo {title} {Persistent quantum beats
  and long-distance entanglement from waveguide-mediated interactions},\ }\href
  {https://doi.org/10.1103/PhysRevLett.110.113601} {\bibfield  {journal}
  {\bibinfo  {journal} {Phys. Rev. Lett.}\ }\textbf {\bibinfo {volume} {110}},\
  \bibinfo {pages} {113601} (\bibinfo {year} {2013})}\BibitemShut {NoStop}%
\bibitem [{\citenamefont {Stannigel}\ \emph {et~al.}(2012)\citenamefont
  {Stannigel}, \citenamefont {Rabl},\ and\ \citenamefont
  {Zoller}}]{Stannigel12NJP}%
  \BibitemOpen
  \bibfield  {author} {\bibinfo {author} {\bibfnamefont {K.}~\bibnamefont
  {Stannigel}}, \bibinfo {author} {\bibfnamefont {P.}~\bibnamefont {Rabl}},\
  and\ \bibinfo {author} {\bibfnamefont {P.}~\bibnamefont {Zoller}},\
  }\bibfield  {title} {\bibinfo {title} {Driven-dissipative preparation of
  entangled states in cascaded quantum-optical networks},\ }\href
  {http://stacks.iop.org/1367-2630/14/i=6/a=063014} {\bibfield  {journal}
  {\bibinfo  {journal} {New J. Phys.}\ }\textbf {\bibinfo {volume} {14}},\
  \bibinfo {pages} {063014} (\bibinfo {year} {2012})}\BibitemShut {NoStop}%
\bibitem [{\citenamefont {Gonz\'alez-Tudela}\ and\ \citenamefont
  {Porras}(2013)}]{GonzalezTudela13}%
  \BibitemOpen
  \bibfield  {author} {\bibinfo {author} {\bibfnamefont {A.}~\bibnamefont
  {Gonz\'alez-Tudela}}\ and\ \bibinfo {author} {\bibfnamefont {D.}~\bibnamefont
  {Porras}},\ }\bibfield  {title} {\bibinfo {title} {Mesoscopic entanglement
  induced by spontaneous emission in solid-state quantum optics},\ }\href
  {https://doi.org/10.1103/PhysRevLett.110.080502} {\bibfield  {journal}
  {\bibinfo  {journal} {Phys. Rev. Lett.}\ }\textbf {\bibinfo {volume} {110}},\
  \bibinfo {pages} {080502} (\bibinfo {year} {2013})}\BibitemShut {NoStop}%
\bibitem [{\citenamefont {Gonz\'alez-Tudela}\ \emph {et~al.}(2015)\citenamefont
  {Gonz\'alez-Tudela}, \citenamefont {Paulisch}, \citenamefont {Chang},
  \citenamefont {Kimble},\ and\ \citenamefont {Cirac}}]{GonzalezTudela15PRL}%
  \BibitemOpen
  \bibfield  {author} {\bibinfo {author} {\bibfnamefont {A.}~\bibnamefont
  {Gonz\'alez-Tudela}}, \bibinfo {author} {\bibfnamefont {V.}~\bibnamefont
  {Paulisch}}, \bibinfo {author} {\bibfnamefont {D.~E.}\ \bibnamefont {Chang}},
  \bibinfo {author} {\bibfnamefont {H.~J.}\ \bibnamefont {Kimble}},\ and\
  \bibinfo {author} {\bibfnamefont {J.~I.}\ \bibnamefont {Cirac}},\ }\bibfield
  {title} {\bibinfo {title} {Deterministic generation of arbitrary photonic
  states assisted by dissipation},\ }\href
  {https://doi.org/10.1103/PhysRevLett.115.163603} {\bibfield  {journal}
  {\bibinfo  {journal} {Phys. Rev. Lett.}\ }\textbf {\bibinfo {volume} {115}},\
  \bibinfo {pages} {163603} (\bibinfo {year} {2015})}\BibitemShut {NoStop}%
\bibitem [{\citenamefont {Zoubi}\ and\ \citenamefont {Ritsch}(2010)}]{Zoubi10}%
  \BibitemOpen
  \bibfield  {author} {\bibinfo {author} {\bibfnamefont {H.}~\bibnamefont
  {Zoubi}}\ and\ \bibinfo {author} {\bibfnamefont {H.}~\bibnamefont {Ritsch}},\
  }\bibfield  {title} {\bibinfo {title} {Metastability and directional emission
  characteristics of excitons in 1{D} optical lattices},\ }\href
  {https://doi.org/10.1209/0295-5075/90/23001} {\bibfield  {journal} {\bibinfo
  {journal} {Europhys. Lett.}\ }\textbf {\bibinfo {volume} {90}},\ \bibinfo
  {pages} {23001} (\bibinfo {year} {2010})}\BibitemShut {NoStop}%
\bibitem [{\citenamefont {Bettles}\ \emph {et~al.}(2015)\citenamefont
  {Bettles}, \citenamefont {Gardiner},\ and\ \citenamefont
  {Adams}}]{Bettles15}%
  \BibitemOpen
  \bibfield  {author} {\bibinfo {author} {\bibfnamefont {R.~J.}\ \bibnamefont
  {Bettles}}, \bibinfo {author} {\bibfnamefont {S.~A.}\ \bibnamefont
  {Gardiner}},\ and\ \bibinfo {author} {\bibfnamefont {C.~S.}\ \bibnamefont
  {Adams}},\ }\bibfield  {title} {\bibinfo {title} {Cooperative ordering in
  lattices of interacting two-level dipoles},\ }\href
  {https://doi.org/10.1103/PhysRevA.92.063822} {\bibfield  {journal} {\bibinfo
  {journal} {Phys. Rev. A}\ }\textbf {\bibinfo {volume} {92}},\ \bibinfo
  {pages} {063822} (\bibinfo {year} {2015})}\BibitemShut {NoStop}%
\bibitem [{\citenamefont {Bettles}\ \emph
  {et~al.}(2016{\natexlab{a}})\citenamefont {Bettles}, \citenamefont
  {Gardiner},\ and\ \citenamefont {Adams}}]{Bettles16PRA}%
  \BibitemOpen
  \bibfield  {author} {\bibinfo {author} {\bibfnamefont {R.~J.}\ \bibnamefont
  {Bettles}}, \bibinfo {author} {\bibfnamefont {S.~A.}\ \bibnamefont
  {Gardiner}},\ and\ \bibinfo {author} {\bibfnamefont {C.~S.}\ \bibnamefont
  {Adams}},\ }\bibfield  {title} {\bibinfo {title} {Cooperative eigenmodes and
  scattering in one-dimensional atomic arrays},\ }\href
  {https://doi.org/10.1103/PhysRevA.94.043844} {\bibfield  {journal} {\bibinfo
  {journal} {Phys. Rev. A}\ }\textbf {\bibinfo {volume} {94}},\ \bibinfo
  {pages} {043844} (\bibinfo {year} {2016}{\natexlab{a}})}\BibitemShut
  {NoStop}%
\bibitem [{\citenamefont {Needham}\ \emph {et~al.}(2019)\citenamefont
  {Needham}, \citenamefont {Lesanovsky},\ and\ \citenamefont
  {Olmos}}]{Needham19}%
  \BibitemOpen
  \bibfield  {author} {\bibinfo {author} {\bibfnamefont {J.~A.}\ \bibnamefont
  {Needham}}, \bibinfo {author} {\bibfnamefont {I.}~\bibnamefont
  {Lesanovsky}},\ and\ \bibinfo {author} {\bibfnamefont {B.}~\bibnamefont
  {Olmos}},\ }\bibfield  {title} {\bibinfo {title} {Subradiance-protected
  excitation transport},\ }\href {https://doi.org/10.1088/1367-2630/ab31e8}
  {\bibfield  {journal} {\bibinfo  {journal} {New J. Phys.}\ }\textbf {\bibinfo
  {volume} {21}},\ \bibinfo {pages} {073061} (\bibinfo {year}
  {2019})}\BibitemShut {NoStop}%
\bibitem [{\citenamefont {Kornovan}\ \emph {et~al.}(2019)\citenamefont
  {Kornovan}, \citenamefont {Corzo}, \citenamefont {Laurat},\ and\
  \citenamefont {Sheremet}}]{Kornovan19}%
  \BibitemOpen
  \bibfield  {author} {\bibinfo {author} {\bibfnamefont {D.~F.}\ \bibnamefont
  {Kornovan}}, \bibinfo {author} {\bibfnamefont {N.~V.}\ \bibnamefont {Corzo}},
  \bibinfo {author} {\bibfnamefont {J.}~\bibnamefont {Laurat}},\ and\ \bibinfo
  {author} {\bibfnamefont {A.~S.}\ \bibnamefont {Sheremet}},\ }\bibfield
  {title} {\bibinfo {title} {Extremely subradiant states in a periodic
  one-dimensional atomic array},\ }\href
  {https://doi.org/10.1103/PhysRevA.100.063832} {\bibfield  {journal} {\bibinfo
   {journal} {Phys. Rev. A}\ }\textbf {\bibinfo {volume} {100}},\ \bibinfo
  {pages} {063832} (\bibinfo {year} {2019})}\BibitemShut {NoStop}%
\bibitem [{\citenamefont {Asenjo-Garcia}\ \emph {et~al.}(2019)\citenamefont
  {Asenjo-Garcia}, \citenamefont {Kimble},\ and\ \citenamefont
  {Chang}}]{Asenjo19}%
  \BibitemOpen
  \bibfield  {author} {\bibinfo {author} {\bibfnamefont {A.}~\bibnamefont
  {Asenjo-Garcia}}, \bibinfo {author} {\bibfnamefont {H.~J.}\ \bibnamefont
  {Kimble}},\ and\ \bibinfo {author} {\bibfnamefont {D.~E.}\ \bibnamefont
  {Chang}},\ }\bibfield  {title} {\bibinfo {title} {Optical waveguiding by
  atomic entanglement in multilevel atom arrays},\ }\href
  {https://doi.org/10.1073/pnas.1911467116} {\bibfield  {journal} {\bibinfo
  {journal} {Proc. Natl. Acad. Sci. USA}\ }\textbf {\bibinfo {volume} {116}},\
  \bibinfo {pages} {25503} (\bibinfo {year} {2019})}\BibitemShut {NoStop}%
\bibitem [{\citenamefont {Carmichael}\ and\ \citenamefont
  {Kim}(2000)}]{Carmichael00}%
  \BibitemOpen
  \bibfield  {author} {\bibinfo {author} {\bibfnamefont {H.~J.}\ \bibnamefont
  {Carmichael}}\ and\ \bibinfo {author} {\bibfnamefont {K.}~\bibnamefont
  {Kim}},\ }\bibfield  {title} {\bibinfo {title} {A quantum trajectory
  unraveling of the superradiance master equation},\ }\href
  {https://doi.org/https://doi.org/10.1016/S0030-4018(99)00694-X} {\bibfield
  {journal} {\bibinfo  {journal} {Opt. Commun.}\ }\textbf {\bibinfo {volume}
  {179}},\ \bibinfo {pages} {417} (\bibinfo {year} {2000})}\BibitemShut
  {NoStop}%
\bibitem [{\citenamefont {Clemens}\ \emph {et~al.}(2003)\citenamefont
  {Clemens}, \citenamefont {Horvath}, \citenamefont {Sanders},\ and\
  \citenamefont {Carmichael}}]{Clemens03}%
  \BibitemOpen
  \bibfield  {author} {\bibinfo {author} {\bibfnamefont {J.~P.}\ \bibnamefont
  {Clemens}}, \bibinfo {author} {\bibfnamefont {L.}~\bibnamefont {Horvath}},
  \bibinfo {author} {\bibfnamefont {B.~C.}\ \bibnamefont {Sanders}},\ and\
  \bibinfo {author} {\bibfnamefont {H.~J.}\ \bibnamefont {Carmichael}},\
  }\bibfield  {title} {\bibinfo {title} {Collective spontaneous emission from a
  line of atoms},\ }\href {https://doi.org/10.1103/PhysRevA.68.023809}
  {\bibfield  {journal} {\bibinfo  {journal} {Phys. Rev. A}\ }\textbf {\bibinfo
  {volume} {68}},\ \bibinfo {pages} {023809} (\bibinfo {year}
  {2003})}\BibitemShut {NoStop}%
\bibitem [{\citenamefont {Bettles}\ \emph
  {et~al.}(2016{\natexlab{b}})\citenamefont {Bettles}, \citenamefont
  {Gardiner},\ and\ \citenamefont {Adams}}]{Bettles16PRL}%
  \BibitemOpen
  \bibfield  {author} {\bibinfo {author} {\bibfnamefont {R.~J.}\ \bibnamefont
  {Bettles}}, \bibinfo {author} {\bibfnamefont {S.~A.}\ \bibnamefont
  {Gardiner}},\ and\ \bibinfo {author} {\bibfnamefont {C.~S.}\ \bibnamefont
  {Adams}},\ }\bibfield  {title} {\bibinfo {title} {Enhanced optical cross
  section via collective coupling of atomic dipoles in a 2{D} array},\ }\href
  {https://doi.org/10.1103/PhysRevLett.116.103602} {\bibfield  {journal}
  {\bibinfo  {journal} {Phys. Rev. Lett.}\ }\textbf {\bibinfo {volume} {116}},\
  \bibinfo {pages} {103602} (\bibinfo {year} {2016}{\natexlab{b}})}\BibitemShut
  {NoStop}%
\bibitem [{\citenamefont {Sutherland}\ and\ \citenamefont
  {Robicheaux}(2016)}]{Sutherland16}%
  \BibitemOpen
  \bibfield  {author} {\bibinfo {author} {\bibfnamefont {R.~T.}\ \bibnamefont
  {Sutherland}}\ and\ \bibinfo {author} {\bibfnamefont {F.}~\bibnamefont
  {Robicheaux}},\ }\bibfield  {title} {\bibinfo {title} {Collective
  dipole-dipole interactions in an atomic array},\ }\href
  {https://doi.org/10.1103/PhysRevA.94.013847} {\bibfield  {journal} {\bibinfo
  {journal} {Phys. Rev. A}\ }\textbf {\bibinfo {volume} {94}},\ \bibinfo
  {pages} {013847} (\bibinfo {year} {2016})}\BibitemShut {NoStop}%
\bibitem [{\citenamefont {Facchinetti}\ \emph {et~al.}(2016)\citenamefont
  {Facchinetti}, \citenamefont {Jenkins},\ and\ \citenamefont
  {Ruostekoski}}]{Facchinetti16}%
  \BibitemOpen
  \bibfield  {author} {\bibinfo {author} {\bibfnamefont {G.}~\bibnamefont
  {Facchinetti}}, \bibinfo {author} {\bibfnamefont {S.~D.}\ \bibnamefont
  {Jenkins}},\ and\ \bibinfo {author} {\bibfnamefont {J.}~\bibnamefont
  {Ruostekoski}},\ }\bibfield  {title} {\bibinfo {title} {Storing light with
  subradiant correlations in arrays of atoms},\ }\href
  {https://doi.org/10.1103/PhysRevLett.117.243601} {\bibfield  {journal}
  {\bibinfo  {journal} {Phys. Rev. Lett.}\ }\textbf {\bibinfo {volume} {117}},\
  \bibinfo {pages} {243601} (\bibinfo {year} {2016})}\BibitemShut {NoStop}%
\bibitem [{\citenamefont {Asenjo-Garcia}\ \emph
  {et~al.}(2017{\natexlab{b}})\citenamefont {Asenjo-Garcia}, \citenamefont
  {Hood}, \citenamefont {Chang},\ and\ \citenamefont {Kimble}}]{Asenjo17PRA}%
  \BibitemOpen
  \bibfield  {author} {\bibinfo {author} {\bibfnamefont {A.}~\bibnamefont
  {Asenjo-Garcia}}, \bibinfo {author} {\bibfnamefont {J.~D.}\ \bibnamefont
  {Hood}}, \bibinfo {author} {\bibfnamefont {D.~E.}\ \bibnamefont {Chang}},\
  and\ \bibinfo {author} {\bibfnamefont {H.~J.}\ \bibnamefont {Kimble}},\
  }\bibfield  {title} {\bibinfo {title} {Atom-light interactions in
  quasi-one-dimensional nanostructures: A Green's-function perspective},\
  }\href {https://doi.org/10.1103/PhysRevA.95.033818} {\bibfield  {journal}
  {\bibinfo  {journal} {Phys. Rev. A}\ }\textbf {\bibinfo {volume} {95}},\
  \bibinfo {pages} {033818} (\bibinfo {year} {2017}{\natexlab{b}})}\BibitemShut
  {NoStop}%
\bibitem [{\citenamefont {Perczel}\ \emph {et~al.}(2017)\citenamefont
  {Perczel}, \citenamefont {Borregaard}, \citenamefont {Chang}, \citenamefont
  {Pichler}, \citenamefont {Yelin}, \citenamefont {Zoller},\ and\ \citenamefont
  {Lukin}}]{Perczel17PRL}%
  \BibitemOpen
  \bibfield  {author} {\bibinfo {author} {\bibfnamefont {J.}~\bibnamefont
  {Perczel}}, \bibinfo {author} {\bibfnamefont {J.}~\bibnamefont {Borregaard}},
  \bibinfo {author} {\bibfnamefont {D.~E.}\ \bibnamefont {Chang}}, \bibinfo
  {author} {\bibfnamefont {H.}~\bibnamefont {Pichler}}, \bibinfo {author}
  {\bibfnamefont {S.~F.}\ \bibnamefont {Yelin}}, \bibinfo {author}
  {\bibfnamefont {P.}~\bibnamefont {Zoller}},\ and\ \bibinfo {author}
  {\bibfnamefont {M.~D.}\ \bibnamefont {Lukin}},\ }\bibfield  {title} {\bibinfo
  {title} {Topological quantum optics in two-dimensional atomic arrays},\
  }\href {https://doi.org/10.1103/PhysRevLett.119.023603} {\bibfield  {journal}
  {\bibinfo  {journal} {Phys. Rev. Lett.}\ }\textbf {\bibinfo {volume} {119}},\
  \bibinfo {pages} {023603} (\bibinfo {year} {2017})}\BibitemShut {NoStop}%
\bibitem [{\citenamefont {Henriet}\ \emph {et~al.}(2019)\citenamefont
  {Henriet}, \citenamefont {Douglas}, \citenamefont {Chang},\ and\
  \citenamefont {Albrecht}}]{Henriet19}%
  \BibitemOpen
  \bibfield  {author} {\bibinfo {author} {\bibfnamefont {L.}~\bibnamefont
  {Henriet}}, \bibinfo {author} {\bibfnamefont {J.~S.}\ \bibnamefont
  {Douglas}}, \bibinfo {author} {\bibfnamefont {D.~E.}\ \bibnamefont {Chang}},\
  and\ \bibinfo {author} {\bibfnamefont {A.}~\bibnamefont {Albrecht}},\
  }\bibfield  {title} {\bibinfo {title} {Critical open-system dynamics in a
  one-dimensional optical-lattice clock},\ }\href
  {https://doi.org/10.1103/PhysRevA.99.023802} {\bibfield  {journal} {\bibinfo
  {journal} {Phys. Rev. A}\ }\textbf {\bibinfo {volume} {99}},\ \bibinfo
  {pages} {023802} (\bibinfo {year} {2019})}\BibitemShut {NoStop}%
\bibitem [{\citenamefont {Moreno-Cardoner}\ \emph {et~al.}(2019)\citenamefont
  {Moreno-Cardoner}, \citenamefont {Plankensteiner}, \citenamefont {Ostermann},
  \citenamefont {Chang},\ and\ \citenamefont {Ritsch}}]{MorenoCardoner19}%
  \BibitemOpen
  \bibfield  {author} {\bibinfo {author} {\bibfnamefont {M.}~\bibnamefont
  {Moreno-Cardoner}}, \bibinfo {author} {\bibfnamefont {D.}~\bibnamefont
  {Plankensteiner}}, \bibinfo {author} {\bibfnamefont {L.}~\bibnamefont
  {Ostermann}}, \bibinfo {author} {\bibfnamefont {D.~E.}\ \bibnamefont
  {Chang}},\ and\ \bibinfo {author} {\bibfnamefont {H.}~\bibnamefont
  {Ritsch}},\ }\bibfield  {title} {\bibinfo {title} {Subradiance-enhanced
  excitation transfer between dipole-coupled nanorings of quantum emitters},\
  }\href {https://doi.org/10.1103/PhysRevA.100.023806} {\bibfield  {journal}
  {\bibinfo  {journal} {Phys. Rev. A}\ }\textbf {\bibinfo {volume} {100}},\
  \bibinfo {pages} {023806} (\bibinfo {year} {2019})}\BibitemShut {NoStop}%
\bibitem [{\citenamefont {Patti}\ \emph {et~al.}(2020)\citenamefont {Patti},
  \citenamefont {Wild}, \citenamefont {Shahmoon}, \citenamefont {Lukin},\ and\
  \citenamefont {Yelin}}]{Patti20arxiv}%
  \BibitemOpen
  \bibfield  {author} {\bibinfo {author} {\bibfnamefont {T.~L.}\ \bibnamefont
  {Patti}}, \bibinfo {author} {\bibfnamefont {D.~S.}\ \bibnamefont {Wild}},
  \bibinfo {author} {\bibfnamefont {E.}~\bibnamefont {Shahmoon}}, \bibinfo
  {author} {\bibfnamefont {M.~D.}\ \bibnamefont {Lukin}},\ and\ \bibinfo
  {author} {\bibfnamefont {S.~F.}\ \bibnamefont {Yelin}},\ }\bibfield  {title}
  {\bibinfo {title} {Controlling interactions between quantum emitters using
  atom arrays},\ }\href@noop {} {\bibfield  {journal} {\bibinfo  {journal}
  {arXiv:2005.03495}\ } (\bibinfo {year} {2020})}\BibitemShut {NoStop}%
\bibitem [{\citenamefont {Holzinger}\ \emph {et~al.}(2020)\citenamefont
  {Holzinger}, \citenamefont {Plankensteiner}, \citenamefont {Ostermann},\ and\
  \citenamefont {Ritsch}}]{Holzinger20PRL}%
  \BibitemOpen
  \bibfield  {author} {\bibinfo {author} {\bibfnamefont {R.}~\bibnamefont
  {Holzinger}}, \bibinfo {author} {\bibfnamefont {D.}~\bibnamefont
  {Plankensteiner}}, \bibinfo {author} {\bibfnamefont {L.}~\bibnamefont
  {Ostermann}},\ and\ \bibinfo {author} {\bibfnamefont {H.}~\bibnamefont
  {Ritsch}},\ }\bibfield  {title} {\bibinfo {title} {Nanoscale coherent light
  source},\ }\href {https://doi.org/10.1103/PhysRevLett.124.253603} {\bibfield
  {journal} {\bibinfo  {journal} {Phys. Rev. Lett.}\ }\textbf {\bibinfo
  {volume} {124}},\ \bibinfo {pages} {253603} (\bibinfo {year}
  {2020})}\BibitemShut {NoStop}%
\bibitem [{\citenamefont {Masson}\ \emph {et~al.}(2020)\citenamefont {Masson},
  \citenamefont {Ferrier-Barbut}, \citenamefont {Orozco}, \citenamefont
  {Browaeys},\ and\ \citenamefont {Asenjo-Garcia}}]{Masson20superradiance}%
  \BibitemOpen
  \bibfield  {author} {\bibinfo {author} {\bibfnamefont {S.}~\bibnamefont
  {Masson}}, \bibinfo {author} {\bibfnamefont {I.}~\bibnamefont
  {Ferrier-Barbut}}, \bibinfo {author} {\bibfnamefont {L.}~\bibnamefont
  {Orozco}}, \bibinfo {author} {\bibfnamefont {A.}~\bibnamefont {Browaeys}},\
  and\ \bibinfo {author} {\bibfnamefont {A.}~\bibnamefont {Asenjo-Garcia}},\
  }\bibfield  {title} {\bibinfo {title} {Many-body signatures of collective
  decay in atomic chains},\ }\href@noop {} {\bibfield  {journal} {\bibinfo
  {journal} {arxiv:2008.08139}\ } (\bibinfo {year} {2020})}\BibitemShut
  {NoStop}%
\bibitem [{\citenamefont {Carmichael}(1993{\natexlab{a}})}]{Carmichael93}%
  \BibitemOpen
  \bibfield  {author} {\bibinfo {author} {\bibfnamefont {H.~J.}\ \bibnamefont
  {Carmichael}},\ }\href@noop {} {\emph {\bibinfo {title} {An Open Systems
  Approach to Quantum Optics}}},\ \bibinfo {series} {Lecture Notes in Physics},
  Vol.~\bibinfo {volume} {18}\ (\bibinfo  {publisher} {Springer-Verlag Berlin
  Heidelberg},\ \bibinfo {year} {1993})\BibitemShut {NoStop}%
\bibitem [{\citenamefont {Gardiner}\ and\ \citenamefont
  {Zoller}(2004)}]{QuantumNoiseBook}%
  \BibitemOpen
  \bibfield  {author} {\bibinfo {author} {\bibfnamefont {C.}~\bibnamefont
  {Gardiner}}\ and\ \bibinfo {author} {\bibfnamefont {P.}~\bibnamefont
  {Zoller}},\ }\href@noop {} {\emph {\bibinfo {title} {Quantum Noise}}}\
  (\bibinfo  {publisher} {Springer-Verlag Berlin Heidelberg},\ \bibinfo {year}
  {2004})\BibitemShut {NoStop}%
\bibitem [{\citenamefont {Clemens}\ \emph {et~al.}(2004)\citenamefont
  {Clemens}, \citenamefont {Horvath}, \citenamefont {Sanders},\ and\
  \citenamefont {Carmichael}}]{Clemens04}%
  \BibitemOpen
  \bibfield  {author} {\bibinfo {author} {\bibfnamefont {J.~P.}\ \bibnamefont
  {Clemens}}, \bibinfo {author} {\bibfnamefont {L.}~\bibnamefont {Horvath}},
  \bibinfo {author} {\bibfnamefont {B.~C.}\ \bibnamefont {Sanders}},\ and\
  \bibinfo {author} {\bibfnamefont {H.~J.}\ \bibnamefont {Carmichael}},\
  }\bibfield  {title} {\bibinfo {title} {Shot-to-shot fluctuations in the
  directed superradiant emission from extended atomic samples},\ }\href
  {https://doi.org/10.1088/1464-4266/6/8/017} {\bibfield  {journal} {\bibinfo
  {journal} {J. Opt. B: Quantum Semiclass. Opt.}\ }\textbf {\bibinfo {volume}
  {6}},\ \bibinfo {pages} {S736} (\bibinfo {year} {2004})}\BibitemShut
  {NoStop}%
\bibitem [{\citenamefont {Stephen}(1964)}]{Stephen64}%
  \BibitemOpen
  \bibfield  {author} {\bibinfo {author} {\bibfnamefont {M.~J.}\ \bibnamefont
  {Stephen}},\ }\bibfield  {title} {\bibinfo {title} {First-order dispersion
  forces},\ }\href@noop {} {\bibfield  {journal} {\bibinfo  {journal} {J. Chem.
  Phys.}\ }\textbf {\bibinfo {volume} {40}},\ \bibinfo {pages} {669} (\bibinfo
  {year} {1964})}\BibitemShut {NoStop}%
\bibitem [{\citenamefont {Lehmberg}(1970)}]{Lehmberg70}%
  \BibitemOpen
  \bibfield  {author} {\bibinfo {author} {\bibfnamefont {R.~H.}\ \bibnamefont
  {Lehmberg}},\ }\bibfield  {title} {\bibinfo {title} {Radiation from an
  $n$-atom system. i. general formalism},\ }\href
  {https://doi.org/10.1103/PhysRevA.2.883} {\bibfield  {journal} {\bibinfo
  {journal} {Phys. Rev. A}\ }\textbf {\bibinfo {volume} {2}},\ \bibinfo {pages}
  {883} (\bibinfo {year} {1970})}\BibitemShut {NoStop}%
\bibitem [{\citenamefont {Gruner}\ and\ \citenamefont
  {Welsch}(1996)}]{Gruner96}%
  \BibitemOpen
  \bibfield  {author} {\bibinfo {author} {\bibfnamefont {T.}~\bibnamefont
  {Gruner}}\ and\ \bibinfo {author} {\bibfnamefont {D.-G.}\ \bibnamefont
  {Welsch}},\ }\bibfield  {title} {\bibinfo {title} {Green-function approach to
  the radiation-field quantization for homogeneous and inhomogeneous
  Kramers-Kronig dielectrics},\ }\href
  {https://doi.org/10.1103/PhysRevA.53.1818} {\bibfield  {journal} {\bibinfo
  {journal} {Phys. Rev. A}\ }\textbf {\bibinfo {volume} {53}},\ \bibinfo
  {pages} {1818} (\bibinfo {year} {1996})}\BibitemShut {NoStop}%
\bibitem [{\citenamefont {Dung}\ \emph {et~al.}(2002)\citenamefont {Dung},
  \citenamefont {Kn\"oll},\ and\ \citenamefont {Welsch}}]{Dung02}%
  \BibitemOpen
  \bibfield  {author} {\bibinfo {author} {\bibfnamefont {H.~T.}\ \bibnamefont
  {Dung}}, \bibinfo {author} {\bibfnamefont {L.}~\bibnamefont {Kn\"oll}},\ and\
  \bibinfo {author} {\bibfnamefont {D.-G.}\ \bibnamefont {Welsch}},\ }\bibfield
   {title} {\bibinfo {title} {Resonant dipole-dipole interaction in the
  presence of dispersing and absorbing surroundings},\ }\href
  {https://doi.org/10.1103/PhysRevA.66.063810} {\bibfield  {journal} {\bibinfo
  {journal} {Phys. Rev. A}\ }\textbf {\bibinfo {volume} {66}},\ \bibinfo
  {pages} {063810} (\bibinfo {year} {2002})}\BibitemShut {NoStop}%
\bibitem [{\citenamefont {Chang}\ \emph {et~al.}(2012)\citenamefont {Chang},
  \citenamefont {Jiang}, \citenamefont {Gorshkov},\ and\ \citenamefont
  {Kimble}}]{Chang12}%
  \BibitemOpen
  \bibfield  {author} {\bibinfo {author} {\bibfnamefont {D.~E.}\ \bibnamefont
  {Chang}}, \bibinfo {author} {\bibfnamefont {L.}~\bibnamefont {Jiang}},
  \bibinfo {author} {\bibfnamefont {A.~V.}\ \bibnamefont {Gorshkov}},\ and\
  \bibinfo {author} {\bibfnamefont {H.~J.}\ \bibnamefont {Kimble}},\ }\bibfield
   {title} {\bibinfo {title} {Cavity {QED} with atomic mirrors},\ }\href
  {http://stacks.iop.org/1367-2630/14/i=6/a=063003} {\bibfield  {journal}
  {\bibinfo  {journal} {New J. Phys.}\ }\textbf {\bibinfo {volume} {14}},\
  \bibinfo {pages} {063003} (\bibinfo {year} {2012})}\BibitemShut {NoStop}%
\bibitem [{\citenamefont {Shi}\ \emph {et~al.}(2015)\citenamefont {Shi},
  \citenamefont {Chang},\ and\ \citenamefont {Cirac}}]{Shi15}%
  \BibitemOpen
  \bibfield  {author} {\bibinfo {author} {\bibfnamefont {T.}~\bibnamefont
  {Shi}}, \bibinfo {author} {\bibfnamefont {D.~E.}\ \bibnamefont {Chang}},\
  and\ \bibinfo {author} {\bibfnamefont {J.~I.}\ \bibnamefont {Cirac}},\
  }\bibfield  {title} {\bibinfo {title} {Multiphoton-scattering theory and
  generalized master equations},\ }\href
  {https://doi.org/10.1103/PhysRevA.92.053834} {\bibfield  {journal} {\bibinfo
  {journal} {Phys. Rev. A}\ }\textbf {\bibinfo {volume} {92}},\ \bibinfo
  {pages} {053834} (\bibinfo {year} {2015})}\BibitemShut {NoStop}%
\bibitem [{\citenamefont {Guimond}\ \emph {et~al.}(2016)\citenamefont
  {Guimond}, \citenamefont {Roulet}, \citenamefont {Le},\ and\ \citenamefont
  {Scarani}}]{Guimond16}%
  \BibitemOpen
  \bibfield  {author} {\bibinfo {author} {\bibfnamefont {P.-O.}\ \bibnamefont
  {Guimond}}, \bibinfo {author} {\bibfnamefont {A.}~\bibnamefont {Roulet}},
  \bibinfo {author} {\bibfnamefont {H.~N.}\ \bibnamefont {Le}},\ and\ \bibinfo
  {author} {\bibfnamefont {V.}~\bibnamefont {Scarani}},\ }\bibfield  {title}
  {\bibinfo {title} {Rabi oscillation in a quantum cavity: Markovian and
  non-Markovian dynamics},\ }\href {https://doi.org/10.1103/PhysRevA.93.023808}
  {\bibfield  {journal} {\bibinfo  {journal} {Phys. Rev. A}\ }\textbf {\bibinfo
  {volume} {93}},\ \bibinfo {pages} {023808} (\bibinfo {year}
  {2016})}\BibitemShut {NoStop}%
\bibitem [{Note1()}]{Note1}%
  \BibitemOpen
  \bibinfo {note} {The dispersion relation for transverse polarization displays
  unusual features for $d \lesssim 0.25\lambda _0$, such that for a particular
  frequency there might be two guided modes of different wave-vector. While we
  do not believe that the physics changes significantly with respect to the
  longitudinal polarization case, we do not treat explicitly the transverse
  polarization in this manuscript.}\BibitemShut {Stop}%
\bibitem [{\citenamefont {Halir}\ \emph {et~al.}(2015)\citenamefont {Halir},
  \citenamefont {Bock}, \citenamefont {Cheben}, \citenamefont
  {Ortega-Mo{\~n}ux}, \citenamefont {Alonso-Ramos}, \citenamefont {Schmid},
  \citenamefont {Lapointe}, \citenamefont {Xu}, \citenamefont
  {Wang{\"u}emert-P{\'e}rez}, \citenamefont {Molina-Fern{\'a}ndez},\ and\
  \citenamefont {Janz}}]{Halir15}%
  \BibitemOpen
  \bibfield  {author} {\bibinfo {author} {\bibfnamefont {R.}~\bibnamefont
  {Halir}}, \bibinfo {author} {\bibfnamefont {P.~J.}\ \bibnamefont {Bock}},
  \bibinfo {author} {\bibfnamefont {P.}~\bibnamefont {Cheben}}, \bibinfo
  {author} {\bibfnamefont {A.}~\bibnamefont {Ortega-Mo{\~n}ux}}, \bibinfo
  {author} {\bibfnamefont {C.}~\bibnamefont {Alonso-Ramos}}, \bibinfo {author}
  {\bibfnamefont {J.~H.}\ \bibnamefont {Schmid}}, \bibinfo {author}
  {\bibfnamefont {J.}~\bibnamefont {Lapointe}}, \bibinfo {author}
  {\bibfnamefont {D.-X.}\ \bibnamefont {Xu}}, \bibinfo {author} {\bibfnamefont
  {J.~G.}\ \bibnamefont {Wang{\"u}emert-P{\'e}rez}}, \bibinfo {author}
  {\bibfnamefont {{\'I}.}~\bibnamefont {Molina-Fern{\'a}ndez}},\ and\ \bibinfo
  {author} {\bibfnamefont {S.}~\bibnamefont {Janz}},\ }\bibfield  {title}
  {\bibinfo {title} {Waveguide sub-wavelength structures: a review of
  principles and applications},\ }\href
  {https://doi.org/10.1002/lpor.201400083} {\bibfield  {journal} {\bibinfo
  {journal} {Laser \& Photon. Rev.}\ }\textbf {\bibinfo {volume} {9}},\
  \bibinfo {pages} {25} (\bibinfo {year} {2015})}\BibitemShut {NoStop}%
\bibitem [{\citenamefont {Blaustein}\ \emph {et~al.}(2007)\citenamefont
  {Blaustein}, \citenamefont {Gozman}, \citenamefont {Samoylova}, \citenamefont
  {Polishchuk},\ and\ \citenamefont {Burin}}]{Blaustein07}%
  \BibitemOpen
  \bibfield  {author} {\bibinfo {author} {\bibfnamefont {G.~S.}\ \bibnamefont
  {Blaustein}}, \bibinfo {author} {\bibfnamefont {M.~I.}\ \bibnamefont
  {Gozman}}, \bibinfo {author} {\bibfnamefont {O.}~\bibnamefont {Samoylova}},
  \bibinfo {author} {\bibfnamefont {I.~Y.}\ \bibnamefont {Polishchuk}},\ and\
  \bibinfo {author} {\bibfnamefont {A.~L.}\ \bibnamefont {Burin}},\ }\bibfield
  {title} {\bibinfo {title} {Guiding optical modes in chains of dielectric
  particles},\ }\href {https://doi.org/10.1364/OE.15.017380} {\bibfield
  {journal} {\bibinfo  {journal} {Opt. Express}\ }\textbf {\bibinfo {volume}
  {15}},\ \bibinfo {pages} {17380} (\bibinfo {year} {2007})}\BibitemShut
  {NoStop}%
\bibitem [{\citenamefont {Al\`u}\ and\ \citenamefont {Engheta}(2006)}]{Alu06}%
  \BibitemOpen
  \bibfield  {author} {\bibinfo {author} {\bibfnamefont {A.}~\bibnamefont
  {Al\`u}}\ and\ \bibinfo {author} {\bibfnamefont {N.}~\bibnamefont
  {Engheta}},\ }\bibfield  {title} {\bibinfo {title} {Theory of linear chains
  of metamaterial/plasmonic particles as subdiffraction optical
  nanotransmission lines},\ }\href {https://doi.org/10.1103/PhysRevB.74.205436}
  {\bibfield  {journal} {\bibinfo  {journal} {Phys. Rev. B}\ }\textbf {\bibinfo
  {volume} {74}},\ \bibinfo {pages} {205436} (\bibinfo {year}
  {2006})}\BibitemShut {NoStop}%
\bibitem [{\citenamefont {Chang}\ \emph {et~al.}(2007)\citenamefont {Chang},
  \citenamefont {S{\o}rensen}, \citenamefont {Demler},\ and\ \citenamefont
  {Lukin}}]{Chang07}%
  \BibitemOpen
  \bibfield  {author} {\bibinfo {author} {\bibfnamefont {D.~E.}\ \bibnamefont
  {Chang}}, \bibinfo {author} {\bibfnamefont {A.~S.}\ \bibnamefont
  {S{\o}rensen}}, \bibinfo {author} {\bibfnamefont {E.~A.}\ \bibnamefont
  {Demler}},\ and\ \bibinfo {author} {\bibfnamefont {M.~D.}\ \bibnamefont
  {Lukin}},\ }\bibfield  {title} {\bibinfo {title} {A single-photon transistor
  using nanoscale surface plasmons},\ }\href {https://doi.org/10.1038/nphys708}
  {\bibfield  {journal} {\bibinfo  {journal} {Nat. Phys.}\ }\textbf {\bibinfo
  {volume} {3}},\ \bibinfo {pages} {807} (\bibinfo {year} {2007})}\BibitemShut
  {NoStop}%
\bibitem [{\citenamefont {S\'anchez-Burillo}\ \emph {et~al.}(2017)\citenamefont
  {S\'anchez-Burillo}, \citenamefont {Zueco}, \citenamefont
  {Mart\'{\i}n-Moreno},\ and\ \citenamefont {Garc\'{\i}a-Ripoll}}]{Sanchez17}%
  \BibitemOpen
  \bibfield  {author} {\bibinfo {author} {\bibfnamefont {E.}~\bibnamefont
  {S\'anchez-Burillo}}, \bibinfo {author} {\bibfnamefont {D.}~\bibnamefont
  {Zueco}}, \bibinfo {author} {\bibfnamefont {L.}~\bibnamefont
  {Mart\'{\i}n-Moreno}},\ and\ \bibinfo {author} {\bibfnamefont {J.~J.}\
  \bibnamefont {Garc\'{\i}a-Ripoll}},\ }\bibfield  {title} {\bibinfo {title}
  {Dynamical signatures of bound states in waveguide {QED}},\ }\href
  {https://doi.org/10.1103/PhysRevA.96.023831} {\bibfield  {journal} {\bibinfo
  {journal} {Phys. Rev. A}\ }\textbf {\bibinfo {volume} {96}},\ \bibinfo
  {pages} {023831} (\bibinfo {year} {2017})}\BibitemShut {NoStop}%
\bibitem [{\citenamefont {Tufarelli}\ \emph {et~al.}(2014)\citenamefont
  {Tufarelli}, \citenamefont {Kim},\ and\ \citenamefont
  {Ciccarello}}]{Tufarelli14}%
  \BibitemOpen
  \bibfield  {author} {\bibinfo {author} {\bibfnamefont {T.}~\bibnamefont
  {Tufarelli}}, \bibinfo {author} {\bibfnamefont {M.~S.}\ \bibnamefont {Kim}},\
  and\ \bibinfo {author} {\bibfnamefont {F.}~\bibnamefont {Ciccarello}},\
  }\bibfield  {title} {\bibinfo {title} {Non-Markovianity of a quantum emitter
  in front of a mirror},\ }\href {https://doi.org/10.1103/PhysRevA.90.012113}
  {\bibfield  {journal} {\bibinfo  {journal} {Phys. Rev. A}\ }\textbf {\bibinfo
  {volume} {90}},\ \bibinfo {pages} {012113} (\bibinfo {year}
  {2014})}\BibitemShut {NoStop}%
\bibitem [{\citenamefont {Pichler}\ and\ \citenamefont
  {Zoller}(2016)}]{Pichler16}%
  \BibitemOpen
  \bibfield  {author} {\bibinfo {author} {\bibfnamefont {H.}~\bibnamefont
  {Pichler}}\ and\ \bibinfo {author} {\bibfnamefont {P.}~\bibnamefont
  {Zoller}},\ }\bibfield  {title} {\bibinfo {title} {Photonic circuits with
  time delays and quantum feedback},\ }\href
  {https://doi.org/10.1103/PhysRevLett.116.093601} {\bibfield  {journal}
  {\bibinfo  {journal} {Phys. Rev. Lett.}\ }\textbf {\bibinfo {volume} {116}},\
  \bibinfo {pages} {093601} (\bibinfo {year} {2016})}\BibitemShut {NoStop}%
\bibitem [{\citenamefont {Sinha}\ \emph {et~al.}(2020)\citenamefont {Sinha},
  \citenamefont {Meystre}, \citenamefont {Goldschmidt}, \citenamefont {Fatemi},
  \citenamefont {Rolston},\ and\ \citenamefont {Solano}}]{Sinha20}%
  \BibitemOpen
  \bibfield  {author} {\bibinfo {author} {\bibfnamefont {K.}~\bibnamefont
  {Sinha}}, \bibinfo {author} {\bibfnamefont {P.}~\bibnamefont {Meystre}},
  \bibinfo {author} {\bibfnamefont {E.~A.}\ \bibnamefont {Goldschmidt}},
  \bibinfo {author} {\bibfnamefont {F.~K.}\ \bibnamefont {Fatemi}}, \bibinfo
  {author} {\bibfnamefont {S.~L.}\ \bibnamefont {Rolston}},\ and\ \bibinfo
  {author} {\bibfnamefont {P.}~\bibnamefont {Solano}},\ }\bibfield  {title}
  {\bibinfo {title} {Non-Markovian collective emission from macroscopically
  separated emitters},\ }\href {https://doi.org/10.1103/PhysRevLett.124.043603}
  {\bibfield  {journal} {\bibinfo  {journal} {Phys. Rev. Lett.}\ }\textbf
  {\bibinfo {volume} {124}},\ \bibinfo {pages} {043603} (\bibinfo {year}
  {2020})}\BibitemShut {NoStop}%
\bibitem [{\citenamefont {Carmele}\ \emph {et~al.}(2020)\citenamefont
  {Carmele}, \citenamefont {Nemet}, \citenamefont {Canela},\ and\ \citenamefont
  {Parkins}}]{Carmele20}%
  \BibitemOpen
  \bibfield  {author} {\bibinfo {author} {\bibfnamefont {A.}~\bibnamefont
  {Carmele}}, \bibinfo {author} {\bibfnamefont {N.}~\bibnamefont {Nemet}},
  \bibinfo {author} {\bibfnamefont {V.}~\bibnamefont {Canela}},\ and\ \bibinfo
  {author} {\bibfnamefont {S.}~\bibnamefont {Parkins}},\ }\bibfield  {title}
  {\bibinfo {title} {Pronounced non-Markovian features in multiply excited,
  multiple emitter waveguide {QED}: Retardation induced anomalous population
  trapping},\ }\href {https://doi.org/10.1103/PhysRevResearch.2.013238}
  {\bibfield  {journal} {\bibinfo  {journal} {Phys. Rev. Research}\ }\textbf
  {\bibinfo {volume} {2}},\ \bibinfo {pages} {013238} (\bibinfo {year}
  {2020})}\BibitemShut {NoStop}%
\bibitem [{\citenamefont {Calaj\'o}\ \emph {et~al.}(2016)\citenamefont
  {Calaj\'o}, \citenamefont {Ciccarello}, \citenamefont {Chang},\ and\
  \citenamefont {Rabl}}]{Calajo16}%
  \BibitemOpen
  \bibfield  {author} {\bibinfo {author} {\bibfnamefont {G.}~\bibnamefont
  {Calaj\'o}}, \bibinfo {author} {\bibfnamefont {F.}~\bibnamefont
  {Ciccarello}}, \bibinfo {author} {\bibfnamefont {D.}~\bibnamefont {Chang}},\
  and\ \bibinfo {author} {\bibfnamefont {P.}~\bibnamefont {Rabl}},\ }\bibfield
  {title} {\bibinfo {title} {Atom-field dressed states in slow-light waveguide
  {QED}},\ }\href {https://doi.org/10.1103/PhysRevA.93.033833} {\bibfield
  {journal} {\bibinfo  {journal} {Phys. Rev. A}\ }\textbf {\bibinfo {volume}
  {93}},\ \bibinfo {pages} {033833} (\bibinfo {year} {2016})}\BibitemShut
  {NoStop}%
\bibitem [{\citenamefont {Shi}\ \emph {et~al.}(2016)\citenamefont {Shi},
  \citenamefont {Wu}, \citenamefont {Gonz\'alez-Tudela},\ and\ \citenamefont
  {Cirac}}]{Shi16}%
  \BibitemOpen
  \bibfield  {author} {\bibinfo {author} {\bibfnamefont {T.}~\bibnamefont
  {Shi}}, \bibinfo {author} {\bibfnamefont {Y.-H.}\ \bibnamefont {Wu}},
  \bibinfo {author} {\bibfnamefont {A.}~\bibnamefont {Gonz\'alez-Tudela}},\
  and\ \bibinfo {author} {\bibfnamefont {J.~I.}\ \bibnamefont {Cirac}},\
  }\bibfield  {title} {\bibinfo {title} {Bound states in boson impurity
  models},\ }\href {https://doi.org/10.1103/PhysRevX.6.021027} {\bibfield
  {journal} {\bibinfo  {journal} {Phys. Rev. X}\ }\textbf {\bibinfo {volume}
  {6}},\ \bibinfo {pages} {021027} (\bibinfo {year} {2016})}\BibitemShut
  {NoStop}%
\bibitem [{\citenamefont {Calaj\'o}\ \emph {et~al.}(2019)\citenamefont
  {Calaj\'o}, \citenamefont {Fang}, \citenamefont {Baranger},\ and\
  \citenamefont {Ciccarello}}]{Calajo19}%
  \BibitemOpen
  \bibfield  {author} {\bibinfo {author} {\bibfnamefont {G.}~\bibnamefont
  {Calaj\'o}}, \bibinfo {author} {\bibfnamefont {Y.-L.~L.}\ \bibnamefont
  {Fang}}, \bibinfo {author} {\bibfnamefont {H.~U.}\ \bibnamefont {Baranger}},\
  and\ \bibinfo {author} {\bibfnamefont {F.}~\bibnamefont {Ciccarello}},\
  }\bibfield  {title} {\bibinfo {title} {Exciting a bound state in the
  continuum through multiphoton scattering plus delayed quantum feedback},\
  }\href {https://doi.org/10.1103/PhysRevLett.122.073601} {\bibfield  {journal}
  {\bibinfo  {journal} {Phys. Rev. Lett.}\ }\textbf {\bibinfo {volume} {122}},\
  \bibinfo {pages} {073601} (\bibinfo {year} {2019})}\BibitemShut {NoStop}%
\bibitem [{\citenamefont {Pichler}\ \emph {et~al.}(2015)\citenamefont
  {Pichler}, \citenamefont {Ramos}, \citenamefont {Daley},\ and\ \citenamefont
  {Zoller}}]{Pichler15}%
  \BibitemOpen
  \bibfield  {author} {\bibinfo {author} {\bibfnamefont {H.}~\bibnamefont
  {Pichler}}, \bibinfo {author} {\bibfnamefont {T.}~\bibnamefont {Ramos}},
  \bibinfo {author} {\bibfnamefont {A.~J.}\ \bibnamefont {Daley}},\ and\
  \bibinfo {author} {\bibfnamefont {P.}~\bibnamefont {Zoller}},\ }\bibfield
  {title} {\bibinfo {title} {Quantum optics of chiral spin networks},\ }\href
  {https://doi.org/10.1103/PhysRevA.91.042116} {\bibfield  {journal} {\bibinfo
  {journal} {Phys. Rev. A}\ }\textbf {\bibinfo {volume} {91}},\ \bibinfo
  {pages} {042116} (\bibinfo {year} {2015})}\BibitemShut {NoStop}%
\bibitem [{\citenamefont {Lodahl}\ \emph {et~al.}(2017)\citenamefont {Lodahl},
  \citenamefont {Mahmoodian}, \citenamefont {Stobbe}, \citenamefont
  {Rauschenbeutel}, \citenamefont {Schneeweiss}, \citenamefont {Volz},
  \citenamefont {Pichler},\ and\ \citenamefont {Zoller}}]{Lodahl17}%
  \BibitemOpen
  \bibfield  {author} {\bibinfo {author} {\bibfnamefont {P.}~\bibnamefont
  {Lodahl}}, \bibinfo {author} {\bibfnamefont {S.}~\bibnamefont {Mahmoodian}},
  \bibinfo {author} {\bibfnamefont {S.}~\bibnamefont {Stobbe}}, \bibinfo
  {author} {\bibfnamefont {A.}~\bibnamefont {Rauschenbeutel}}, \bibinfo
  {author} {\bibfnamefont {P.}~\bibnamefont {Schneeweiss}}, \bibinfo {author}
  {\bibfnamefont {J.}~\bibnamefont {Volz}}, \bibinfo {author} {\bibfnamefont
  {H.}~\bibnamefont {Pichler}},\ and\ \bibinfo {author} {\bibfnamefont
  {P.}~\bibnamefont {Zoller}},\ }\bibfield  {title} {\bibinfo {title} {Chiral
  quantum optics},\ }\href {http://dx.doi.org/10.1038/nature21037} {\bibfield
  {journal} {\bibinfo  {journal} {Nature}\ }\textbf {\bibinfo {volume} {541}},\
  \bibinfo {pages} {473} (\bibinfo {year} {2017})}\BibitemShut {NoStop}%
\bibitem [{\citenamefont {Jones}\ \emph {et~al.}(2020)\citenamefont {Jones},
  \citenamefont {Buonaiuto}, \citenamefont {Lang}, \citenamefont {Lesanovsky},\
  and\ \citenamefont {Olmos}}]{Jones20}%
  \BibitemOpen
  \bibfield  {author} {\bibinfo {author} {\bibfnamefont {R.}~\bibnamefont
  {Jones}}, \bibinfo {author} {\bibfnamefont {G.}~\bibnamefont {Buonaiuto}},
  \bibinfo {author} {\bibfnamefont {B.}~\bibnamefont {Lang}}, \bibinfo {author}
  {\bibfnamefont {I.}~\bibnamefont {Lesanovsky}},\ and\ \bibinfo {author}
  {\bibfnamefont {B.}~\bibnamefont {Olmos}},\ }\bibfield  {title} {\bibinfo
  {title} {Collectively enhanced chiral photon emission from an atomic array
  near a nanofiber},\ }\href {https://doi.org/10.1103/PhysRevLett.124.093601}
  {\bibfield  {journal} {\bibinfo  {journal} {Phys. Rev. Lett.}\ }\textbf
  {\bibinfo {volume} {124}},\ \bibinfo {pages} {093601} (\bibinfo {year}
  {2020})}\BibitemShut {NoStop}%
\bibitem [{\citenamefont {Carmichael}(1993{\natexlab{b}})}]{Carmichael93PRL}%
  \BibitemOpen
  \bibfield  {author} {\bibinfo {author} {\bibfnamefont {H.~J.}\ \bibnamefont
  {Carmichael}},\ }\bibfield  {title} {\bibinfo {title} {Quantum trajectory
  theory for cascaded open systems},\ }\href
  {https://doi.org/10.1103/PhysRevLett.70.2273} {\bibfield  {journal} {\bibinfo
   {journal} {Phys. Rev. Lett.}\ }\textbf {\bibinfo {volume} {70}},\ \bibinfo
  {pages} {2273} (\bibinfo {year} {1993}{\natexlab{b}})}\BibitemShut {NoStop}%
\bibitem [{\citenamefont {Gardiner}(1993)}]{Gardiner93}%
  \BibitemOpen
  \bibfield  {author} {\bibinfo {author} {\bibfnamefont {C.~W.}\ \bibnamefont
  {Gardiner}},\ }\bibfield  {title} {\bibinfo {title} {Driving a quantum system
  with the output field from another driven quantum system},\ }\href
  {https://doi.org/10.1103/PhysRevLett.70.2269} {\bibfield  {journal} {\bibinfo
   {journal} {Phys. Rev. Lett.}\ }\textbf {\bibinfo {volume} {70}},\ \bibinfo
  {pages} {2269} (\bibinfo {year} {1993})}\BibitemShut {NoStop}%
\bibitem [{\citenamefont {Ramos}\ \emph {et~al.}(2016)\citenamefont {Ramos},
  \citenamefont {Vermersch}, \citenamefont {Hauke}, \citenamefont {Pichler},\
  and\ \citenamefont {Zoller}}]{Ramos16}%
  \BibitemOpen
  \bibfield  {author} {\bibinfo {author} {\bibfnamefont {T.}~\bibnamefont
  {Ramos}}, \bibinfo {author} {\bibfnamefont {B.}~\bibnamefont {Vermersch}},
  \bibinfo {author} {\bibfnamefont {P.}~\bibnamefont {Hauke}}, \bibinfo
  {author} {\bibfnamefont {H.}~\bibnamefont {Pichler}},\ and\ \bibinfo {author}
  {\bibfnamefont {P.}~\bibnamefont {Zoller}},\ }\bibfield  {title} {\bibinfo
  {title} {Non-Markovian dynamics in chiral quantum networks with spins and
  photons},\ }\href {https://doi.org/10.1103/PhysRevA.93.062104} {\bibfield
  {journal} {\bibinfo  {journal} {Phys. Rev. A}\ }\textbf {\bibinfo {volume}
  {93}},\ \bibinfo {pages} {062104} (\bibinfo {year} {2016})}\BibitemShut
  {NoStop}%
\bibitem [{\citenamefont {Albrecht}\ \emph {et~al.}(2019)\citenamefont
  {Albrecht}, \citenamefont {Henriet}, \citenamefont {Asenjo-Garcia},
  \citenamefont {Dieterle}, \citenamefont {Painter},\ and\ \citenamefont
  {Chang}}]{Albrecht19}%
  \BibitemOpen
  \bibfield  {author} {\bibinfo {author} {\bibfnamefont {A.}~\bibnamefont
  {Albrecht}}, \bibinfo {author} {\bibfnamefont {L.}~\bibnamefont {Henriet}},
  \bibinfo {author} {\bibfnamefont {A.}~\bibnamefont {Asenjo-Garcia}}, \bibinfo
  {author} {\bibfnamefont {P.~B.}\ \bibnamefont {Dieterle}}, \bibinfo {author}
  {\bibfnamefont {O.}~\bibnamefont {Painter}},\ and\ \bibinfo {author}
  {\bibfnamefont {D.~E.}\ \bibnamefont {Chang}},\ }\bibfield  {title} {\bibinfo
  {title} {Subradiant states of quantum bits coupled to a one-dimensional
  waveguide},\ }\href {https://doi.org/10.1088/1367-2630/ab0134} {\bibfield
  {journal} {\bibinfo  {journal} {New J. Phys.}\ }\textbf {\bibinfo {volume}
  {21}},\ \bibinfo {pages} {025003} (\bibinfo {year} {2019})}\BibitemShut
  {NoStop}%
\bibitem [{\citenamefont {Zhang}\ and\ \citenamefont
  {M\o{}lmer}(2019)}]{Zhang19}%
  \BibitemOpen
  \bibfield  {author} {\bibinfo {author} {\bibfnamefont {Y.-X.}\ \bibnamefont
  {Zhang}}\ and\ \bibinfo {author} {\bibfnamefont {K.}~\bibnamefont
  {M\o{}lmer}},\ }\bibfield  {title} {\bibinfo {title} {Theory of subradiant
  states of a one-dimensional two-level atom chain},\ }\href
  {https://doi.org/10.1103/PhysRevLett.122.203605} {\bibfield  {journal}
  {\bibinfo  {journal} {Phys. Rev. Lett.}\ }\textbf {\bibinfo {volume} {122}},\
  \bibinfo {pages} {203605} (\bibinfo {year} {2019})}\BibitemShut {NoStop}%
\bibitem [{\citenamefont {Milburn}(1989)}]{Milburn89}%
  \BibitemOpen
  \bibfield  {author} {\bibinfo {author} {\bibfnamefont {G.~J.}\ \bibnamefont
  {Milburn}},\ }\bibfield  {title} {\bibinfo {title} {Quantum optical fredkin
  gate},\ }\href {https://doi.org/10.1103/PhysRevLett.62.2124} {\bibfield
  {journal} {\bibinfo  {journal} {Phys. Rev. Lett.}\ }\textbf {\bibinfo
  {volume} {62}},\ \bibinfo {pages} {2124} (\bibinfo {year}
  {1989})}\BibitemShut {NoStop}%
\bibitem [{\citenamefont {Ma\ifmmode~\check{s}\else \v{s}\fi{}alas}\ and\
  \citenamefont {Fleischhauer}(2004)}]{Masalas04}%
  \BibitemOpen
  \bibfield  {author} {\bibinfo {author} {\bibfnamefont {M.}~\bibnamefont
  {Ma\ifmmode~\check{s}\else \v{s}\fi{}alas}}\ and\ \bibinfo {author}
  {\bibfnamefont {M.}~\bibnamefont {Fleischhauer}},\ }\bibfield  {title}
  {\bibinfo {title} {Scattering of dark-state polaritons in optical lattices
  and quantum phase gate for photons},\ }\href
  {https://doi.org/10.1103/PhysRevA.69.061801} {\bibfield  {journal} {\bibinfo
  {journal} {Phys. Rev. A}\ }\textbf {\bibinfo {volume} {69}},\ \bibinfo
  {pages} {061801(R)} (\bibinfo {year} {2004})}\BibitemShut {NoStop}%
\bibitem [{\citenamefont {Gorshkov}\ \emph {et~al.}(2011)\citenamefont
  {Gorshkov}, \citenamefont {Otterbach}, \citenamefont {Fleischhauer},
  \citenamefont {Pohl},\ and\ \citenamefont {Lukin}}]{Gorshkov11}%
  \BibitemOpen
  \bibfield  {author} {\bibinfo {author} {\bibfnamefont {A.~V.}\ \bibnamefont
  {Gorshkov}}, \bibinfo {author} {\bibfnamefont {J.}~\bibnamefont {Otterbach}},
  \bibinfo {author} {\bibfnamefont {M.}~\bibnamefont {Fleischhauer}}, \bibinfo
  {author} {\bibfnamefont {T.}~\bibnamefont {Pohl}},\ and\ \bibinfo {author}
  {\bibfnamefont {M.~D.}\ \bibnamefont {Lukin}},\ }\bibfield  {title} {\bibinfo
  {title} {Photon-photon interactions via {R}ydberg blockade},\ }\href
  {https://doi.org/10.1103/PhysRevLett.107.133602} {\bibfield  {journal}
  {\bibinfo  {journal} {Phys. Rev. Lett.}\ }\textbf {\bibinfo {volume} {107}},\
  \bibinfo {pages} {133602} (\bibinfo {year} {2011})}\BibitemShut {NoStop}%
\bibitem [{\citenamefont {Brod}\ \emph {et~al.}(2016)\citenamefont {Brod},
  \citenamefont {Combes},\ and\ \citenamefont {Gea-Banacloche}}]{Brod16PRA}%
  \BibitemOpen
  \bibfield  {author} {\bibinfo {author} {\bibfnamefont {D.~J.}\ \bibnamefont
  {Brod}}, \bibinfo {author} {\bibfnamefont {J.}~\bibnamefont {Combes}},\ and\
  \bibinfo {author} {\bibfnamefont {J.}~\bibnamefont {Gea-Banacloche}},\
  }\bibfield  {title} {\bibinfo {title} {Two photons co- and counterpropagating
  through $n$ cross-kerr sites},\ }\href
  {https://doi.org/10.1103/PhysRevA.94.023833} {\bibfield  {journal} {\bibinfo
  {journal} {Phys. Rev. A}\ }\textbf {\bibinfo {volume} {94}},\ \bibinfo
  {pages} {023833} (\bibinfo {year} {2016})}\BibitemShut {NoStop}%
\bibitem [{\citenamefont {Chuang}\ and\ \citenamefont
  {Yamamoto}(1995)}]{Chuang95}%
  \BibitemOpen
  \bibfield  {author} {\bibinfo {author} {\bibfnamefont {I.~L.}\ \bibnamefont
  {Chuang}}\ and\ \bibinfo {author} {\bibfnamefont {Y.}~\bibnamefont
  {Yamamoto}},\ }\bibfield  {title} {\bibinfo {title} {Simple quantum
  computer},\ }\href {https://doi.org/10.1103/PhysRevA.52.3489} {\bibfield
  {journal} {\bibinfo  {journal} {Phys. Rev. A}\ }\textbf {\bibinfo {volume}
  {52}},\ \bibinfo {pages} {3489} (\bibinfo {year} {1995})}\BibitemShut
  {NoStop}%
\bibitem [{\citenamefont {Hutchinson}\ and\ \citenamefont
  {Milburn}(2004)}]{Hutchinson04}%
  \BibitemOpen
  \bibfield  {author} {\bibinfo {author} {\bibfnamefont {G.~D.}\ \bibnamefont
  {Hutchinson}}\ and\ \bibinfo {author} {\bibfnamefont {G.~J.}\ \bibnamefont
  {Milburn}},\ }\bibfield  {title} {\bibinfo {title} {Nonlinear quantum optical
  computing via measurement},\ }\bibfield  {booktitle} {\emph {\bibinfo
  {booktitle} {Journal of Modern Optics}},\ }\href
  {https://doi.org/10.1080/09500340408230417} {\bibfield  {journal} {\bibinfo
  {journal} {J. Mod. Opt.}\ }\textbf {\bibinfo {volume} {51}},\ \bibinfo
  {pages} {1211} (\bibinfo {year} {2004})}\BibitemShut {NoStop}%
\bibitem [{\citenamefont {Greif}\ \emph {et~al.}(2016)\citenamefont {Greif},
  \citenamefont {Parsons}, \citenamefont {Mazurenko}, \citenamefont {Chiu},
  \citenamefont {Blatt}, \citenamefont {Huber}, \citenamefont {Ji},\ and\
  \citenamefont {Greiner}}]{Greif16}%
  \BibitemOpen
  \bibfield  {author} {\bibinfo {author} {\bibfnamefont {D.}~\bibnamefont
  {Greif}}, \bibinfo {author} {\bibfnamefont {M.~F.}\ \bibnamefont {Parsons}},
  \bibinfo {author} {\bibfnamefont {A.}~\bibnamefont {Mazurenko}}, \bibinfo
  {author} {\bibfnamefont {C.~S.}\ \bibnamefont {Chiu}}, \bibinfo {author}
  {\bibfnamefont {S.}~\bibnamefont {Blatt}}, \bibinfo {author} {\bibfnamefont
  {F.}~\bibnamefont {Huber}}, \bibinfo {author} {\bibfnamefont
  {G.}~\bibnamefont {Ji}},\ and\ \bibinfo {author} {\bibfnamefont
  {M.}~\bibnamefont {Greiner}},\ }\bibfield  {title} {\bibinfo {title}
  {Site-resolved imaging of a fermionic {M}ott insulator},\ }\href
  {https://doi.org/10.1126/science.aad9041} {\bibfield  {journal} {\bibinfo
  {journal} {Science}\ }\textbf {\bibinfo {volume} {351}},\ \bibinfo {pages}
  {953} (\bibinfo {year} {2016})}\BibitemShut {NoStop}%
\bibitem [{\citenamefont {Kumar}\ \emph {et~al.}(2018)\citenamefont {Kumar},
  \citenamefont {Wu}, \citenamefont {Giraldo},\ and\ \citenamefont
  {Weiss}}]{Kumar18}%
  \BibitemOpen
  \bibfield  {author} {\bibinfo {author} {\bibfnamefont {A.}~\bibnamefont
  {Kumar}}, \bibinfo {author} {\bibfnamefont {T.-Y.}\ \bibnamefont {Wu}},
  \bibinfo {author} {\bibfnamefont {F.}~\bibnamefont {Giraldo}},\ and\ \bibinfo
  {author} {\bibfnamefont {D.~S.}\ \bibnamefont {Weiss}},\ }\bibfield  {title}
  {\bibinfo {title} {Sorting ultracold atoms in a three-dimensional optical
  lattice in a realization of {M}axwell's demon},\ }\href
  {https://doi.org/10.1038/s41586-018-0458-7} {\bibfield  {journal} {\bibinfo
  {journal} {Nature}\ }\textbf {\bibinfo {volume} {561}},\ \bibinfo {pages}
  {83} (\bibinfo {year} {2018})}\BibitemShut {NoStop}%
\bibitem [{\citenamefont {Rui}\ \emph {et~al.}(2020)\citenamefont {Rui},
  \citenamefont {Wei}, \citenamefont {Rubio-Abadal}, \citenamefont {Hollerith},
  \citenamefont {Zeiher}, \citenamefont {Stamper-Kurn}, \citenamefont {Gross},\
  and\ \citenamefont {Bloch}}]{Rui20}%
  \BibitemOpen
  \bibfield  {author} {\bibinfo {author} {\bibfnamefont {J.}~\bibnamefont
  {Rui}}, \bibinfo {author} {\bibfnamefont {D.}~\bibnamefont {Wei}}, \bibinfo
  {author} {\bibfnamefont {A.}~\bibnamefont {Rubio-Abadal}}, \bibinfo {author}
  {\bibfnamefont {S.}~\bibnamefont {Hollerith}}, \bibinfo {author}
  {\bibfnamefont {J.}~\bibnamefont {Zeiher}}, \bibinfo {author} {\bibfnamefont
  {D.~M.}\ \bibnamefont {Stamper-Kurn}}, \bibinfo {author} {\bibfnamefont
  {C.}~\bibnamefont {Gross}},\ and\ \bibinfo {author} {\bibfnamefont
  {I.}~\bibnamefont {Bloch}},\ }\bibfield  {title} {\bibinfo {title} {A
  subradiant optical mirror formed by a single structured atomic layer},\
  }\href {https://doi.org/10.1038/s41586-020-2463-x} {\bibfield  {journal}
  {\bibinfo  {journal} {Nature}\ }\textbf {\bibinfo {volume} {583}},\ \bibinfo
  {pages} {369} (\bibinfo {year} {2020})}\BibitemShut {NoStop}%
\bibitem [{\citenamefont {Kim}\ \emph {et~al.}(2016)\citenamefont {Kim},
  \citenamefont {Lee}, \citenamefont {Lee}, \citenamefont {Jo}, \citenamefont
  {Song},\ and\ \citenamefont {Ahn}}]{Kim16}%
  \BibitemOpen
  \bibfield  {author} {\bibinfo {author} {\bibfnamefont {H.}~\bibnamefont
  {Kim}}, \bibinfo {author} {\bibfnamefont {W.}~\bibnamefont {Lee}}, \bibinfo
  {author} {\bibfnamefont {H.-G.}\ \bibnamefont {Lee}}, \bibinfo {author}
  {\bibfnamefont {H.}~\bibnamefont {Jo}}, \bibinfo {author} {\bibfnamefont
  {Y.}~\bibnamefont {Song}},\ and\ \bibinfo {author} {\bibfnamefont
  {J.}~\bibnamefont {Ahn}},\ }\bibfield  {title} {\bibinfo {title} {In situ
  single-atom array synthesis using dynamic holographic optical tweezers},\
  }\href {https://doi.org/10.1038/ncomms13317} {\bibfield  {journal} {\bibinfo
  {journal} {Nat. Commun.}\ }\textbf {\bibinfo {volume} {7}},\ \bibinfo {pages}
  {13317} (\bibinfo {year} {2016})}\BibitemShut {NoStop}%
\bibitem [{\citenamefont {Endres}\ \emph {et~al.}(2016)\citenamefont {Endres},
  \citenamefont {Bernien}, \citenamefont {Keesling}, \citenamefont {Levine},
  \citenamefont {Anschuetz}, \citenamefont {Krajenbrink}, \citenamefont
  {Senko}, \citenamefont {Vuletic}, \citenamefont {Greiner},\ and\
  \citenamefont {Lukin}}]{Endres16}%
  \BibitemOpen
  \bibfield  {author} {\bibinfo {author} {\bibfnamefont {M.}~\bibnamefont
  {Endres}}, \bibinfo {author} {\bibfnamefont {H.}~\bibnamefont {Bernien}},
  \bibinfo {author} {\bibfnamefont {A.}~\bibnamefont {Keesling}}, \bibinfo
  {author} {\bibfnamefont {H.}~\bibnamefont {Levine}}, \bibinfo {author}
  {\bibfnamefont {E.~R.}\ \bibnamefont {Anschuetz}}, \bibinfo {author}
  {\bibfnamefont {A.}~\bibnamefont {Krajenbrink}}, \bibinfo {author}
  {\bibfnamefont {C.}~\bibnamefont {Senko}}, \bibinfo {author} {\bibfnamefont
  {V.}~\bibnamefont {Vuletic}}, \bibinfo {author} {\bibfnamefont
  {M.}~\bibnamefont {Greiner}},\ and\ \bibinfo {author} {\bibfnamefont {M.~D.}\
  \bibnamefont {Lukin}},\ }\bibfield  {title} {\bibinfo {title} {Atom-by-atom
  assembly of defect-free one-dimensional cold atom arrays},\ }\href
  {https://doi.org/10.1126/science.aah3752} {\bibfield  {journal} {\bibinfo
  {journal} {Science}\ }\textbf {\bibinfo {volume} {354}},\ \bibinfo {pages}
  {1024} (\bibinfo {year} {2016})}\BibitemShut {NoStop}%
\bibitem [{\citenamefont {Barredo}\ \emph {et~al.}(2016)\citenamefont
  {Barredo}, \citenamefont {de~L{\'e}s{\'e}leuc}, \citenamefont {Lienhard},
  \citenamefont {Lahaye},\ and\ \citenamefont {Browaeys}}]{Barredo16}%
  \BibitemOpen
  \bibfield  {author} {\bibinfo {author} {\bibfnamefont {D.}~\bibnamefont
  {Barredo}}, \bibinfo {author} {\bibfnamefont {S.}~\bibnamefont
  {de~L{\'e}s{\'e}leuc}}, \bibinfo {author} {\bibfnamefont {V.}~\bibnamefont
  {Lienhard}}, \bibinfo {author} {\bibfnamefont {T.}~\bibnamefont {Lahaye}},\
  and\ \bibinfo {author} {\bibfnamefont {A.}~\bibnamefont {Browaeys}},\
  }\bibfield  {title} {\bibinfo {title} {An atom-by-atom assembler of
  defect-free arbitrary two-dimensional atomic arrays},\ }\href
  {https://doi.org/10.1126/science.aah3778} {\bibfield  {journal} {\bibinfo
  {journal} {Science}\ }\textbf {\bibinfo {volume} {354}},\ \bibinfo {pages}
  {1021} (\bibinfo {year} {2016})}\BibitemShut {NoStop}%
\bibitem [{\citenamefont {Bloch}(2005)}]{Bloch05}%
  \BibitemOpen
  \bibfield  {author} {\bibinfo {author} {\bibfnamefont {I.}~\bibnamefont
  {Bloch}},\ }\bibfield  {title} {\bibinfo {title} {Ultracold quantum gases in
  optical lattices},\ }\href {https://doi.org/10.1038/nphys138} {\bibfield
  {journal} {\bibinfo  {journal} {Nat. Phys.}\ }\textbf {\bibinfo {volume}
  {1}},\ \bibinfo {pages} {23} (\bibinfo {year} {2005})}\BibitemShut {NoStop}%
\bibitem [{\citenamefont {Bakr}\ \emph {et~al.}(2009)\citenamefont {Bakr},
  \citenamefont {Gillen}, \citenamefont {Peng}, \citenamefont {F{\"o}lling},\
  and\ \citenamefont {Greiner}}]{Bakr09}%
  \BibitemOpen
  \bibfield  {author} {\bibinfo {author} {\bibfnamefont {W.~S.}\ \bibnamefont
  {Bakr}}, \bibinfo {author} {\bibfnamefont {J.~I.}\ \bibnamefont {Gillen}},
  \bibinfo {author} {\bibfnamefont {A.}~\bibnamefont {Peng}}, \bibinfo {author}
  {\bibfnamefont {S.}~\bibnamefont {F{\"o}lling}},\ and\ \bibinfo {author}
  {\bibfnamefont {M.}~\bibnamefont {Greiner}},\ }\bibfield  {title} {\bibinfo
  {title} {A quantum gas microscope for detecting single atoms in a
  {H}ubbard-regime optical lattice},\ }\href
  {https://doi.org/10.1038/nature08482} {\bibfield  {journal} {\bibinfo
  {journal} {Nature}\ }\textbf {\bibinfo {volume} {462}},\ \bibinfo {pages}
  {74} (\bibinfo {year} {2009})}\BibitemShut {NoStop}%
\bibitem [{\citenamefont {Bakr}\ \emph {et~al.}(2010)\citenamefont {Bakr},
  \citenamefont {Peng}, \citenamefont {Tai}, \citenamefont {Ma}, \citenamefont
  {Simon}, \citenamefont {Gillen}, \citenamefont {F{\"o}lling}, \citenamefont
  {Pollet},\ and\ \citenamefont {Greiner}}]{Bakr10}%
  \BibitemOpen
  \bibfield  {author} {\bibinfo {author} {\bibfnamefont {W.~S.}\ \bibnamefont
  {Bakr}}, \bibinfo {author} {\bibfnamefont {A.}~\bibnamefont {Peng}}, \bibinfo
  {author} {\bibfnamefont {M.~E.}\ \bibnamefont {Tai}}, \bibinfo {author}
  {\bibfnamefont {R.}~\bibnamefont {Ma}}, \bibinfo {author} {\bibfnamefont
  {J.}~\bibnamefont {Simon}}, \bibinfo {author} {\bibfnamefont {J.~I.}\
  \bibnamefont {Gillen}}, \bibinfo {author} {\bibfnamefont {S.}~\bibnamefont
  {F{\"o}lling}}, \bibinfo {author} {\bibfnamefont {L.}~\bibnamefont
  {Pollet}},\ and\ \bibinfo {author} {\bibfnamefont {M.}~\bibnamefont
  {Greiner}},\ }\bibfield  {title} {\bibinfo {title} {Probing the
  superfluid{\textendash}to{\textendash}{M}ott insulator transition at the
  single-atom level},\ }\href {https://doi.org/10.1126/science.1192368}
  {\bibfield  {journal} {\bibinfo  {journal} {Science}\ }\textbf {\bibinfo
  {volume} {329}},\ \bibinfo {pages} {547} (\bibinfo {year}
  {2010})}\BibitemShut {NoStop}%
\bibitem [{\citenamefont {Lester}\ \emph {et~al.}(2015)\citenamefont {Lester},
  \citenamefont {Luick}, \citenamefont {Kaufman}, \citenamefont {Reynolds},\
  and\ \citenamefont {Regal}}]{Lester15}%
  \BibitemOpen
  \bibfield  {author} {\bibinfo {author} {\bibfnamefont {B.~J.}\ \bibnamefont
  {Lester}}, \bibinfo {author} {\bibfnamefont {N.}~\bibnamefont {Luick}},
  \bibinfo {author} {\bibfnamefont {A.~M.}\ \bibnamefont {Kaufman}}, \bibinfo
  {author} {\bibfnamefont {C.~M.}\ \bibnamefont {Reynolds}},\ and\ \bibinfo
  {author} {\bibfnamefont {C.~A.}\ \bibnamefont {Regal}},\ }\bibfield  {title}
  {\bibinfo {title} {Rapid production of uniformly filled arrays of neutral
  atoms},\ }\href {https://doi.org/10.1103/PhysRevLett.115.073003} {\bibfield
  {journal} {\bibinfo  {journal} {Phys. Rev. Lett.}\ }\textbf {\bibinfo
  {volume} {115}},\ \bibinfo {pages} {073003} (\bibinfo {year}
  {2015})}\BibitemShut {NoStop}%
\bibitem [{\citenamefont {Labuhn}\ \emph {et~al.}(2016)\citenamefont {Labuhn},
  \citenamefont {Barredo}, \citenamefont {Ravets}, \citenamefont
  {de~L{\'e}s{\'e}leuc}, \citenamefont {Macr{\`\i}}, \citenamefont {Lahaye},\
  and\ \citenamefont {Browaeys}}]{Labuhn16}%
  \BibitemOpen
  \bibfield  {author} {\bibinfo {author} {\bibfnamefont {H.}~\bibnamefont
  {Labuhn}}, \bibinfo {author} {\bibfnamefont {D.}~\bibnamefont {Barredo}},
  \bibinfo {author} {\bibfnamefont {S.}~\bibnamefont {Ravets}}, \bibinfo
  {author} {\bibfnamefont {S.}~\bibnamefont {de~L{\'e}s{\'e}leuc}}, \bibinfo
  {author} {\bibfnamefont {T.}~\bibnamefont {Macr{\`\i}}}, \bibinfo {author}
  {\bibfnamefont {T.}~\bibnamefont {Lahaye}},\ and\ \bibinfo {author}
  {\bibfnamefont {A.}~\bibnamefont {Browaeys}},\ }\bibfield  {title} {\bibinfo
  {title} {Tunable two-dimensional arrays of single {R}ydberg atoms for
  realizing quantum {I}sing models},\ }\href
  {https://doi.org/10.1038/nature18274} {\bibfield  {journal} {\bibinfo
  {journal} {Nature}\ }\textbf {\bibinfo {volume} {534}},\ \bibinfo {pages}
  {667} (\bibinfo {year} {2016})}\BibitemShut {NoStop}%
\bibitem [{\citenamefont {Bernien}\ \emph {et~al.}(2017)\citenamefont
  {Bernien}, \citenamefont {Schwartz}, \citenamefont {Keesling}, \citenamefont
  {Levine}, \citenamefont {Omran}, \citenamefont {Pichler}, \citenamefont
  {Choi}, \citenamefont {Zibrov}, \citenamefont {Endres}, \citenamefont
  {Greiner}, \citenamefont {Vuleti{\'c}},\ and\ \citenamefont
  {Lukin}}]{Bernien17}%
  \BibitemOpen
  \bibfield  {author} {\bibinfo {author} {\bibfnamefont {H.}~\bibnamefont
  {Bernien}}, \bibinfo {author} {\bibfnamefont {S.}~\bibnamefont {Schwartz}},
  \bibinfo {author} {\bibfnamefont {A.}~\bibnamefont {Keesling}}, \bibinfo
  {author} {\bibfnamefont {H.}~\bibnamefont {Levine}}, \bibinfo {author}
  {\bibfnamefont {A.}~\bibnamefont {Omran}}, \bibinfo {author} {\bibfnamefont
  {H.}~\bibnamefont {Pichler}}, \bibinfo {author} {\bibfnamefont
  {S.}~\bibnamefont {Choi}}, \bibinfo {author} {\bibfnamefont {A.~S.}\
  \bibnamefont {Zibrov}}, \bibinfo {author} {\bibfnamefont {M.}~\bibnamefont
  {Endres}}, \bibinfo {author} {\bibfnamefont {M.}~\bibnamefont {Greiner}},
  \bibinfo {author} {\bibfnamefont {V.}~\bibnamefont {Vuleti{\'c}}},\ and\
  \bibinfo {author} {\bibfnamefont {M.~D.}\ \bibnamefont {Lukin}},\ }\bibfield
  {title} {\bibinfo {title} {Probing many-body dynamics on a 51-atom quantum
  simulator},\ }\href@noop {} {\bibfield  {journal} {\bibinfo  {journal}
  {Nature}\ }\textbf {\bibinfo {volume} {551}},\ \bibinfo {pages} {579}
  (\bibinfo {year} {2017})}\BibitemShut {NoStop}%
\bibitem [{\citenamefont {Barredo}\ \emph {et~al.}(2018)\citenamefont
  {Barredo}, \citenamefont {Lienhard}, \citenamefont {de~L{\'e}s{\'e}leuc},
  \citenamefont {Lahaye},\ and\ \citenamefont {Browaeys}}]{Barredo18}%
  \BibitemOpen
  \bibfield  {author} {\bibinfo {author} {\bibfnamefont {D.}~\bibnamefont
  {Barredo}}, \bibinfo {author} {\bibfnamefont {V.}~\bibnamefont {Lienhard}},
  \bibinfo {author} {\bibfnamefont {S.}~\bibnamefont {de~L{\'e}s{\'e}leuc}},
  \bibinfo {author} {\bibfnamefont {T.}~\bibnamefont {Lahaye}},\ and\ \bibinfo
  {author} {\bibfnamefont {A.}~\bibnamefont {Browaeys}},\ }\bibfield  {title}
  {\bibinfo {title} {Synthetic three-dimensional atomic structures assembled
  atom by atom},\ }\href {https://doi.org/10.1038/s41586-018-0450-2} {\bibfield
   {journal} {\bibinfo  {journal} {Nature}\ }\textbf {\bibinfo {volume}
  {561}},\ \bibinfo {pages} {79} (\bibinfo {year} {2018})}\BibitemShut
  {NoStop}%
\bibitem [{\citenamefont {Ohl~de Mello}\ \emph {et~al.}(2019)\citenamefont
  {Ohl~de Mello}, \citenamefont {Sch\"affner}, \citenamefont {Werkmann},
  \citenamefont {Preuschoff}, \citenamefont {Kohfahl}, \citenamefont
  {Schlosser},\ and\ \citenamefont {Birkl}}]{OhlDeMello19}%
  \BibitemOpen
  \bibfield  {author} {\bibinfo {author} {\bibfnamefont {D.}~\bibnamefont
  {Ohl~de Mello}}, \bibinfo {author} {\bibfnamefont {D.}~\bibnamefont
  {Sch\"affner}}, \bibinfo {author} {\bibfnamefont {J.}~\bibnamefont
  {Werkmann}}, \bibinfo {author} {\bibfnamefont {T.}~\bibnamefont
  {Preuschoff}}, \bibinfo {author} {\bibfnamefont {L.}~\bibnamefont {Kohfahl}},
  \bibinfo {author} {\bibfnamefont {M.}~\bibnamefont {Schlosser}},\ and\
  \bibinfo {author} {\bibfnamefont {G.}~\bibnamefont {Birkl}},\ }\bibfield
  {title} {\bibinfo {title} {Defect-free assembly of 2{D} clusters of more than
  100 single-atom quantum systems},\ }\href
  {https://doi.org/10.1103/PhysRevLett.122.203601} {\bibfield  {journal}
  {\bibinfo  {journal} {Phys. Rev. Lett.}\ }\textbf {\bibinfo {volume} {122}},\
  \bibinfo {pages} {203601} (\bibinfo {year} {2019})}\BibitemShut {NoStop}%
\bibitem [{\citenamefont {Glicenstein}\ \emph {et~al.}(2020)\citenamefont
  {Glicenstein}, \citenamefont {Ferioli}, \citenamefont {\ifmmode
  \check{S}\else \v{S}\fi{}ibali\'{c}},
  \citenamefont {Brossard}, \citenamefont {Ferrier-Barbut},\ and\ \citenamefont
  {Browaeys}}]{Glicenstein20}%
  \BibitemOpen
  \bibfield  {author} {\bibinfo {author} {\bibfnamefont {A.}~\bibnamefont
  {Glicenstein}}, \bibinfo {author} {\bibfnamefont {G.}~\bibnamefont
  {Ferioli}}, \bibinfo {author} {\bibfnamefont {N.}~\bibnamefont {\ifmmode
  \check{S}\else \v{S}\fi{}ibali\ifmmode~\acute{c}\else \'{c}\fi{}}}, \bibinfo
  {author} {\bibfnamefont {L.}~\bibnamefont {Brossard}}, \bibinfo {author}
  {\bibfnamefont {I.}~\bibnamefont {Ferrier-Barbut}},\ and\ \bibinfo {author}
  {\bibfnamefont {A.}~\bibnamefont {Browaeys}},\ }\bibfield  {title} {\bibinfo
  {title} {Collective shift in resonant light scattering by a one-dimensional
  atomic chain},\ }\href {https://doi.org/10.1103/PhysRevLett.124.253602}
  {\bibfield  {journal} {\bibinfo  {journal} {Phys. Rev. Lett.}\ }\textbf
  {\bibinfo {volume} {124}},\ \bibinfo {pages} {253602} (\bibinfo {year}
  {2020})}\BibitemShut {NoStop}%
\bibitem [{\citenamefont {Bekenstein}\ \emph {et~al.}(2020)\citenamefont
  {Bekenstein}, \citenamefont {Pikovski}, \citenamefont {Pichler},
  \citenamefont {Shahmoon}, \citenamefont {Yelin},\ and\ \citenamefont
  {Lukin}}]{Bekenstein20}%
  \BibitemOpen
  \bibfield  {author} {\bibinfo {author} {\bibfnamefont {R.}~\bibnamefont
  {Bekenstein}}, \bibinfo {author} {\bibfnamefont {I.}~\bibnamefont
  {Pikovski}}, \bibinfo {author} {\bibfnamefont {H.}~\bibnamefont {Pichler}},
  \bibinfo {author} {\bibfnamefont {E.}~\bibnamefont {Shahmoon}}, \bibinfo
  {author} {\bibfnamefont {S.~F.}\ \bibnamefont {Yelin}},\ and\ \bibinfo
  {author} {\bibfnamefont {M.~D.}\ \bibnamefont {Lukin}},\ }\bibfield  {title}
  {\bibinfo {title} {Quantum metasurfaces with atom arrays},\ }\href
  {https://doi.org/10.1038/s41567-020-0845-5} {\bibfield  {journal} {\bibinfo
  {journal} {Nat. Phys.}\ }\textbf {\bibinfo {volume} {16}},\ \bibinfo {pages}
  {676} (\bibinfo {year} {2020})}\BibitemShut {NoStop}%
\bibitem [{\citenamefont {Olmos}\ \emph {et~al.}(2013)\citenamefont {Olmos},
  \citenamefont {Yu}, \citenamefont {Singh}, \citenamefont {Schreck},
  \citenamefont {Bongs},\ and\ \citenamefont {Lesanovsky}}]{Olmos13}%
  \BibitemOpen
  \bibfield  {author} {\bibinfo {author} {\bibfnamefont {B.}~\bibnamefont
  {Olmos}}, \bibinfo {author} {\bibfnamefont {D.}~\bibnamefont {Yu}}, \bibinfo
  {author} {\bibfnamefont {Y.}~\bibnamefont {Singh}}, \bibinfo {author}
  {\bibfnamefont {F.}~\bibnamefont {Schreck}}, \bibinfo {author} {\bibfnamefont
  {K.}~\bibnamefont {Bongs}},\ and\ \bibinfo {author} {\bibfnamefont
  {I.}~\bibnamefont {Lesanovsky}},\ }\bibfield  {title} {\bibinfo {title}
  {Long-range interacting many-body systems with alkaline-earth-metal atoms},\
  }\href {https://doi.org/10.1103/PhysRevLett.110.143602} {\bibfield  {journal}
  {\bibinfo  {journal} {Phys. Rev. Lett.}\ }\textbf {\bibinfo {volume} {110}},\
  \bibinfo {pages} {143602} (\bibinfo {year} {2013})}\BibitemShut {NoStop}%
\bibitem [{\citenamefont {Hebenstreit}\ \emph {et~al.}(2017)\citenamefont
  {Hebenstreit}, \citenamefont {Kraus}, \citenamefont {Ostermann},\ and\
  \citenamefont {Ritsch}}]{Hebenstreit17}%
  \BibitemOpen
  \bibfield  {author} {\bibinfo {author} {\bibfnamefont {M.}~\bibnamefont
  {Hebenstreit}}, \bibinfo {author} {\bibfnamefont {B.}~\bibnamefont {Kraus}},
  \bibinfo {author} {\bibfnamefont {L.}~\bibnamefont {Ostermann}},\ and\
  \bibinfo {author} {\bibfnamefont {H.}~\bibnamefont {Ritsch}},\ }\bibfield
  {title} {\bibinfo {title} {Subradiance via entanglement in atoms with several
  independent decay channels},\ }\href
  {https://doi.org/10.1103/PhysRevLett.118.143602} {\bibfield  {journal}
  {\bibinfo  {journal} {Phys. Rev. Lett.}\ }\textbf {\bibinfo {volume} {118}},\
  \bibinfo {pages} {143602} (\bibinfo {year} {2017})}\BibitemShut {NoStop}%
\bibitem [{\citenamefont {Covey}\ \emph {et~al.}(2019)\citenamefont {Covey},
  \citenamefont {Sipahigil}, \citenamefont {Szoke}, \citenamefont {Sinclair},
  \citenamefont {Endres},\ and\ \citenamefont {Painter}}]{Covey19}%
  \BibitemOpen
  \bibfield  {author} {\bibinfo {author} {\bibfnamefont {J.~P.}\ \bibnamefont
  {Covey}}, \bibinfo {author} {\bibfnamefont {A.}~\bibnamefont {Sipahigil}},
  \bibinfo {author} {\bibfnamefont {S.}~\bibnamefont {Szoke}}, \bibinfo
  {author} {\bibfnamefont {N.}~\bibnamefont {Sinclair}}, \bibinfo {author}
  {\bibfnamefont {M.}~\bibnamefont {Endres}},\ and\ \bibinfo {author}
  {\bibfnamefont {O.}~\bibnamefont {Painter}},\ }\bibfield  {title} {\bibinfo
  {title} {Telecom-band quantum optics with ytterbium atoms and silicon
  nanophotonics},\ }\href {https://doi.org/10.1103/PhysRevApplied.11.034044}
  {\bibfield  {journal} {\bibinfo  {journal} {Phys. Rev. Applied}\ }\textbf
  {\bibinfo {volume} {11}},\ \bibinfo {pages} {034044} (\bibinfo {year}
  {2019})}\BibitemShut {NoStop}%
\bibitem [{\citenamefont {Guimond}\ \emph {et~al.}(2019)\citenamefont
  {Guimond}, \citenamefont {Grankin}, \citenamefont {Vasilyev}, \citenamefont
  {Vermersch},\ and\ \citenamefont {Zoller}}]{Guimond19}%
  \BibitemOpen
  \bibfield  {author} {\bibinfo {author} {\bibfnamefont {P.-O.}\ \bibnamefont
  {Guimond}}, \bibinfo {author} {\bibfnamefont {A.}~\bibnamefont {Grankin}},
  \bibinfo {author} {\bibfnamefont {D.~V.}\ \bibnamefont {Vasilyev}}, \bibinfo
  {author} {\bibfnamefont {B.}~\bibnamefont {Vermersch}},\ and\ \bibinfo
  {author} {\bibfnamefont {P.}~\bibnamefont {Zoller}},\ }\bibfield  {title}
  {\bibinfo {title} {Subradiant {B}ell states in distant atomic arrays},\
  }\href {https://doi.org/10.1103/PhysRevLett.122.093601} {\bibfield  {journal}
  {\bibinfo  {journal} {Phys. Rev. Lett.}\ }\textbf {\bibinfo {volume} {122}},\
  \bibinfo {pages} {093601} (\bibinfo {year} {2019})}\BibitemShut {NoStop}%
\bibitem [{\citenamefont {Plankensteiner}\ \emph {et~al.}(2015)\citenamefont
  {Plankensteiner}, \citenamefont {Ostermann}, \citenamefont {Ritsch},\ and\
  \citenamefont {Genes}}]{Plankensteiner15}%
  \BibitemOpen
  \bibfield  {author} {\bibinfo {author} {\bibfnamefont {D.}~\bibnamefont
  {Plankensteiner}}, \bibinfo {author} {\bibfnamefont {L.}~\bibnamefont
  {Ostermann}}, \bibinfo {author} {\bibfnamefont {H.}~\bibnamefont {Ritsch}},\
  and\ \bibinfo {author} {\bibfnamefont {C.}~\bibnamefont {Genes}},\ }\bibfield
   {title} {\bibinfo {title} {Selective protected state preparation of coupled
  dissipative quantum emitters},\ }\href {https://doi.org/10.1038/srep16231}
  {\bibfield  {journal} {\bibinfo  {journal} {Sci. Rep.}\ }\textbf {\bibinfo
  {volume} {5}},\ \bibinfo {pages} {16231} (\bibinfo {year}
  {2015})}\BibitemShut {NoStop}%
\bibitem [{\citenamefont {He}\ \emph {et~al.}(2019)\citenamefont {He},
  \citenamefont {Ji}, \citenamefont {Wang}, \citenamefont {Qiu}, \citenamefont
  {Zhao}, \citenamefont {Ma}, \citenamefont {Huang}, \citenamefont {Chang},\
  and\ \citenamefont {Wu}}]{He19arxiv}%
  \BibitemOpen
  \bibfield  {author} {\bibinfo {author} {\bibfnamefont {Y.}~\bibnamefont
  {He}}, \bibinfo {author} {\bibfnamefont {L.}~\bibnamefont {Ji}}, \bibinfo
  {author} {\bibfnamefont {Y.}~\bibnamefont {Wang}}, \bibinfo {author}
  {\bibfnamefont {L.}~\bibnamefont {Qiu}}, \bibinfo {author} {\bibfnamefont
  {J.}~\bibnamefont {Zhao}}, \bibinfo {author} {\bibfnamefont {Y.}~\bibnamefont
  {Ma}}, \bibinfo {author} {\bibfnamefont {X.}~\bibnamefont {Huang}}, \bibinfo
  {author} {\bibfnamefont {D.~E.}\ \bibnamefont {Chang}},\ and\ \bibinfo
  {author} {\bibfnamefont {S.}~\bibnamefont {Wu}},\ }\bibfield  {title}
  {\bibinfo {title} {Geometric control of collective spontaneous emission},\
  }\href@noop {} {\bibfield  {journal} {\bibinfo  {journal} {arXiv:1910.02289}\
  } (\bibinfo {year} {2019})}\BibitemShut {NoStop}%
\bibitem [{\citenamefont {Stellmer}\ and\ \citenamefont
  {Schreck}(2014)}]{Stellmer14}%
  \BibitemOpen
  \bibfield  {author} {\bibinfo {author} {\bibfnamefont {S.}~\bibnamefont
  {Stellmer}}\ and\ \bibinfo {author} {\bibfnamefont {F.}~\bibnamefont
  {Schreck}},\ }\bibfield  {title} {\bibinfo {title} {Reservoir spectroscopy of
  $5s5p$ ${}^{3}p{}_{2}$--$5snd$ ${}^{3}{D}_{1,2,3}$ transitions in
  strontium},\ }\href {https://doi.org/10.1103/PhysRevA.90.022512} {\bibfield
  {journal} {\bibinfo  {journal} {Phys. Rev. A}\ }\textbf {\bibinfo {volume}
  {90}},\ \bibinfo {pages} {022512} (\bibinfo {year} {2014})}\BibitemShut
  {NoStop}%
\bibitem [{\citenamefont {Anderegg}\ \emph {et~al.}(2019)\citenamefont
  {Anderegg}, \citenamefont {Cheuk}, \citenamefont {Bao}, \citenamefont
  {Burchesky}, \citenamefont {Ketterle}, \citenamefont {Ni},\ and\
  \citenamefont {Doyle}}]{Anderegg19}%
  \BibitemOpen
  \bibfield  {author} {\bibinfo {author} {\bibfnamefont {L.}~\bibnamefont
  {Anderegg}}, \bibinfo {author} {\bibfnamefont {L.~W.}\ \bibnamefont {Cheuk}},
  \bibinfo {author} {\bibfnamefont {Y.}~\bibnamefont {Bao}}, \bibinfo {author}
  {\bibfnamefont {S.}~\bibnamefont {Burchesky}}, \bibinfo {author}
  {\bibfnamefont {W.}~\bibnamefont {Ketterle}}, \bibinfo {author}
  {\bibfnamefont {K.-K.}\ \bibnamefont {Ni}},\ and\ \bibinfo {author}
  {\bibfnamefont {J.~M.}\ \bibnamefont {Doyle}},\ }\bibfield  {title} {\bibinfo
  {title} {An optical tweezer array of ultracold molecules},\ }\href
  {https://doi.org/10.1126/science.aax1265} {\bibfield  {journal} {\bibinfo
  {journal} {Science}\ }\textbf {\bibinfo {volume} {365}},\ \bibinfo {pages}
  {1156} (\bibinfo {year} {2019})}\BibitemShut {NoStop}%
\bibitem [{\citenamefont {Sipahigil}\ \emph {et~al.}(2016)\citenamefont
  {Sipahigil}, \citenamefont {Evans}, \citenamefont {Sukachev}, \citenamefont
  {Burek}, \citenamefont {Borregaard}, \citenamefont {Bhaskar}, \citenamefont
  {Nguyen}, \citenamefont {Pacheco}, \citenamefont {Atikian}, \citenamefont
  {Meuwly}, \citenamefont {Camacho}, \citenamefont {Jelezko}, \citenamefont
  {Bielejec}, \citenamefont {Park}, \citenamefont {Lon{\v c}ar},\ and\
  \citenamefont {Lukin}}]{Sipahigil16}%
  \BibitemOpen
  \bibfield  {author} {\bibinfo {author} {\bibfnamefont {A.}~\bibnamefont
  {Sipahigil}}, \bibinfo {author} {\bibfnamefont {R.~E.}\ \bibnamefont
  {Evans}}, \bibinfo {author} {\bibfnamefont {D.~D.}\ \bibnamefont {Sukachev}},
  \bibinfo {author} {\bibfnamefont {M.~J.}\ \bibnamefont {Burek}}, \bibinfo
  {author} {\bibfnamefont {J.}~\bibnamefont {Borregaard}}, \bibinfo {author}
  {\bibfnamefont {M.~K.}\ \bibnamefont {Bhaskar}}, \bibinfo {author}
  {\bibfnamefont {C.~T.}\ \bibnamefont {Nguyen}}, \bibinfo {author}
  {\bibfnamefont {J.~L.}\ \bibnamefont {Pacheco}}, \bibinfo {author}
  {\bibfnamefont {H.~A.}\ \bibnamefont {Atikian}}, \bibinfo {author}
  {\bibfnamefont {C.}~\bibnamefont {Meuwly}}, \bibinfo {author} {\bibfnamefont
  {R.~M.}\ \bibnamefont {Camacho}}, \bibinfo {author} {\bibfnamefont
  {F.}~\bibnamefont {Jelezko}}, \bibinfo {author} {\bibfnamefont
  {E.}~\bibnamefont {Bielejec}}, \bibinfo {author} {\bibfnamefont
  {H.}~\bibnamefont {Park}}, \bibinfo {author} {\bibfnamefont {M.}~\bibnamefont
  {Lon{\v c}ar}},\ and\ \bibinfo {author} {\bibfnamefont {M.~D.}\ \bibnamefont
  {Lukin}},\ }\bibfield  {title} {\bibinfo {title} {An integrated diamond
  nanophotonics platform for quantum-optical networks},\ }\href
  {https://doi.org/10.1126/science.aah6875} {\bibfield  {journal} {\bibinfo
  {journal} {Science}\ }\textbf {\bibinfo {volume} {354}},\ \bibinfo {pages}
  {847} (\bibinfo {year} {2016})}\BibitemShut {NoStop}%
\bibitem [{\citenamefont {Kornher}\ \emph {et~al.}(2016)\citenamefont
  {Kornher}, \citenamefont {Xia}, \citenamefont {Kolesov}, \citenamefont
  {Kukharchyk}, \citenamefont {Reuter}, \citenamefont {Siyushev}, \citenamefont
  {St{\"o}hr}, \citenamefont {Schreck}, \citenamefont {Becker}, \citenamefont
  {Villa}, \citenamefont {Wieck},\ and\ \citenamefont {Wrachtrup}}]{Kornher16}%
  \BibitemOpen
  \bibfield  {author} {\bibinfo {author} {\bibfnamefont {T.}~\bibnamefont
  {Kornher}}, \bibinfo {author} {\bibfnamefont {K.}~\bibnamefont {Xia}},
  \bibinfo {author} {\bibfnamefont {R.}~\bibnamefont {Kolesov}}, \bibinfo
  {author} {\bibfnamefont {N.}~\bibnamefont {Kukharchyk}}, \bibinfo {author}
  {\bibfnamefont {R.}~\bibnamefont {Reuter}}, \bibinfo {author} {\bibfnamefont
  {P.}~\bibnamefont {Siyushev}}, \bibinfo {author} {\bibfnamefont
  {R.}~\bibnamefont {St{\"o}hr}}, \bibinfo {author} {\bibfnamefont
  {M.}~\bibnamefont {Schreck}}, \bibinfo {author} {\bibfnamefont {H.-W.}\
  \bibnamefont {Becker}}, \bibinfo {author} {\bibfnamefont {B.}~\bibnamefont
  {Villa}}, \bibinfo {author} {\bibfnamefont {A.~D.}\ \bibnamefont {Wieck}},\
  and\ \bibinfo {author} {\bibfnamefont {J.}~\bibnamefont {Wrachtrup}},\
  }\bibfield  {title} {\bibinfo {title} {Production yield of rare-earth ions
  implanted into an optical crystal},\ }\href
  {https://doi.org/10.1063/1.4941403} {\bibfield  {journal} {\bibinfo
  {journal} {Appl. Phys. Lett.}\ }\textbf {\bibinfo {volume} {108}},\ \bibinfo
  {pages} {053108} (\bibinfo {year} {2016})}\BibitemShut {NoStop}%
\bibitem [{\citenamefont {Jacob}\ \emph {et~al.}(2016)\citenamefont {Jacob},
  \citenamefont {Groot-Berning}, \citenamefont {Wolf}, \citenamefont {Ulm},
  \citenamefont {Couturier}, \citenamefont {Dawkins}, \citenamefont
  {Poschinger}, \citenamefont {Schmidt-Kaler},\ and\ \citenamefont
  {Singer}}]{Jacob16}%
  \BibitemOpen
  \bibfield  {author} {\bibinfo {author} {\bibfnamefont {G.}~\bibnamefont
  {Jacob}}, \bibinfo {author} {\bibfnamefont {K.}~\bibnamefont
  {Groot-Berning}}, \bibinfo {author} {\bibfnamefont {S.}~\bibnamefont {Wolf}},
  \bibinfo {author} {\bibfnamefont {S.}~\bibnamefont {Ulm}}, \bibinfo {author}
  {\bibfnamefont {L.}~\bibnamefont {Couturier}}, \bibinfo {author}
  {\bibfnamefont {S.~T.}\ \bibnamefont {Dawkins}}, \bibinfo {author}
  {\bibfnamefont {U.~G.}\ \bibnamefont {Poschinger}}, \bibinfo {author}
  {\bibfnamefont {F.}~\bibnamefont {Schmidt-Kaler}},\ and\ \bibinfo {author}
  {\bibfnamefont {K.}~\bibnamefont {Singer}},\ }\bibfield  {title} {\bibinfo
  {title} {Transmission microscopy with nanometer resolution using a
  deterministic single ion source},\ }\href
  {https://doi.org/10.1103/PhysRevLett.117.043001} {\bibfield  {journal}
  {\bibinfo  {journal} {Phys. Rev. Lett.}\ }\textbf {\bibinfo {volume} {117}},\
  \bibinfo {pages} {043001} (\bibinfo {year} {2016})}\BibitemShut {NoStop}%
\bibitem [{\citenamefont {Palacios-Berraquero}\ \emph
  {et~al.}(2017)\citenamefont {Palacios-Berraquero}, \citenamefont {Kara},
  \citenamefont {Montblanch}, \citenamefont {Barbone}, \citenamefont
  {Latawiec}, \citenamefont {Yoon}, \citenamefont {Ott}, \citenamefont
  {Loncar}, \citenamefont {Ferrari},\ and\ \citenamefont
  {Atat{\"u}re}}]{Palacios17}%
  \BibitemOpen
  \bibfield  {author} {\bibinfo {author} {\bibfnamefont {C.}~\bibnamefont
  {Palacios-Berraquero}}, \bibinfo {author} {\bibfnamefont {D.~M.}\
  \bibnamefont {Kara}}, \bibinfo {author} {\bibfnamefont {A.~R.~P.}\
  \bibnamefont {Montblanch}}, \bibinfo {author} {\bibfnamefont
  {M.}~\bibnamefont {Barbone}}, \bibinfo {author} {\bibfnamefont
  {P.}~\bibnamefont {Latawiec}}, \bibinfo {author} {\bibfnamefont
  {D.}~\bibnamefont {Yoon}}, \bibinfo {author} {\bibfnamefont {A.~K.}\
  \bibnamefont {Ott}}, \bibinfo {author} {\bibfnamefont {M.}~\bibnamefont
  {Loncar}}, \bibinfo {author} {\bibfnamefont {A.~C.}\ \bibnamefont
  {Ferrari}},\ and\ \bibinfo {author} {\bibfnamefont {M.}~\bibnamefont
  {Atat{\"u}re}},\ }\bibfield  {title} {\bibinfo {title} {Large-scale
  quantum-emitter arrays in atomically thin semiconductors},\ }\href
  {https://doi.org/10.1038/ncomms15093} {\bibfield  {journal} {\bibinfo
  {journal} {Nat. Commun.}\ }\textbf {\bibinfo {volume} {8}},\ \bibinfo {pages}
  {15093} (\bibinfo {year} {2017})}\BibitemShut {NoStop}%
\bibitem [{\citenamefont {Proscia}\ \emph {et~al.}(2018)\citenamefont
  {Proscia}, \citenamefont {Shotan}, \citenamefont {Jayakumar}, \citenamefont
  {Reddy}, \citenamefont {Cohen}, \citenamefont {Dollar}, \citenamefont
  {Alkauskas}, \citenamefont {Doherty}, \citenamefont {Meriles},\ and\
  \citenamefont {Menon}}]{Proscia18}%
  \BibitemOpen
  \bibfield  {author} {\bibinfo {author} {\bibfnamefont {N.~V.}\ \bibnamefont
  {Proscia}}, \bibinfo {author} {\bibfnamefont {Z.}~\bibnamefont {Shotan}},
  \bibinfo {author} {\bibfnamefont {H.}~\bibnamefont {Jayakumar}}, \bibinfo
  {author} {\bibfnamefont {P.}~\bibnamefont {Reddy}}, \bibinfo {author}
  {\bibfnamefont {C.}~\bibnamefont {Cohen}}, \bibinfo {author} {\bibfnamefont
  {M.}~\bibnamefont {Dollar}}, \bibinfo {author} {\bibfnamefont
  {A.}~\bibnamefont {Alkauskas}}, \bibinfo {author} {\bibfnamefont
  {M.}~\bibnamefont {Doherty}}, \bibinfo {author} {\bibfnamefont {C.~A.}\
  \bibnamefont {Meriles}},\ and\ \bibinfo {author} {\bibfnamefont {V.~M.}\
  \bibnamefont {Menon}},\ }\bibfield  {title} {\bibinfo {title}
  {Near-deterministic activation of room-temperature quantum emitters in
  hexagonal boron nitride},\ }\href {https://doi.org/10.1364/OPTICA.5.001128}
  {\bibfield  {journal} {\bibinfo  {journal} {Optica}\ }\textbf {\bibinfo
  {volume} {5}},\ \bibinfo {pages} {1128} (\bibinfo {year} {2018})}\BibitemShut
  {NoStop}%
\bibitem [{\citenamefont {Domokos}\ and\ \citenamefont
  {Ritsch}(2002)}]{Domokos02}%
  \BibitemOpen
  \bibfield  {author} {\bibinfo {author} {\bibfnamefont {P.}~\bibnamefont
  {Domokos}}\ and\ \bibinfo {author} {\bibfnamefont {H.}~\bibnamefont
  {Ritsch}},\ }\bibfield  {title} {\bibinfo {title} {Collective cooling and
  self-organization of atoms in a cavity},\ }\href
  {https://doi.org/10.1103/PhysRevLett.89.253003} {\bibfield  {journal}
  {\bibinfo  {journal} {Phys. Rev. Lett.}\ }\textbf {\bibinfo {volume} {89}},\
  \bibinfo {pages} {253003} (\bibinfo {year} {2002})}\BibitemShut {NoStop}%
\bibitem [{\citenamefont {Black}\ \emph {et~al.}(2003)\citenamefont {Black},
  \citenamefont {Chan},\ and\ \citenamefont {Vuleti
  \'{c}}}]{Black03}%
  \BibitemOpen
  \bibfield  {author} {\bibinfo {author} {\bibfnamefont {A.~T.}\ \bibnamefont
  {Black}}, \bibinfo {author} {\bibfnamefont {H.~W.}\ \bibnamefont {Chan}},\
  and\ \bibinfo {author} {\bibfnamefont {V.}~\bibnamefont
  {Vuleti\ifmmode~\acute{c}\else \'{c}\fi{}}},\ }\bibfield  {title} {\bibinfo
  {title} {Observation of collective friction forces due to spatial
  self-organization of atoms: From Rayleigh to Bragg scattering},\ }\href
  {https://doi.org/10.1103/PhysRevLett.91.203001} {\bibfield  {journal}
  {\bibinfo  {journal} {Phys. Rev. Lett.}\ }\textbf {\bibinfo {volume} {91}},\
  \bibinfo {pages} {203001} (\bibinfo {year} {2003})}\BibitemShut {NoStop}%
\bibitem [{\citenamefont {Chang}\ \emph {et~al.}(2013)\citenamefont {Chang},
  \citenamefont {Cirac},\ and\ \citenamefont {Kimble}}]{Chang13}%
  \BibitemOpen
  \bibfield  {author} {\bibinfo {author} {\bibfnamefont {D.~E.}\ \bibnamefont
  {Chang}}, \bibinfo {author} {\bibfnamefont {J.~I.}\ \bibnamefont {Cirac}},\
  and\ \bibinfo {author} {\bibfnamefont {H.~J.}\ \bibnamefont {Kimble}},\
  }\bibfield  {title} {\bibinfo {title} {Self-organization of atoms along a
  nanophotonic waveguide},\ }\href
  {https://doi.org/10.1103/PhysRevLett.110.113606} {\bibfield  {journal}
  {\bibinfo  {journal} {Phys. Rev. Lett.}\ }\textbf {\bibinfo {volume} {110}},\
  \bibinfo {pages} {113606} (\bibinfo {year} {2013})}\BibitemShut {NoStop}%
\bibitem [{\citenamefont {Cantu}\ \emph {et~al.}(2020)\citenamefont {Cantu},
  \citenamefont {Venkatramani}, \citenamefont {Xu}, \citenamefont {Zhou},
  \citenamefont {Jelenkovi{\'c}}, \citenamefont {Lukin},\ and\ \citenamefont
  {Vuleti{\'c}}}]{Cantu20}%
  \BibitemOpen
  \bibfield  {author} {\bibinfo {author} {\bibfnamefont {S.~H.}\ \bibnamefont
  {Cantu}}, \bibinfo {author} {\bibfnamefont {A.~V.}\ \bibnamefont
  {Venkatramani}}, \bibinfo {author} {\bibfnamefont {W.}~\bibnamefont {Xu}},
  \bibinfo {author} {\bibfnamefont {L.}~\bibnamefont {Zhou}}, \bibinfo {author}
  {\bibfnamefont {B.}~\bibnamefont {Jelenkovi{\'c}}}, \bibinfo {author}
  {\bibfnamefont {M.~D.}\ \bibnamefont {Lukin}},\ and\ \bibinfo {author}
  {\bibfnamefont {V.}~\bibnamefont {Vuleti{\'c}}},\ }\bibfield  {title}
  {\bibinfo {title} {Repulsive photons in a quantum nonlinear medium},\
  }\bibfield  {journal} {\bibinfo  {journal} {Nat. Phys.}\ }\href
  {https://doi.org/10.1038/s41567-020-0917-6}
  (\bibinfo {year} {2020})\BibitemShut {NoStop}%
\bibitem [{\citenamefont {Wiseman}\ and\ \citenamefont
  {Milburn}(1994)}]{Wiseman94}%
  \BibitemOpen
  \bibfield  {author} {\bibinfo {author} {\bibfnamefont {H.~M.}\ \bibnamefont
  {Wiseman}}\ and\ \bibinfo {author} {\bibfnamefont {G.~J.}\ \bibnamefont
  {Milburn}},\ }\bibfield  {title} {\bibinfo {title} {Squeezing via feedback},\
  }\href {https://doi.org/10.1103/PhysRevA.49.1350} {\bibfield  {journal}
  {\bibinfo  {journal} {Phys. Rev. A}\ }\textbf {\bibinfo {volume} {49}},\
  \bibinfo {pages} {1350} (\bibinfo {year} {1994})}\BibitemShut {NoStop}%
\bibitem [{\citenamefont {Klimov}\ and\ \citenamefont
  {Ducloy}(2004)}]{Klimov04}%
  \BibitemOpen
  \bibfield  {author} {\bibinfo {author} {\bibfnamefont {V.~V.}\ \bibnamefont
  {Klimov}}\ and\ \bibinfo {author} {\bibfnamefont {M.}~\bibnamefont
  {Ducloy}},\ }\bibfield  {title} {\bibinfo {title} {Spontaneous emission rate
  of an excited atom placed near a nanofiber},\ }\href
  {https://doi.org/10.1103/PhysRevA.69.013812} {\bibfield  {journal} {\bibinfo
  {journal} {Phys. Rev. A}\ }\textbf {\bibinfo {volume} {69}},\ \bibinfo
  {pages} {013812} (\bibinfo {year} {2004})}\BibitemShut {NoStop}%
\end{thebibliography}
\end{document}